
\magnification\magstep1
\parskip 5pt
\vsize 8.9 true in
\hsize 6.4 true in
\magnification\magstep1
\newdimen\itemindent \itemindent=32pt
\def\textindent#1{\parindent=\itemindent\let\par=\resetpar%
\indent\llap{#1\enspace}\ignorespaces}

\let\oldpar=\par
\def\resetpar{\oldpar\parindent=25pt\let\par=\oldpar}

\font\ninerm=cmr9 \font\ninesy=cmsy9
\font\eightrm=cmr8 \font\sixrm=cmr6
\font\eighti=cmmi8 \font\sixi=cmmi6
\font\eightsy=cmsy8 \font\sixsy=cmsy6
\font\eightbf=cmbx8 \font\sixbf=cmbx6
\font\eightit=cmti8
\def\eightpoint{\def\rm{\fam0\eightrm}
  \textfont0=\eightrm \scriptfont0=\sixrm \scriptscriptfont0=\fiverm
  \textfont1=\eighti  \scriptfont1=\sixi  \scriptscriptfont1=\fivei
  \textfont2=\eightsy \scriptfont2=\sixsy \scriptscriptfont2=\fivesy
  \textfont3=\tenex   \scriptfont3=\tenex \scriptscriptfont3=\tenex
  \textfont\itfam=\eightit  \def\it{\fam\itfam\eightit}%
  \textfont\bffam=\eightbf  \scriptfont\bffam=\sixbf
  \scriptscriptfont\bffam=\fivebf  \def\bf{\fam\bffam\eightbf}%
  \normalbaselineskip=9pt
  \setbox\strutbox=\hbox{\vrule height7pt depth2pt width0pt}%
  \let\big=\eightbig  \normalbaselines\rm}
\catcode`@=11 %
\def\eightbig#1{{\hbox{$\textfont0=\ninerm\textfont2=\ninesy
  \left#1\vbox to6.5pt{}\right.\n@space$}}}
\def\vfootnote#1{\insert\footins\bgroup\eightpoint
  \interlinepenalty=\interfootnotelinepenalty
  \splittopskip=\ht\strutbox %
  \splitmaxdepth=\dp\strutbox %
  \leftskip=0pt \rightskip=0pt \spaceskip=0pt \xspaceskip=0pt
  \textindent{#1}\footstrut\futurelet\next\fo@t}
\catcode`@=12 %

\parskip 5pt
\def \de{\delta}
\def \si{\sigma}
\def \Ga{\Gamma}

\def \bsi{\bar \sigma}
\def \ga{\gamma}
\def \hD{{\hat {D }}}

\def \hga{{\hat {\gamma}}}
\def \nab{\nabla}
\def \hnab{{\hat \nabla}}
\def \Del{\Delta}
\def \pr{\partial}
\def \hx{{\hat x}}
\def \bhx{{\hat {\bf x}}}

\def \hsi{{\hat \sigma}}
\def \D{{\cal D}}

\def \P{{\cal P}}
\def \V{{\cal V }}
\def \T{{\cal T}}
\def \Y{{\cal Y}}
\def \Z{{\cal Z}}
\def \hDel{{\hat {\Del}}}
\def \hg{{\hat{g}}}

\def \hB{{\hat {\cal B}}}

\def \d2{{\textstyle {1 \over 2}} d}
\def \K{{\kappa}}
\def \M{{\cal M}}

\def \G{{\cal G}}
\def \Tr{{\rm Tr \, }}
\def \tr{{\rm tr \, }}
\def \bG{{\bar {G}}}

\def \O{{\rm O}}
\def \vep{\varepsilon}
\def \half{{\textstyle {1 \over 2}}}
\def \quar{{\textstyle {1 \over 4}}}
\def \eight{{\textstyle {1 \over 8}}}
\def \tw{{\textstyle {1 \over 12}}}
\def \thir{{\textstyle {1 \over 3}}}
\def \six{{\textstyle {1 \over 6}}}
\def \ts{ \textstyle}
\def \24{{\textstyle {1\over 24}}}
\def \hR{{ \hat R}}

\def \bx{{\bf x}}
\def \vphi{{\varphi}}

\def \P{{\cal P}}
\def \hI{{\hat I}}
\def \hh{{1\over 2}}

\def \te{{\tilde e}}
\def \tf{{\tilde f}}
\def \tg{{\tilde g}}
\def \th{{\tilde h}}

\def \vphi{\varphi}

\def \hI{{\hat I}}
\def \hh{{1\over 2}}
\def \hphi { {\hat {\varphi}}}
\def \vphi { {\varphi}}
\def \hbx  {{\hat {\bf x}}}
\def \hS { {\hat {S}}}
\def \hB { {\hat {\cal {B}}}}
\def \eight{{\ts {1\over 8}}}

\def \H{{\cal H}}
\def \C{{\cal C}}
\def \c{{\rm c}}
\def \K{{\cal {K}}}
\def \B{{\cal B}}
\def \X{{\cal X}}
\def \ep{\epsilon}
\def \K{{\cal K}}

\def \hphi{{\hat {\varphi}}}
\font \bigbf=cmbx10 scaled \magstep1
\rightline{DAMTP/92-31}
\vskip 2truecm
\centerline {\bigbf Quantum Field Theories on Manifolds with Curved
Boundaries:}
\vskip 5pt
\centerline {\bigbf Scalar Fields}
\vskip 2.0 true cm
\centerline {D.M. McAvity and H. Osborn}
\vskip 10pt
\centerline {Department of Applied Mathematics and Theoretical Physics,}
\centerline {Silver Street, Cambridge, UK}
\vskip 2.0 true cm

A framework allowing for perturbative calculations to be carried out
for quantum field theories with arbitrary
smoothly curved boundaries is described. It is based on an expansion of the
Green function for second order differential operators valid in the
neighbourhood of the boundary and which is obtained from a corresponding
expansion of of the associated heat kernel derived earlier for arbitrary mixed
Dirichlet and Neumann boundary conditions. The first few leading terms in
the expansion are sufficient to calculate all additional divergences present
in a perturbative loop expansion as a consequence of the presence of the
boundary. The method is applied to a general renormalisable scalar field
theory in four dimensions using dimensional regularisation to two loops
and expanding about arbitrary background fields. Detailed results are also
specialised to an $O(n)$ symmetric model with a single coupling constant.
Extra boundary terms are introduced into the action which give rise to either
Dirichlet or generalised Neumann boundary conditions for the quantum fields.
For plane boundaries the resulting renormalisation group functions are in
accord with earlier results but here the additional terms depending on the
extrinsic curvature of the boundary are found. Various consistency relations
are also checked and the implications of conformal invariance at the critical
point where the $\beta$ function vanishes are also derived. For a general
scalar field  theory, where the fields $\phi$ attain specified values $\hphi$
on the boundary, the local Scr\"odinger equation for the wave functional
defined by the functional integral
under deformations of the boundary is also verified
to two loops. The perturbative expansion for the wave functional is defined by
expansion around the solution of the classical field equations satisfying the
required boundary values and the counterterms necessary to derive a finite
hamiltonian operator, which includes a functional laplace operator on the
fields $\hphi$, are found to the order considered. Consistency of the local
Schr\"odinger equation with the renormalisation group to all orders in
perturbation theory is also discussed.
\vfill\eject

\leftline{\bf 1 Introduction}

Although quantum field theory has an enormous literature, papers
devoted to the properties of and perturbative calculations for quantum fields
in the vicinity of a boundary are relatively sparse. Nevertheless considering
quantum field theory on a manifold with boundary is not without interest for
the following reasons:
\item{(i)} The Casimir energy [1,2,3] is obviously an effect which depends on
the presence of boundaries and exemplifies the crucial zero point
energy of quantum fields. However explicit calculation for non trivial
geometries and for interactions are difficult and the usual
renormalisation framework requires extension.
\item{(ii)} In statistical physics there are additional critical indices
associated with phase transitions for physical observables measured near
a boundary [4]. These
may be calculated in the framework of the $\vep$ expansion of the same
quantum field theory used to obtain the bulk exponents but in the presence
of a boundary [5].
\item{(iii)} In bag models, which may possibly be regarded as an
approximation to QCD with some phenomenological validity, it is necessary to
consider quantum fields inside cavities with suitable boundary conditions [6].
Various perturbative calculations in bag models have been undertaken based on
using a multiple scattering expansion for the propagator in the presence of a
boundary [7].
\item {(iv)} Open strings correspond to conformal field theories with
additional operators attached to the boundary of the two dimensional
world sheet [8].
\item{(v)} Cardy has shown how additional information on the structure of two
dimensional conformal field theory may be obtained by considering the usual
machinery of operator product expansions with boundary operators [9].
\item{(vi)} An alternative formulation of quantum field theory
to the conventional approach may be given
in terms of a Schr\"odinger like representation with wave functionals
$\Psi ({\hat \vphi})$ defined by the functional integral over quantum fields
on a manifold $\M$ which attain a specified value ${\hat \vphi}$ on the
boundary $\pr \M$. Such wave functionals obey formally a Schr\"odinger like
equation where the time differentiation corresponds to deformations of the
boundary [10,11]. However the second order functional derivatives in the
Hamiltonian operator involve singularities which require careful discussion.
The solutions of the functional Schr\"odinger equation provided by the
functional integral over fields on manifolds with a boundary are essentially
identical to formal constructions of wave functionals satisfying the
Wheeler-DeWitt equation in quantum gravity [12] which have been extensively
explored recently.

However, in the presence of a boundary the usual efficient calculational
techniques based on momentum space representations of the propagators are no
longer valid, although in the case of plane boundaries there are possible
extensions [5]. Here we wish to discuss quantum field theory on curved space
with an arbitrary smoothly curved boundary. For simplicity at this stage we
restrict our attention to a general renormalisable scalar field theory on a
Euclidean space with a positive definite metric in four dimensions using
dimensional regularisation, developing further the treatment some time ago
of Symanzik [10]. However, our methods should be feasible for more
interesting field theories,
such as $\sigma$ models in two dimensions which are relevant for strings.

In order to carry out calculations for the short distance behaviour of
amplitudes for a quantum field theory in the neighbourhood of a boundary
it is crucial to be able to determine besides the structure of the
leading singularities at coincident points $x'\to x$
of the Green function $G_\Del (x,x')$ for the
elliptic operator $\Del$, defined when the action is expanded to quadratic
order around some background, also the behaviour as $x,x'$ approach
the boundary. Recently we have developed [13] an alternative to the multiple
scattering expansion based on an analysis of the heat kernel
$\G_\Del(x,x';\tau)$ corresponding to $e^{-\tau \Del}$ as $\tau \to 0$ in
the neighbourhood of a boundary for either Dirichlet and Neumann boundary
conditions. The results depend on an extension of the usual DeWitt ansatz [14]
to include an effectively semi-classical expansion about geodesics which
undergo reflection at the boundary. The details of this expansion are
summarised in appendix A. Throughout we maintain manifest
reparameterisation invariance with respect to the coordinates ${\hat x}^i$ for
$\pr \M$ and for the purposes of calculation, in the neighbourhood of $\pr \M$,
adopt the coordinates $x^\mu= (x^i,y)$ where $y$
is the geodesic distance from $x^\mu \in \M$ to $x^i = {\hat x}^i\in
\pr \M$ defined by the geodesic tangent to the unit normal $n^\mu$ at $\hx^i$.
In these coordinates the metric $g_{\mu \nu}$ on $\M$ is given by
$$ ds^2 = \ga_{ij} (\bx,y) dx^i dx^j + dy^2~, ~~~~~n^\mu = ({\bf 0},1) ~,
\eqno(1.1)
$$
where $\ga_{ij}(\bx,0) = \hga_{ij}(\bhx)$ is thus the induced metric on
$\pr M$.

We assume that the operator $\Del$, acting on sections of some vector bundle
over $\M$, is of the form
$$ \Del = -D^2 + X~,~~~~ D^2 = {1\over \sqrt{g}} D_\mu g^{\mu \nu} \sqrt{g}
D_\nu ~,~ D_\mu = \pr_\mu + A_\mu ~,
\eqno(1.2)
$$
for $X$, $A_\mu$ matrix valued fields on $\M$. $A_\mu$ can be regarded as
an external background gauge field but may also be taken to include any
appropriate spin connection. Assuming also that $\Del$ is a symmetric
operator acting on vector fields $\xi(x)$ over $\M$ the general form of the
boundary conditions considered here, which is sufficient for most field
theoretic applications, involve $\xi$ and its normal derivative on $\pr \M$
in the form
$$
\bigl ( \P n^\mu D_\mu + \psi \bigl ) \xi \bigl |_{\pr \M} = 0~,~~
(1-\P) \xi \bigl |_{\pr \M} = 0 ~, ~~~
\P^2 = \P ~,~~ \P\psi= \psi \P = \psi ~,
\eqno(1.3) $$
for $\P(\bhx )$ a projection operator and $\psi(\bhx )$ a matrix valued
function
on $\pr \M$. Clearly from (1.3)
taking $\P=0$ corresponds to pure Dirichlet boundary conditions while
$\P=1$ is the Neumann case. With these boundary conditions the Green
function $G_\Del$ is defined by
$$  \Del_x G_\Del (x,x') = \de^d(x,x') \equiv {1\over \sqrt{g}}
\de^d (x-x') ~.
\eqno (1.4) $$

In the interior of $\M$ the DeWitt asymptotic expansion of the heat kernel
$\G_\Del$ determines a corresponding expansion of $G_\Del $ which for
discussing renormalisable field theories in four dimensions may be restricted
to the form [15]
$$ G_\Del = \sum^2_{n=0} G_n a_n^\Del + {\bar {G}}_\Del ~,
\eqno(1.5)
$$
for $a_n^\Del (x,x')$  the Seeley-DeWitt coefficients, nonsingular for $x
\approx x'$, and the singular part of $G_\Del$ is given entirely in terms of
$G_n(x,x')$ where
$$\eqalign {
G_0= {}&{\Ga(\half d -1 )\over 4 \pi^{\hh d}} {\Del^\hh \over (2 \si )^{
\hh d-1}}~, \cr
G_1= {}&{\Ga(\half d -2 )\over 16 \pi^{\hh d}}
{\Del^\hh \over (2 \si )^{\hh d-2}}
+ {\mu ^{-\vep} \over \vep} {\Del^\hh \over 8 \pi^2} ~, \cr
G_2 = {}&{\Ga(\half d -3 )\over 64 \pi^{\hh d}} {\Del^\hh \over (2 \si )^{
\hh d-3}} - {\mu ^{-\vep} \over \vep} {\Del^\hh \over 16 \pi^2} \,
\si ~, \cr }
\eqno(1.6)
$$
for $\si(x,x')$ the geodetic interval given by $\half \int_0^1\, ds\, {\dot
x} (s)^2$ for $x^\mu(s)$ the geodesic path from $x^\mu(0) = x'$ to $x^\mu(1)=
x$. $\Del^\hh (x,x')$ is a nonsingular symmetric biscalar which may be
absorbed in $a_n^\Del$ but the expressions (1.6) are convenient for
correspondence with previous results. The representation provided by (1.5,6)
is valid for an arbitrary dimension $d$ of $\M$, with $\pr \M$ of dimension
$d-1$. However in the expressions for $G_1, \, G_2$
the pole in the first term of each on the r.h.s. of (1.6) at $\vep =
4-d \to 0$ has been subtracted so that there is a well defined limit for
$d=4$, this ensures a convenient form for ${\bar G}_\Delta$ which is regular
for $\vep \to 0$ as well as $x\approx x'$.
The subtraction also represents the removal of infra-red divergences of the
corresponding flat space Fourier transforms $k^{-4}, k^{-6}$ when $k \to 0$.
In these pole terms an arbitrary mass scale $\mu$ has been introduced
for dimensional consistency, as usual in dimensional regularisation.
$G_\Delta$ is independent of $\mu$, any variation in $\mu$ may be compensated
by a corresponding change in ${\bar G}_\Delta$ in (1.5).

In discussing a renormalisable field theory it is also
sufficient to resort to just the first few terms in a covariant Taylor
expansion of $a_n(x,x')$ about $x'=x$,
$$\eqalign {
a^\Del_0 = {}& I ~, ~~~~ \Del^\hh \approx 1 + {\ts {1\over 12}} R_{\mu \nu}
\si^\mu \si^\nu -
{\ts {1\over 24}} \nab_\rho R_{\mu \nu} \si^\rho \si^\mu \si^\nu ~, \cr
a^\Del_1 \approx {}& \bigl ({\ts {1\over 6}} R - X \bigl ) I -
\bigl ( {\ts {1\over 12}}  \pr_{\mu} R -
{\ts {1\over 6}} D^{\nu} F_{\mu \nu} -{\ts {1\over 2}} D_{\mu} X \bigl )
\si^\mu I \cr
& + \bigl ( -{\ts {1\over 6}} D_\mu D_\nu X + {\ts {1\over 12}} F_{\mu
\si} F_\nu {}^\si - {\ts {1\over 12}} D_\mu D^\si F_{\nu \si} +
{\ts {1\over 40}} \nab_\mu \nab_\nu R + {\ts {1\over 120}} \nab^2 R_{\mu
\nu} \cr
& ~+ {\ts {1\over 180}}( R^{\si \rho} R_{\mu \si \nu \rho} + R_{\mu \si \rho
\tau} R_\nu {}^{\si \rho \tau} - 2 R_{\mu \si} R_\nu {}^\si ) \bigl )
\si^\mu \si^\nu I ~, \cr
a^\Del_2 \approx{}&
\bigl ( {\ts {1\over 2}} ( {\ts {1\over 6}} R - X )^2 -
{\ts {1\over 6 }} D^2 X + {\ts {1\over 12}} F_{\mu \nu} F^{\mu \nu} \cr
& + {\ts {1\over 180}} ( R_{\mu \nu \sigma \rho} R^{\mu \nu \sigma \rho}
- R_{\mu \nu} R^{\mu \nu} ) + {\ts {1\over 30}} \nab^2 R \bigl ) I
\;, \cr }
\eqno (1.7)
$$
where $\si^\mu(x,x') = g^{\mu \nu} (x) \pr_\nu \si(x,x') $ and $I(x,x')$
is the matrix giving parallel transport along the geodesic from
$x'$ to $x$ defined by $\si^\mu D_\mu I = 0$, $I(x,x) =1$. $F_{\mu \nu}$ is
the usual field strength formed from $A_\mu$. With the
decomposition of $G_\Del$ given in (1.5) extensive calculations at two and
more loops have been undertaken using formulae for the singular parts of the
products of $G_n$ such as
$$
G_0^2 \sim {} {\mu^{-\vep} \over 8\pi^2\vep} \, \delta^d ~,~~~
G_0^3 \sim {} {\mu^{-2\vep} \over (16\pi^2)^2} \, {1\over \vep}\half (\nab^2 +
\six R ) \delta^d ~, ~~
G_0^2 G_1 \sim {} {\mu^{-2\vep} \over (16\pi^2)^2} \, \Bigl (
{2\over \vep^2} + {1\over \vep} \Bigl ) \delta^d ~.
\eqno(1.8) $$
Similar relations involving derivatives may be found in ref. (15), in general
at $\ell$ loops dimensional consistency requires a factor $\mu^{-\ell \vep}$.

In the neighbourhood of $\pr \M$ the expansion (1.6) is no longer
sufficient for revealing all singular contributions in amplitudes at one or
more loops. If $\hsi (\bhx,\bhx')$  is the geodetic interval on $\pr \M$, as
determined by the metric $\hga_{ij} (\bhx)$, then, for coordinates
corresponding to the metric (1.1), assuming
$$
\hsi^i ,\, y,\, y' = \O(\epsilon)~,~~~~\hsi = \half \hga_{ij} \hsi^i \hsi^j
= \O(\epsilon^2) ~,
\eqno(1.9)
$$
and with the choice of gauge
$$ A_\mu = ( {\bf A},0) ~,
\eqno (1.10) $$
$G_\Del$ may now expanded as
$$
G_\Del = \sum^3_{n=0} \bigl ( G_n^B +{\bar G}_n^B \bigl ) {} +
{\bar {G}}_\Del^B ~, ~~~~~ G_n^B , \, {\bar G}_n^B = \O ( \epsilon^{n+2-d}) ~,
\eqno (1.11)$$
with ${\bar G}_\Del^B $ regular both for $x'\to x$ and also on $\pr \M$ for
$d=4$. $G_n^B $ correspond to the singular part of $G_\Del$ as $x'\to x$ while
${\bar G}_n^B $, which may be determined from the heat kernel expansion
described in appendix A, are non singular when $x'=x$ but are
necessary in order to satisfy the boundary conditions and lead to additional
divergences as the boundary is approached. If we define
$$ \eqalign {
G_r(u)= {}& {\Ga(\half d -r )\over 4 \pi^{\hh d}} \bigl ( u^2 + 2\hsi
\bigl )^{r- \hh d} ~, ~~~~ G_{r,s} (u) = \int_0^\infty dz\, z^s \,
G_r (z+ u) ~, \cr
& G_{r,s+1} (u) + u G_{r,s} (u) = \half s G_{r+1, s-1} (u) +
\half \delta_{s0} G_{r+1} (u) ~, \cr}
\eqno (1.12) $$
then explicitly for $n=0,1$ we may obtain for general $\Delta$ and boundary
conditions as in (1.2,3)
$$ \eqalign {
G_0^B = {}& G_1(v) \, \hI ~,~~~~ G_1^B = G_0(v) u\, K_{ij} \hsi^i \hsi^j \,
\hI ~ , ~~~~ v = y-y' ~, \cr
{\bar G}_0^B = {}& G_1(u) \P_- \, \hI ~, ~~~~~~~~~\P_- = 2\P-1~, ~~~~~~~~~
u = y+y' ~, \cr
{\bar G}_1^B = {}& G_{1,0}(u) \, 2\psi \, \hI + G_0(u) u\, K_{ij} \hsi^i
\hsi^j \P_- \hI \cr
& + 2yy' \bigl ( G_{0,0} (u) \, K - 2 G_{-1,0} (u) \, K_{ij} \hsi^i \hsi^j
\bigl ) \P_- \hI \cr
& + \bigl ( G_{1,0} (u) - G_{0,2} (u) \bigl ) K \P \, \hI +
2 G_{-1,2}(u) K_{ij} \hsi^i \hsi^j \, \P \hI \cr
& - G_1 (u) \, 2 \P \hsi^i \hD_i \P \, \hI + G_{0,0} (u)
\bigl ( 4y' \P \hsi^i \hD_i \P  - 4y \, \hsi^i \hD_i \P \, \P \bigl ) \hI ~,
\cr}
\eqno(1.13) $$
where the coefficients, $\P$, $\psi$, $K_{ij}$, are evaluated at $\hx$.
$\hI ( \bhx,\bhx')$, defined similarly to $I(x,x')$ earlier, is the matrix
corresponding to parallel transport along the geodesic in $\pr \M$ from $\bhx'$
to $\bhx$ for the connection ${\hat{\bf A}}={\bf A}|_{y=0}$ and $\hD_i$ denotes
the corresponding covariant derivative on the boundary, $\hD_i \P = \pr_i \P +
[{\hat A}_i , \P]$. $K_{ij}$ is here the extrinsic curvature of the boundary,
with coordinates given by (1.1) $K_{ij} = - \half \pr_y \ga_{ij} |_{y=0}$,
{}~ $K= \hga^{ij} K_{ij}$.
Clearly the presence of a curved boundary, so that $K_{ij}$ is non zero, is a
significant complication. For $\hsi=0$ the integrals in the definition of
$G_{r,s}$ in (1.12) are easily evaluated but in general the form given is
sufficient for further calculations. To obtain (1.13) the terms
$\sum_n G_n^B$ in (1.11) may be derived by expanding the singular
contributions in (1.5) using (1.6) and (1.7) and the remaining terms $\sum_n
{\bar G}_n^B$ can then be verified to be in accord with the essential Green
function equation (1.4) and the boundary conditions by expanding about $y=0$,
$\Delta = \Del_0 + \Del_1 + \dots , ~ \Del_n = \O (\epsilon^{n-2})$,
regarding $\pr_y , \, {\hat D}_i$ as $\O (\epsilon^{-1})$ in addition to
$y=\O (\epsilon)$ as in (1.9), and using results such as
$$ \eqalign {
\pr_i G_r (u) = {}& - 2 G_{r-1}(u) \hsi_i ~,~~~\pr_y G_r (u) = - 2 G_{r-1}
(u)u ~, \cr
\pr_y G_{r,s} (u) = {}& - s G_{r,s-1} (u) + \delta_{s0} G_{r} (u) ~, \cr
\Del_0 G_r(u) = {}& 2\bigl ( 2(r-1) + f \bigl ) G_{r-1} (u) ~,~~~~
f= \hnab^2 \hsi - d +1 ~, \cr }
\eqno (1.14) $$
where $\Del_0 = - {\hat D}^2 - \pr_y^{\, 2} $ and $\hnab_i \hsi^j = \delta_i
{}^{\! j} - {\ts {1\over 3}} {\hat R}_{ki\ell}{}^j \hsi^k \hsi^\ell + \dots$,
for ${\hat R}_{ijk\ell}$ the Riemann tensor formed from the induced
metric on $\pr \M$. The complete expression for the $n=2$ terms in (1.11) may
be found from the results obtained elsewhere for the heat kernel to this
order, in appendix A, in (A.10) we give the results for $n=2$ involving $X$
and $\psi$. These involve functions such as $G_2(u)$ where it is
necessary to subtract a pole term at $\vep=0$ as in (1.6).

Instead of the result given by (1.13) an alternate equivalent form for
${\bar G}_1^B$ is given by
$$ \eqalign {
{\bar G}_1^B = {}& G_0(u) u \Bigl ( 1 + {yy'\over \hsi} \Bigl ) K_{ij} \hsi^i
\hsi^j \P_- \hI \cr
& + \Bigl ( 2yy' G_{0,0} (u) \P_- + {1\over d-1} G_{1,0} (u)\P - G_{0,2}(u)\P
\Bigl ) \Bigl ( K - (d-1) {K_{ij} \hsi^i \hsi^j \over 2\hsi} \Bigl ) \hI \cr
& + G_{1,0}(u) \Bigl (  2\psi + {d-2\over d-1} \, K \P \Bigl ) \hI \cr
& - G_1 (u) \, 2 \P \hsi^i \hD_i \P \, \hI + G_{0,0} (u)
\bigl ( 4y' \P \hsi^i \hD_i \P  - 4y \, \hsi^i \hD_i \P \, \P \bigl ) \hI ~.
\cr}
\eqno (1.15) $$
This expression is in accord with the result that for a sphere of
radius $a$ when $K_{ij} = {\hat \gamma}_{ij}/a$ and also in the Neumann case
for $\psi=- \half (d-2)/a$ it is possible to find the Green function for
$-\pr^2$ in terms of the elementary flat space result for no boundary by the
method of images. In this particular case (1.15) is independent of the special
functions $G_{r,s}$.

In any amplitude with superficial degree of divergence $D$ then expanding
$G_\Del$ associated with any internal line reduces $D$ to $D-2n$ for
the contribution of the $G_n \,a_n^\Del$ term in (1.5)
and to $D-n$ for the terms $G_n^B ,\, {\bar G}_n^B $ in (1.11). For vacuum
diagrams for renormalisable theories in four dimensions $D=4$ while for
divergences associated with the boundary $D=3$.

At one loop amplitudes are determined by the functional
determinant of $\Del$ and standard methods give
$$ - {\rm {ln \, det}} \, \Del = \int_0^\infty {d\tau \over \tau} \, {\rm Tr}
\bigl ( e^{-\tau \Del} \bigl ) ~,
\eqno (1.16) $$
and the singular part for $d\to 4$ is then, using previous results [13,16] for
the asymptotic form of ${\rm Tr} \bigl ( e^{-\tau \Del} \bigl )$ as $\tau \to
0$,
$$ \eqalignno {
- {\rm {ln \, det}} \, \Del \sim {2\over \vep} \, {\mu^{-\vep}\over 16\pi^2}
\biggl \{ & \int_{\M} \! dv \, {\rm tr}
\bigl ( {\ts {1\over 2}} ( {\ts {1\over 6}} R - X )^2
+ {\ts {1\over 12}} F_{\mu \nu} F^{\mu \nu}
+ {\ts {1\over 180}} ( R_{\mu \nu \sigma \rho} R^{\mu \nu \sigma \rho}
- R_{\mu \nu} R^{\mu \nu} ) \bigl ) \cr
+ {}& \int _{\pr \M} \! \! dS \, {\rm tr} \Bigl (
\half \P_- (\six \pr_n R -\pr_n X ) + (2\psi +\thir K) (\six\hR - X) \cr
& + {\ts {2 \over 3}} \psi R_{nn} + {\ts {7 \over 15}} \psi
K_{ij}K^{ij} + {\ts {1 \over 15}}\psi K^2 + {\ts {4\over 3}}\psi^2 (K +
\psi ) \cr
&+ {\ts {1 \over 10}} K R_{nn} +{\ts {1\over 30}} K^{ij}R_{injn}
 -{\ts{1\over 90}} K^{ij}\hR_{ij} +\thir \P_- \hD_i \P \, F^i{}_n  \cr
& +{\ts {1\over 63}} \bigl ( (2\P -{\ts{19 \over 6}})K^3 + ({\ts
{18\over 5}} \P +2) K K_{ij}K^{ij} +( {\ts {59
\over 30}}-{\ts{4\over 5}} \P ) K_{ij}K^{jk} K_k{}^i\bigl ) \cr
& -{\ts {7\over 15}} \hD_i \P \hD^i \P \, K +  {\ts {1 \over 15}}
\hD_i \P \hD_j \P \, K^{ij} -{\ts {4 \over 3 }} \hD_i \P \hD^i\P \,
\psi \Bigl ) \biggl \} ~, & (1.17) \cr}
$$
with $dv = d^d x \sqrt g$ and $dS = d^{d-1} \bhx \sqrt {\hat \gamma}$.

Beyond one loop calculations become more involved.
Using the representation outlined above for the singular part
of $G_\Del$ in the neighbourhood of a boundary we have undertaken
calculations at two loops for the simplest renormalisable field
theory in four dimensions, i.e. purely scalar field theory with a
multi-component scalar field $\phi_i$ represented by an action
$$ S(\phi,g) = \int_\M \! dv \, \bigl ( \half \pr_\mu \phi_i \pr^\mu \phi_i +
V(\phi) \bigl ) ~, ~~~ V(\phi) = {\ts {1\over 24}} g_{ijk\ell} \phi_i \phi_j
\phi_k \phi_\ell + \O (\phi^3) ~,
\eqno (1.18) $$
with $g$ denoting generically the complete set of couplings, including mass
terms, on which $V$ depends, for $v_I(\phi)$ a basis of polynomials of degree
4 we may write $V(\phi) = \sum_I g_I v_I(\phi)$. In order to take account the
effects of the boundary $\pr \M$ we also add to $S$ an extra surface term
${\hat S}(\phi,{\hat g})$ which is a integral over the boundary of a local
scalar formed from $\phi$ and its normal derivative $\pr_n \phi$ and which
depends on additional couplings, or sources, ${\hat g}$. We also follow the
usual background field technique expanding
$$ \phi = \vphi + f ~,
\eqno (1.19) $$
for some fixed background field $\vphi$ and assume boundary conditions on
$\vphi$ so that $S+{\hat S}$ has no terms linear in $f$ when $\vphi$ satisfies
the classical equation of motion $ -\nab^2 \vphi + V'(\vphi) =0$. The same
boundary conditions are then imposed on $\phi$, which therefore entails
corresponding boundary conditions for the fluctuations $f$. The background
field
method allows an efficient procedure for calculating divergent parts of
vacuum self energy diagrams as local dimension 4 functions of $\vphi$ and
hence, since $\phi$ and $\vphi$ satisfy the same boundary conditions,
obtaining the necessary counterterms
$S_{\rm c.t.} (\phi,g),~ {\hat S}_{\rm c.t.} (\phi,g,{\hat g})$ to ensure that
the functional integral
$$ \int d[\phi] \, e^{- S_0 (\phi,g,{\hat g})} ~, ~~~~
S_0 = \mu^{-\vep}\bigl ( S + S_{\rm c.t.} + {\hat S} + {\hat S}_{\rm c.t.}
\bigl ) ~,
\eqno (1.20) $$
defines a finite measure over fields $\phi$ satisfying the
required boundary conditions. The factor $\mu^{-\vep}$ in $S_0$ ensures that
the couplings $g,\, {\hat g}$, as well as the fields $\phi$, retain their
canonical dimensions although the quantum field theory is formally extended
to $d$ dimensions.

In the next section we apply these methods to the situation where ${\hat S}$
is chosen so that $\phi |_{\pr \M} = {\hat \vphi}$ for ${\hat \vphi}_i(\bhx)$
an arbitrary smooth boundary field. This corresponds to requiring that the
quantum field $f$ in (1.19) satisfies Dirichlet boundary conditions on $\pr
\M$.
In this case it is essential, besides the usual counterterms on $\M$
necessary for finiteness, to introduce additional local counterterms on
$\pr \M$ linear in $\pr_n \phi$ and also in $K$ which are constrained by power
counting to have overall dimension 3. These are calculated to two loops
making use of the expansion (1.11) for the Green function $G_\Del$. The
renormalisation group equations which now include terms involving the boundary
operator $\pr_n \phi$ are also discussed. Similar calculations are also
undertaken in section 3 for the case when ${\hat S}$ is chosen such that
$\pr_n \phi$ is related to $\phi$ on $\pr \M$ and $f$ satisfies generalised
Neumann boundary conditions. In both cases the final results are also
specialised to the $O(n)$ symmetric case, where there is a single coupling
$g$ and some results were also obtained by Diehl and Diettrich [5]. As usual
with dimensional regularisation the renormalisation group equations determine
the higher order poles in $\vep$ of the counterterms in terms of the simple
poles and these conditions are verified by our two loop results. The details
of the calculation, for either the Dirichlet or Neumann case, are relegated
to appendix B. They depend on the detailed form of the expansion given by
(1.11), (1.12) and (1.13) or (1.15). A brief summary of the important results
for the heat kernel for a general second order operator $\Del$ of the form
(1.2) with boundary conditions (1.3) in our treatment is given in appendix A.
In section 4 consistency conditions are derived under local Weyl rescalings of
the metric which in effect determine the $K$ dependent counterterms in terms
of those necessary, on flat space and for a plane boundary, to define local
composite operators on $\pr \M$. These results allow the derivation of a local
renormalisation group equation which is discussed for the $O(n)$ symmetric
field theory at the critical point where ${\hat \beta}(g)=0$.

In section 5 we return to the situation for a general scalar field theory
where the quantum field $\phi$ attains a fixed
boundary value $\hphi$ on $\pr \M$ and the functional integral (1.20) defines
a wave functional ${\hat \Psi} (\hphi)$. Following earlier work by Symanzik
[10],
and in two dimensions by L\"uscher {\it {et al}} [11], this is shown, to two
loops in a semiclassical expansion with the background field $\vphi$ in (1.19)
solving the classical equations of motion with the boundary condition
$\vphi = \hphi$ on $\pr \M$, to satisfy a local Schr\"odinger equation of
the form
$$
\Bigl ( \hbar {\de \over \de t (\bhx)} + \H_0 (\bhx) \Bigl ) {\hat \Psi}
(\hphi) =0 ~.
\eqno (1.21) $$
$\de t (\bhx)$ represents an arbitrary local deformation of the boundary
$\pr \M$ along the normal $n^\mu (\bhx)$ and $\H_0(\bhx)$ is a local
Hamiltonian operator. To leading order in the semiclassical expansion
$\H_0 \to \H$ where the functional differential operator is
$$
\H = - \half \, {\de^2 \over \de \hphi^2} + V(\hphi) +
\half {\hat \gamma}^{pq} \pr_p \hphi_i \pr_q \hphi_i ~,
\eqno (1.22) $$
although $\H_0$ contains counterterms reflecting those occurring in $S_0$
and also since the second functional derivative
$\de^2/\de\hphi(\bhx) \de \hphi(\bhx')$ is ill defined in general
as $\bhx'\to \bhx$. Such divergences are demonstrated here to be fully
resolved by use of dimensional regularisation. Furthermore the structure
of $\H_0$ is shown to be determined in terms of $S_0$ to all orders
by use of the renormalisation group. Some mathematical details necessary
to verify (1.21) are contained in appendix C. Finally in section 6
a few concluding remarks are made.
\vfill\eject

\leftline{\bf 2 Quantum Scalar Field Theory with Prescribed Boundary
Values}

In this section we consider the quantum field theory defined by the
action (1.18) with the additional surface contribution
$$
\hS (\phi, \hphi) = \int_{\pr \M} \! \!  dS\,(\phi - \hphi)_i \pr_n
\phi_i  ~ ,
\eqno(2.1)
$$
where $\hphi(\bhx)$ may be regarded as a source defining the surface operator
$\pr_n \phi(\bhx) = \pr_y \phi(x)|_{y=0}$.

When $\phi$ is expanded about a background $\vphi$ as in (1.19) we may
write $S + \hS = S_\c + S_1 + S_2 + \dots$ with $ S_n = \O(f^n)$ and
$S_\c = S(\vphi,g) + {\hat S}(\vphi,\hphi)$. It is
then straightforward to see that
$$ S_1 = \int_{\M} \! dv\, f_i \bigl (-\nab^2 \vphi_i + V'\! {}_i (\vphi)
\bigl ){} + \int_{\pr \M} \! \! dS\, \pr_n f_i (\vphi - \hphi)_i  ~,
\eqno(2.2)
$$
so that the appropriate boundary conditions to impose in a perturbative
expansion are
$$ \phi \bigl |_{\pr \M} = \hphi ~ , ~~~~ f \bigl |_{\pr \M}=0  ~ .
\eqno(2.3)
$$
Also with this boundary condition
$$S_2 = \half \int_\M \! dv\,f_i(\Del f)_i ~ ,~~~~
\Del = -\nab^2 + V''(\vphi) ~,
\eqno(2.4)
$$
so that, acting on fields $f$ satisfying (2.4), $\Del$ is a symmetric operator
of the necessary form (1.2), with $X=V''(\vphi)$,
and the associated Green function $G_{ij}(x,x')$, which corresponds to the
propagator  for internal lines in a perturbative expansion, is
therefore defined by requiring Dirichlet boundary conditions. The higher
order terms in the expansion in $f$ as usual generate the vertices for
Feynman diagrams in a perturbative expansion,
in this theory they arise just from $S_3$, depending
on $V'''(\vphi)$, and $S_4$ with no interactions restricted just to $\pr \M$.

For considering correlation functions of $\phi$ in this quantum field theory
with boundary conditions (2.3) the essential functional integral is
$$ e^{W ( \hphi ; J )} = \int_{\phi|_{\pr \M} = \hphi} \! \! \! \! \! \! \! \!
\! \! \! \! \! \! d[\phi]\, \, e^{ -S_0(\phi) + \int_\M \! dv\, J_i \phi_i } ~.
\eqno(2.5)
$$
The counterterms in $S_0$ are chosen so that $W ( \hphi ; J)$
is finite for arbitrary external sources $J_i(x)$. For calculation it is
convenient to define
$$ \Gamma (\hphi;\vphi) = W (\hphi;J) - \int_\M \! dv \, J_i \vphi_i ~,
\eqno (2.6) $$
where $\vphi$ is assumed to
be an arbitrary smooth field on $\M$, with boundary value $\hphi$,
while $J$ is constrained to be defined in terms of $\vphi$, with an
invertible relationship $J \leftrightarrow \vphi$.
For a quantum field theory without any boundary it is conventional to
choose $J$ so that $\langle \phi \rangle = \vphi$, to all orders in the
loop expansion, and then $\Gamma(\vphi)$
is the generating functional of connected one particle irreducible
amplitudes. This contains the only primitive
divergences requiring renormalisation by appropriate counterterms.
In a perturbative expansion $\Gamma$
may be calculated via the background field method, taking $\phi = \vphi + f$
with $d[\phi] = d[f]$, when to ensure $\langle f \rangle = 0$
it is necessary to require $J = J_\c + \dots $ for
$J_\c(\vphi) \equiv \mu^{-\vep}(-\nab^2 \vphi + V'(\vphi))$.
However, in the presence of a boundary this condition on $\langle f \rangle$
necessitates that $J(x)$ is singular as $x$ approaches the boundary,
assuming the background $\vphi$ is smooth in the neighbourhood of $\pr \M$.
This is a consequence of the fact that one particle reducible
amplitudes also have primitive divergences, in the presence of a boundary,
requiring corresponding contributions to $\hS_{\rm{c.t.}}$. For the purpose
of the calculations of the counterterms required by the presence of a boundary
undertaken here, it is simpler to choose $J=J_\c$, without any higher order
corrections, to define $\Gamma (\hphi;\vphi)$ in (2.5) and (2.6). This choice,
together with the boundary condition on $\vphi$, cancels the linear term
$S_1$ given by (2.2) in the expansion of the classical action although
in this case it is necessary to include one particle irreducible graphs in the
perturbative expansion beyond one loop.
Since with dimensional regularisation there is a unique separation
$\langle \vphi \rangle = \langle \vphi \rangle_{\rm {n.s.}} + \langle
\vphi \rangle_{\rm s.}$, where $\langle \vphi\rangle_{\rm {n.s.}}$,
$\langle \vphi\rangle_{\rm s.}$ are respectively non-singular, singular as
the boundary is approached, an alternative possibility would be,to require
$\langle \vphi \rangle_{\rm {n.s.}}=0$ but this is not pursued here.
In this framework it
straightforward to define a Scr\"odinger wave functional ${\hat{\Psi}}(\hphi)$
by ${\hat{\Psi}}(\hphi)= \exp W ( \hphi;0) = \exp \Gamma (\hphi;\vphi_\c)$
where $\vphi_\c$ is the classical solution to $-\nab^2 \vphi_\c +
V'(\vphi_\c)=0$ prescribed by the boundary value $\hphi$.

Following standard procedure the structure of possible counterterms
is determined by power counting. As usual
$$
S_{\rm{c.t.}}(\phi,g) = \int_\M \! dv\, \bigl(\half \pr_\mu \phi_i A_{ij}
\pr^\mu \phi_j + V_{\rm{c.t.}} (\phi) + C(\lambda) \bigl )~,
\eqno(2.7)
$$
with $V_{\rm{c.t.}}(\phi)$ a polynomial of degree 4 in $\phi$ which is also
taken to include terms linear in $R$ which are quadratic in $\phi$.
$C(\lambda)$ contains all additional counterterms which involve no operators,
or dependence on $\phi$, but are
necessary for a curved space background, it may be written as
$$ \eqalign {
C(\lambda) ={}& - ( aF + bG + cR^2 \bigl ) ~, ~~~~~
\lambda = (a,b,c) ~, \cr
&F= R^{\mu \nu \rho \si} R_{\mu \nu \rho \si}-{4\over d-2}R^{\mu \nu}
R_{\mu \nu} + {2 \over (d-2)(d-1)}R^2 ~, \cr
& G= R^{\mu \nu \rho \si}R_{\mu \nu \rho \si}-4 R^{\mu \nu}R_{\mu \nu} +
R^2 ~, \cr }
$$
where $\lambda$ is determined perturbatively in terms of the dimensional
coupling $g$. The additional counterterms restricted to
the boundary are of dimension 3 and have the general form
$$ \hS_{\rm{c.t.}}(\phi,g,\hphi ) = \int_{\pr \M} \! \! dS\, \bigl (
-\hphi_{\rm{c.t.}}{}_i(\hphi) \pr_n \phi_i + \rho (\hphi) K +
{\hat  C} ({\hat \lambda}) \bigl ) ~ ,
\eqno(2.8)
$$
using $\phi |_{\pr \M}=\hphi$ and also that with dimensional regularisation
there are no linearly divergent terms which are ${\rm O} (\hphi^2)$.
Assuming minimal subtraction $S_{\rm {c.t.}}$ and ${\hat S}_{\rm {c.t.}}$
contain only poles in $\vep$. $\hat C$, with $\hat \lambda$ denoting the
coefficients in some basis, includes all terms
which are proportional to any dimension 3 scalar, independent of $\hphi$,
formed solely from the extrinsic and intrinsic curvature on $\pr \M$,
such as appear in the surface terms in (1.17) not involving $\psi$
and $X$.

Due to the complexity  of the calculations necessary to determine such
contributions we do not consider them here but focus primarily on
$\hphi_{\rm{c.t.}}$, which has in effect been calculated previously to two
loop order by Diehl and Dietrich, and also on $\rho$. Of course $S_{\rm{c.t.}}$
in the form (2.7) has been calculated extensively before at two and
higher loops although it is necessary to redetermine all contributions
to $A$ since any integration  by parts of derivatives may lead to
surface terms. It is also necessary to determine any divergences
proportional to total derivatives which are usually discarded. Using the
background field method $S_{\rm{c.t.}}$ may be calculated with
$\pr_n \phi \to \pr_n \vphi$. It is easy to see that
${\hat \vphi}_{\rm c.t.}({\hphi})$ is linear in $\hphi$ and depends
in general on $V'''({\hat \vphi})$ while $\rho(\hphi)$ is quadratic in
$\hphi$ and is proportional to $V''({\hphi})$ or $V'''({\hphi})^2$ and no
other $\hphi$ dependent terms are required in (2.8).

In general for a perturbative expansion of the functional integral (2.5)
with a background field $\vphi$
$$ \Gamma (\hphi;\vphi) = - S_0 (\vphi)  +
\sum_{\ell =1} \Gamma^{(\ell)} (\hphi;\vphi) ~,
\eqno (2.9) $$
where $\Gamma^{(\ell)}(\vphi)$ is the connected amplitude at $\ell$ loops
and $S_0(\vphi)$ contains the necessary counterterms to ensure that
$\Gamma (\hphi;\vphi)$ is finite. At one loop
$$
\Gamma^{(1)}= -\half \, {\rm ln}\, {\rm det} \,
\Del ~, \eqno(2.10)
$$
and using (1.17) it is easy to see that to ensure cancellation of the poles
in $\vep$ it is sufficient to take in (2.8)
$$
\hphi^{(1)}_{\rm{c.t.}}{}_i (\hphi) = - {1\over 16 \pi^2 \vep} \,\half
V'''_{ijj}(\hphi) ~, ~~~~~ \rho^{(1)}(\hphi) = -{1\over 16 \pi^2 \vep} \,
{\ts {1\over 3}} V''_{jj}(\hphi) ~.
\eqno(2.11)
$$

Beyond one loop the calculational details are non trivial and are mostly
relegated to appendix B. The essential  two loop vacuum diagrams are shown in
fig.1 and accordingly we decompose $\Gamma^{(2)} = \Gamma_a + \Gamma_b +\dots$.
Of course at two loops there are one loop subdivergences which are
removed by one loop counterterms. For the scalar field theory defined by
(1.18) and (2.1) these are given solely by
$$ V_{\rm {c.t.}}^{(1)} - c^{(1)} R^2
= {1\over 16\pi^2 \vep}\, \half \, {\tilde V}_{ij}
{\tilde V}_{ij} ~, ~~~~ {\tilde V}_{ij}(\phi) = V''_{ij}(\phi) - \six R \,
\delta_{ij} ~.
\eqno (2.12)$$
The amplitude for fig.1a, with appropriate counterterms, is then
$$ \eqalign {
\Gamma_a = {}& - {\mu^\vep \over 8} \, \int dv \, g_{ijk\ell} \, G_{ij}\bigl|
G_{k\ell} \bigl| {}- {1\over 16\pi^2 \vep} \, {1\over 2} \int dv \,
g_{ijk\ell} \, {\tilde V}_{ij}(\vphi) G_{k\ell} \bigl | \cr
= {}& {1\over 8} \, \int dv \, g_{ijk\ell} \, \Bigl (
{4\mu^{-\vep}\over (16\pi^2 \vep)^2} \, {\tilde V}_{ij}(\vphi)
{\tilde V}_{k\ell}(\vphi)
- \mu^\vep {\bar G}_{ij}\bigl| {\bar G}_{k\ell} \bigl| \Bigl ){} ~, \cr}
\eqno (2.13) $$
where in general $G|(x)$ denotes the the coincident limit, $x'\to x$, of
the Green function $G(x,x')$. In (2.13) we have used
$$ G_{ij}\bigl | {} = - {2\mu^{-\vep} \over 16\pi^2\vep } \, {\tilde V}_{ij}
(\vphi) + {\bar G}_{ij} \bigl | ~,
\eqno (2.14) $$
where, for manifolds without boundary, ${\bar G}_{ij} |$ is finite for $d=4$
from (1.5), (1.6) and (1.7). Similarly for fig.1b, with its counterterms,
$$
\Gamma_b = {\mu^\vep\over 12} \, \int \! \int dv dv' \, V'''_{ijk}(\vphi)
G_{ii'} G_{jj'} G_{kk'} V'''_{i'j'k'} (\vphi')
- {1\over 16\pi^2 \vep} \, {1\over 2} \int dv \,
V'''_{ijk}(\vphi)  V'''_{ij\ell}(\vphi)  G_{k\ell} \bigl | ~ .
\eqno (2.15) $$
Using (1.5) and (1.8) it is straightforward to see that
$$ \eqalign {
\Gamma_b \sim {}& {\mu^{-\vep}\over (16\pi^2)^2 \vep} \, {1\over 24} \, \int
dv \, V'''_{ijk}(\vphi) \bigl( \nab^2 + \six R\bigl) V'''_{ijk} (\vphi) \cr
& ~~~ + {\mu^{-\vep}\over (16\pi^2\vep)^2} \, {1\over 2} (1-\half \vep) \int
dv \, V'''_{ijk} (\vphi) V'''_{ij\ell} (\vphi) {\tilde V}_{k\ell}(\vphi) ~,
\cr}
\eqno (2.16) $$
and together with $1/\vep^2$ term in (2.13) this determines the required two
loop counterterms for $S_{\rm{c.t.}}$ as usual.

In addition it is essential to consider also the one particle reducible
diagram fig.1c for which
$$ \Gamma_c = {\mu^\vep\over 8} \, \int \! \int dvdv'\, {\bar G}_{jk}\bigl |
V'''_{ijk}(\vphi) \, G_{ii'} V'''_{i'j'k'}(\vphi') {\bar G}_{j'k'} \bigl | ~.
\eqno (2.17) $$
For no boundary $\Gamma_c$ is finite while here it is necessary to allow for
the extra one loop boundary counterterms given by (2.8) and (2.11) which give
the one particle reducible diagrams in fig.1d and
$$ \eqalign {
\Gamma_d = {}& {1\over 4} \, {1\over 16\pi^2 \vep} \, \int \! \int dSdv'\,
V'''_{ijj}({\hat \vphi}) \, \pr_n G_{ii'} \bigl |_{y=0} V'''_{i'j'k'}(\vphi')
{\bar G}_{j'k'} \bigl | \cr
& + {1\over 8} \, {\mu^{-\vep}\over (16\pi^2 \vep)^2} \, \int \! \int dSdS'\,
V'''_{ijj}({\hat \vphi}) \,\pr_n G_{ii'} \!{\overleftarrow \pr}{}'
{}_{\! \! \! n} \bigl |_{y=y'=0} V'''_{i'j'j'}({\hat \vphi}') ~. \cr}
\eqno (2.18) $$
When subdivergences are subtracted in one particle reducible amplitudes then
according to the usual lore of renormalisation theory there is no remaining
overall divergence. If $G_{ii'} (x,x')$ in (2.17) and (2.18) is separated
into a singular and a regular part then by integrating by parts from (2.17)
and (2.18) we find
$$ \eqalign {
(\Gamma_c + \Gamma_d)^{\rm {reg}} = {}& {\mu^\vep \over 8} \, \int \! \int
dvdv'\, D_{jk}
V'''_{ijk}(\vphi) \, G_{ii'}^{\rm {reg}} V'''_{i'j'k'}(\vphi') D_{j'k'} ~, \cr
D_{jk} = {}& {\bar G}_{jk}\bigl | {} + {\mu^{-\vep}\over 16 \pi^2 \vep} \,
\delta_{jk} \bigl ( - \delta'(y) + {\ts {1\over 3}} K \,
\delta (y) \bigl ) ~, \cr}
\eqno (2.19) $$
where $D_{jk}(x)$, as in (C.20), is the dimensionally regularised form of
$G_{jk}(x,x)$ with poles in $\vep$ representing both the usual local and
also boundary divergences subtracted (the $\delta(y)$ term is irrelevant
in this case assuming $G_{ii'}^{\rm {reg}}$ also obeys Dirichlet boundary
conditions and vanishes at $y=0$ or $y'=0$) and hence, so long as
$ G_{ii'}^{\rm {reg}}(x,x')$ is a sufficiently smooth function on
$\M \times \M$, (2.19) is finite as $\vep \to 0$. However, as pointed out by
Symanzik [10], for the full Green function $G_{ii'} (x,x')$ there are
additional
divergences in (2.17) not removed by the counterterms (2.18) and in general
the manipulations leading to (2.19) are not justified. An additional
complication in this case is that in (2.18) $\pr_n G_{ii'}
\!{\overleftarrow \pr}{}'{}_{\! \! \! n}$ has a non integrable singularity as
$\bhx'\to\bhx$ which is not regularised by analytic continuation in $d$ as
usual with dimensional regularisation. Nevertheless elsewhere [17] we have
shown that there is a well defined prescription for handling this which we
follow in detailed calculations in appendix B.

The results of appendix B allow the divergent terms as $\vep \to 0$ on the
boundary to be readily calculated for each contribution. For $\Gamma_a$
from (B.4,7), and also using (B.8,10), we get, keeping only terms relevant for
determining ${\hat \vphi}_{\rm {c.t.}}$ and $\rho$ in (2.8),
$$
\Gamma_a \sim {\mu^{-\vep}\over (16\pi^2 \vep)^2} \, {1\over 4}\, \int dS \,
g_{ijkk} \Bigl \{ (1-\half \vep) \pr_n V''_{ij} (\vphi) \bigl |_{y=0} {} -
{\ts {2\over 3}}(1 - {\ts {2\over 3}} \vep ) V''_{ij} ( {\hat \vphi} ) \, K
\Bigl \} ~.
\eqno (2.20) $$
The amplitude $\Gamma_b$ also has, apart from (2.16), a contribution to
the divergence on the boundary given by, from (B.19) and (B.20a,..f),
$$ \eqalign {
\Gamma_b \sim &
- {\mu^{-\vep}\over (16\pi^2)^2 \vep} \, {1\over 24}
\int dv \, \nab_\mu V'''_{ijk} \nab^\mu V'''_{ijk} \cr
& + {\mu^{-\vep}\over (16\pi^2 \vep)^2} \, {1\over 2}
\int dS \, \Bigl \{ (1-\six \vep) V'''_{ijk} ({\hat \vphi}) \pr_n V'''_{ijk}
(\vphi) \bigl |_{y=0} {} - {\ts {1\over 3}} (1-\quar \vep)
V'''_{ijk} ( {\hat \vphi} ) V'''_{ijk} ( {\hat \vphi} ) \, K \Bigl \} ~.\cr}
\eqno (2.21) $$
For the remaining one particle reducible amplitudes we may use (B.33) to give
$$
\Gamma_c + \Gamma_d \sim -{\mu^{-\vep}\over (16\pi^2 \vep)^2} \, {1\over 8}
\int dS \, \Bigl \{ (1-\vep) V'''_{ijj} ({\hat \vphi}) \pr_n V'''_{ikk} (\vphi)
\bigl |_{y=0} {} - {\ts {2\over 3}}(1- {\ts {1\over 6}}\vep)
V'''_{ijj} ( {\hat \vphi} ) V'''_{ikk} ( {\hat \vphi} ) \, K \Bigl \} ~.
\eqno (2.22) $$

{}From these results we may then read off
$$ \eqalign {
\hphi^{(2)}_{\rm{c.t.}}{}_i = {}& - {1\over (16 \pi^2 \vep)^2} \Bigl \{
{1\over 4} (1-\half \vep) V'''_{ijk} g_{jk\ell \ell} + {1\over 2}
(1-\six \vep) g_{ijk\ell} V'''_{jk\ell} - {1\over 8}
(1-\vep) g_{ijjk} V'''_{k\ell \ell} \Bigl \} ~, \cr
\rho^{(2)} = {}& - {1\over (16 \pi^2 \vep)^2} \Bigl \{ {1\over 6}
(1 - {\ts {2\over 3}} \vep ) g_{ijkk} V''_{ij} +
{1\over 6} (1-\quar \vep) V'''_{ijk} V'''_{ijk} -
{1\over 12}(1- {\ts {1\over 6}} \vep) V'''_{ijj} V'''_{ikk} \Bigl \} ~. \cr}
\eqno (2.23) $$
With $S_0$ defined as in (1.20) we may write
$$ \eqalign {
S_0 (\phi) = {}& S(\phi_0, g_0) + {\hat S} ( \phi_0, \hphi_0 ) +
\C (\rho) + {\tilde \C} (\lambda , {\hat \lambda}) ~, \cr
\C (\rho) = {}& \mu^{-\vep} \int_{\pr \M} \! \! dS\, \rho(\hphi) K ~,~~~
{\tilde \C} (\lambda , {\hat \lambda}) = \mu^{-\vep} \int_\M \! dv \,
C(\lambda) +
\mu^{-\vep} \int_{\pr \M} \! \! dS\,{\hat  C} ({\hat \lambda}) ~ , \cr}
\eqno (2.24) $$
in terms of the expressions for $S$, ${\hat S}$ given by (1.18), (2.1).
${\tilde \C}$ includes all necessary $\phi,\, \hphi$ independent counterterms
not calculated here. To obtain (2.24) it is necessary to require\footnote{*}
{Our treatment at this point differs from Symanzik [10] and also Diehl and
Dietrich [5], although results to two loops are equivalent with the latter.}
$$ \eqalign {
\phi_{0i} = {}& \mu^{-\hh \vep} Z_{ij} \phi_j ~,~~~~~ Z = (1+ A)^\hh ~, \cr
\hphi_{0i} = {}& \mu^{-\hh \vep}\bigl ( Z_{ij} \hphi_j + Z^{-1} {}_{\! ij}
\hphi_{{\rm c.t.} j} \bigl ) ~. \cr}
\eqno (2.25) $$

The renormalisation group equations are derived as usual by requiring
invariance of physical amplitudes under rescalings of the arbitrary mass
$\mu$. To obtain these it is sufficient to require
$$ \eqalign {
\mu {d\over d\mu} S_0 = {}& \Bigl ( \D - \int_\M \! dv \,
({\hat \gamma} \phi)_i {\delta \over \delta \phi_i} \Bigl ) S_0
= - \C (\beta^\rho)- {\tilde \C} (\beta^\lambda, \beta^{\hat \lambda} ) ~, \cr
& \D = \mu {\pr \over \pr \mu} + {\hat \beta}^V \! \cdot {\pr \over \pr V} +
\int_{\pr \M} \! \! dS \, {\hat \beta}^\hphi{}_{\! \! i}
{\delta \over \delta \hphi_i} ~. \cr}
\eqno (2.26) $$
where ${\hat \beta}^V \! \cdot \pr / \pr V \equiv {\hat \beta}^g_I\pr /
\pr g_I$ and $\beta^\rho, \beta^\lambda, \beta^{\hat \lambda}$ are finite.
By using (2.26) with (2.5), and discarding a total functional derivative in
the functional integral, it is easy to find the renormalisation group
equation,
$$ \Bigl ( \D + \int_\M \! dv \, ({\hat \gamma} J)_i {\delta \over \delta J_i}
\Bigl ) W (\hphi;J) =  \C (\beta^\rho) + {\tilde \C}
(\beta^\lambda , \beta^{\hat \lambda}) ~.
\eqno (2.27) $$
As is conventional for field theories without boundary in
dimensional regularisation on the basis of minimal subtraction
$$ {\hat \beta}^V (\phi) = \vep \bigl ( V(\phi) - \half V'{}_{\! i} (\phi)
\phi_i \bigl ){} + \beta^V(\phi) ~,~~~{\hat \gamma}_{ij} = - \half \vep
\delta_{ij} + \gamma_{ij} ~,~~ {\hat \beta}^V \! \cdot {\pr \over \pr V} Z
= Z \gamma ~.
\eqno (2.28) $$
The $\beta$ functions associated with the boundary that may be obtained by
the calculations here
are ${\hat \beta}^\hphi$ and $\beta^\rho$ which from (2.26) are determined by
$$ \eqalignno {
\Bigl ( {\hat \beta}^V \! \cdot {\pr \over \pr V} +
{\hat \beta}^\hphi{}_{\! \! j}
(\hphi) {\pr \over \pr \hphi_j} - \half \vep \Bigl )
\hphi_{0i} (\hphi) = {} &0 ~, ~~~~~ {\hat \beta}^\hphi (\hphi) = \half \vep
\hphi + \beta^\hphi (\hphi) ~, & (2.29a) \cr
\Bigl ( {\hat \beta}^V \! \cdot {\pr \over \pr V} +
{\hat \beta}^\hphi{}_{\! \! j} (\hphi) {\pr \over \pr \hphi_j} - \vep \Bigl )
\rho(\hphi) = {}& - \beta^\rho (\hphi) ~. & (2.29b) \cr}
$$
{}From (2.29a) we therefore find to two loops, since $\beta^{V(1)} =
\hh V''_{ij} V''_{ij}/(16\pi^2)$,
$$ \eqalign {
\beta^\hphi {}_{\! \! i} (\hphi) &{} = - {1\over 16 \pi^2} \,\half
V'''_{ijj}(\hphi) \cr
& \, + {1\over (16 \pi^2 )^2}\, {1\over 12} \Bigl \{
3 V'''_{ijk}(\hphi) g_{jk\ell \ell}^{\vphantom D} + 2
g_{ijk\ell}^{\vphantom D} V'''_{jk\ell}(\hphi) - 3 g_{ijjk} V'''_{k\ell \ell}
(\hphi) -
g_{ik\ell m}^{\vphantom D} g_{jk\ell m}^{\vphantom D} \hphi_j \Bigl \} ~, \cr}
\eqno (2.30) $$
using (2.11) and (2.23) with the two loop result for $Z^{(2)}_{ij} = -
{1\over 24} \, g_{ik\ell m} g_{jk\ell m}/(16\pi^2)^2\vep$. As usual with
dimensional regularisation the double poles in $\hphi^{(2)}_{\rm{c.t.}}$ are
determined by (2.29a) which provides a significant consistency check on
our calculations. In a similar fashion
$$ \eqalign {
\beta^\rho  (\hphi) &{} = - {1\over 16 \pi^2} \, {\ts {1\over 3}}
V''_{jj}(\hphi) \cr
& \, + {1\over (16 \pi^2 )^2}\, {1\over 36} \Bigl \{ 8 g_{ijkk}^{\vphantom D}
V''_{ij}(\hphi) + 3 V'''_{ijk}(\hphi) V'''_{ijk}(\hphi) -
 V'''_{ijj} (\hphi) V'''_{ikk} (\hphi)  \Bigl \} ~. \cr}
\eqno (2.31) $$

For specific application of these results we focus on the $O(n)$ symmetric
potential $V(\phi)={1\over 24}g(\phi^2)^2 + \half m^2 \phi^2$, for $\phi_i$
an $n$ component field. In this case we may write
$$
\beta^\hphi {}_{\! \! i} (\hphi) = {\hat \eta} \, \hphi_i ~,~~~~
\beta^\rho  (\hphi) = \half {\hat \rho}\, \hphi^2 + {\hat \rho}_m m^2 ~ ,
\eqno (2.32) $$
so that ${\hat \eta}$ is the anomalous dimension of the boundary operator
$\pr_n \phi$. From (2.30) and (2.31) to two loop order
$$ \eqalign {
{\hat \eta} = {}& - {\ts {1\over 6}} (n+2) u + {\ts {1\over 36}}
(n+2) u^2 ~, \cr
{\hat \rho} = {}& - {\ts {1\over 9}} (n+2) u + {\ts {1\over 54}}
(n+2)(n+5) u^2 \, , ~
{\hat \rho}_m = {1\over 16\pi^2} \, {n\over 3}\bigl ( -1 + {\textstyle
{2\over 9}}(n+2)u\bigl) ~, \cr}
\eqno (2.33) $$
for $u = g/(16\pi^2)$. For comparison with Diehl and Dietrich [5] ${\hat \eta}
= \half \eta_1 + \gamma$ where for this model $\gamma_{ij} =
\gamma \de_{ij}$ and $\gamma^{(2)}={1\over 36}(n+2) u^2$.

As an illustration of the renormalisation group equation we apply (2.27) in
this $O(n)$ model to the expansion of the field operator $\phi(x)$ in the
vicinity of the boundary. Defining
$$ \langle \phi (x)\rangle_J^{\vphantom H}={\delta W (\hphi;J)\over \delta
J(x)}
{}~,~~~~ \mu^{-\vep} \langle \pr_n \phi (\bhx)\rangle_J^{\vphantom H} =
{\delta W (\hphi;J)\over \delta \hphi(\bhx)} ~,
\eqno (2.34) $$
then we may write, for $y\to 0$ and neglecting $m$,
$$ \langle \phi (\bx,y)\rangle_J^{\vphantom H} \sim C_0 (\mu y) \, \hphi (\bx)
+ C_1 (\mu y) \, y \langle \pr_n \phi (\bx)\rangle_J^{\vphantom H}
+ C_2(\mu y) \, y K(\bx) \hphi (\bx) ~.
\eqno (2.35) $$
{}From (2.27) the leading singular parts of the coefficient functions satisfy
$$ \eqalign {
\Bigl ( \mu {\pr \over \pr \mu} + {\hat \beta} (g) {\pr \over \pr g} +
\gamma(g) + {\hat \eta}(g) \Bigl ) C_0 (\mu y) = {}& 0 ~, \cr
\Bigl ( \mu {\pr \over \pr \mu} + {\hat \beta} (g) {\pr \over \pr g} +
\gamma(g) - {\hat \eta}(g) \Bigl ) C_1 (\mu y) = {}& 0 ~, \cr
\Bigl ( \mu {\pr \over \pr \mu} + {\hat \beta} (g) {\pr \over \pr g} +
\gamma(g) + {\hat \eta}(g) \Bigl ) C_2 (\mu y) = {}& - {\hat \rho}(g)
C_1(\mu y) ~, \cr}
\eqno (2.36) $$
where manifestly for $g=0$ $C_0 = C_1 = 1, \, C_2 =0$. This result defines
an expansion for the operator $\phi(x)$, analogous to the usual operator
product expansion, in the neighbourhood of the boundary.
At the infra-red fixed point corresponding to ${\hat \beta} (g_*) =0$
$$ C_0(\mu y) \propto (\mu y)^{-\gamma_* - {\hat \eta}_*} \, ,~~
C_1(\mu y) \propto (\mu y)^{-\gamma_* + {\hat \eta}_*} \, ,~~~
C_2(\mu y) \sim - {{\hat \rho}_* \over 2{\hat \eta}_*} \, C_1(\mu y)
+ {\rm O}\bigl ( (\mu y)^{-\gamma_* - {\hat \eta}_*}\bigl)  ~.
\eqno (2.37) $$
{}From (2.33), and using that with minimal subtraction the critical coupling
is given by
$$ u_* = {3\over (n+8)} \, \vep + {9\over (n+8)^3}(3n+14) \, \vep^2 +
{\rm O}(\vep^3) ~,
$$
then the critical indices are as usual expressed as an expansion in $\vep$
$$ \eqalign {
{\hat \gamma}_* = {}& {1\over 4}\, {(n+2)\over(n+8)^2} \, \vep^2 ~,~~~
{\hat \eta}_* = -{1\over 2}\, {n+2\over n+8} \, \vep
- {1\over 4}\, {(n+2)\over(n+8)^3}(17n+76) \, \vep^2 ~, \cr
{{\hat \rho}_* \over {\hat \eta}_*} = {}& {2\over 3} - {1\over 3}\, {n+4\over
n+8} \, \vep ~. \cr}
\eqno (2.38) $$
\vfill\eject

\leftline{\bf 3 Quantum Scalar Field Theory with Free Boundary
Values}

In this section the essential quantum field theory given by the
action (1.18) is extended by the surface term
$$
\hS (\phi, {\hat g}) = \int_{\pr \M} \! \! dS\, Q(\phi) ~,
\eqno(3.1)
$$
where $Q(\phi)$, depending on couplings $\hat g$, is a scalar and there is no
dependence on the normal
derivative $\pr_n \phi$ as in (2.1). For renormalisability $Q(\phi)$ is at
most cubic in $\phi$ although if $\phi \to - \phi$ symmetry is imposed at
leading order $Q(\phi) = \O (\phi^2)$ and this is stable under
renormalisation.

When $\phi$ is expanded about a background $\vphi$ then to first order
$$ S_1 = \int_{\M} \! dv\, f_i \bigl (-\nab^2 \vphi_i + V'{}_{\! i} (\vphi)
\bigl ){} + \int_{\pr \M} \! \! dS\, f_i \bigl ( -\pr_n \vphi +
Q'(\vphi) \bigl) {}_i  ~,
\eqno(3.2) $$
so that the appropriate boundary conditions, assuming that they apply to
$\phi$ as required for invariance under the shift symmetry generated by
$\delta \vphi = - \delta f$, which should be imposed are now
$$ \bigl ( \pr_n \phi - Q'(\phi) \bigl) {}_i \bigl |_{\pr \M} = 0 ~.
\eqno(3.3) $$
To second order in $f$ the action is again of the form (2.4) where
$\Del$ is now a symmetric operator acting on fields obeying the generalised
Neumann boundary condition
$$ \bigl ( \pr_n f_i - Q''{}_{\! \! \! ij}(\vphi) f_j \bigl) \bigl |_{\pr \M}
= 0 ~.
\eqno(3.4) $$
If $Q(\phi) = {\ts {1\over 6}} {\hat g}_{ijk} \phi_i \phi_j \phi_k + \dots $
there is an additional surface interaction in this case
$$ {\hat S}_3 = - {1\over 12} \int_{\pr \M} \! \! dS \,
{\hat g}_{ijk} f_i f_j f_k ~,
\eqno (3.5) $$
arising from (3.1) and also $\O (f^2)$ terms in the expansion of (3.3) but
we do not undertake any calculations involving this term here.

The basic functional integral remains as in (2.5) but now integrating over
fields satisfying the boundary condition (3.3) and defining now $W(J)$.
The necessary local counterterms on $\M$ for finiteness of $W$ are just as
usual for the scalar field theory without boundary, as in (2.7),  but on
the boundary $\pr \M$ they may now be written as
$$ \hS_{\rm{c.t.}}(\phi,g,{\hat g}) = \int_{\pr \M} \! \! dS\,
\bigl ( Q_{\rm{c.t.}}(\phi) + {\hat C} ({\hat \lambda}) \bigl ) ~,
\eqno(3.6) $$
where ${\hat C} ({\hat \lambda})$ represents contributions independent of
$\phi$ which include all terms of the same form as those present in
$\hat C$ for the Dirichlet case in (2.8) but also, for $\hat g_{ijk} =0$, any
terms involving dimension two scalars formed from the metric, such as
$R_{nn}$, $K_{ij} K^{ij}$ or $K^2$, and proportional to $Q''$.
$Q_{\rm{c.t.}}(\phi)$ then contains the remaining counterterms independent of
the metric and also those linear in $K$.
In the background field method $Q_{\rm c.t.}(\vphi)$ may be calculated
directly and involves, for ${\hat g}_{ijk} =0$, just terms with the
general structure $V'''(\vphi)Q'(\vphi)$, $V''(\vphi)Q''$,
$V'''(\vphi)^2 Q''$, $Q''^3$ while terms proportional
to $K$ may contain $V''(\vphi)$, $V'''(\vphi)^2$ or $Q''^2$. Clearly
$Q_{\rm c.t.}(\vphi)$ remains $\O(\vphi^2)$ in this case.

The one loop amplitude is still determined by the functional determinant of
$\Delta$ as in (2.9), although now with the alternative boundary conditions
(3.4), and (1.17) gives, after using (3.3) to eliminate $\pr_n \vphi$ terms,
$$
Q^{(1)}_{\rm{c.t.}} = {1\over 16 \pi^2 \vep} \, \bigl ( - \half
V'''_{iij} Q'_j + 2 (V'' Q'')_{ii} - {\ts {4\over 3}} (Q''^3)_{ii}
+ {\ts {4\over 3}} (Q''^2)_{ii} \, K  - {\ts {1\over 3}}V''_{ii} \, K \bigl )
{}~.
\eqno(3.7) $$
The corresponding result for ${\hat C}^{(1)}$ is easy to read off from (1.17).

At two loops the essential amplitudes are still given by (2.12), (2.14) and
(2.16), if ${\hat g}_{ijk} =0$, although $G_{ij}(x,x')$ is now the Green
function for $\Delta$ subject to boundary conditions given by (3.4). The
one particle reducible diagrams in fig.1d, with
the one loop boundary counterterms determined by (3.7), now give
$$ \eqalign {
\Gamma_d = {}& {1\over 2} \int \! \int dSdv'\, Q^{(1)\prime}_{{\rm c.t.}\, i}
(\vphi) \, G_{ii'} \bigl |_{y=0} V'''_{i'j'k'}(\vphi') {\bar G}_{j'k'}
\bigl | \cr
& + {\mu^{-\vep}\over 2} \, \int \! \int dSdS'\,
Q^{(1)\prime}_{{\rm c.t.}\, i} (\vphi)
\, G_{ii'} Q^{(1)\prime}_{{\rm c.t.}\, i'} (\vphi') \bigl |_{y=y'=0} ~. \cr}
\eqno (3.8) $$
In a similar fashion to (2.19), if $G_{ii'}(x,x') \to G_{ii'}^{\rm {reg}}
(x,x')$ which is assumed to satisfy the same boundary conditions, then
combining (2.17) and (3.8) gives the finite expression
$$ \eqalign {
(\Gamma_c + \Gamma_d)^{\rm {reg}} = {}& {\mu^\vep \over 8} \, \int \! \int
dvdv'\, D_{jk}
V'''_{ijk}(\vphi) \, G_{ii'}^{\rm {reg}} V'''_{i'j'k'}(\vphi') D_{j'k'} ~, \cr
D_{jk} = {}& {\bar G}_{jk}\bigl | {} + {\mu^{-\vep}\over 16 \pi^2 \vep} \,
\bigl ( \delta_{jk} \, \delta'(y) + ( 4Q''_{jk} -
\delta_{jk} {\ts {5\over 3}} K) \, \delta (y) \bigl ) ~, \cr}
\eqno (3.9) $$
since $D_{jk}$ is in accord with (C.20) for this case as well.
Since $Q^{(1)}_{\rm{c.t.}}(\phi)$ is quadratic in $\phi$ there is an
additional diagram fig.1e which corresponds to
$$ \eqalign {
\Gamma_e = {}& -{\mu^{-\vep}\over 2} \, \int dS \,
Q^{(1)\prime\prime}_{{\rm c.t.}\, ij} \, G_{ij} \bigl | \cr
= {}& - {1\over 4} \, {\mu^{-\vep}\over 16 \pi^2 \vep} \, \int dv \,
g_{ijk\ell} \, \bigl ( \delta_{k\ell} \, \delta'(y) + ( 4Q''_{k\ell} -
\delta_{k\ell} {\ts {5\over 3}} K) \, \delta (y) \bigl ) G_{ij} \bigl
|~. \cr}
\eqno (3.10) $$

The results of appendix B again allow the divergent contributions on the
boundary to be calculated for terms independent of and linear in $K$.
{}From (B.4,6) we get
$$ \eqalign {
\Gamma_a + \Gamma_e \sim {\mu^{-\vep}\over (16\pi^2 \vep)^2} \, & \int dS \,
\Bigl \{ g_{ijkk} \Bigl ( 2 (Q''^3)_{ij} - (1-\vep) (Q''V'')_{ij} \cr
& -{\ts{1\over 8}} (2+\vep) V'''_{ij\ell} Q'_\ell
- {\ts{2\over3}} (1-\vep) (Q''^2)_{ij} \, K -
\half (1 + {\ts {2\over 9}} \vep ) V''_{ij} \, K \Bigl ) \cr
& + g_{ijk\ell} \Bigl ( (3-\vep) Q''_{ij} V''_{k\ell} - 4 Q''_{ij}
(Q''^2)_{k\ell} + {\ts {4\over 3}} ( 2 - {\ts {1\over 3}} \vep) Q''_{ij}
Q''_{k\ell} \, K \Bigl ) \Bigl \} ~. \cr}
\eqno (3.11) $$
Also from (B.19) and (B.20a,..f)
$$ \eqalign {
\Gamma_b \sim {}& - {\mu^{-\vep}\over (16\pi^2)^2 \vep} \, {1\over 24}
\int dv \, \nab_\mu V'''_{ijk} \nab^\mu V'''_{ijk} \cr
& + {\mu^{-\vep}\over (16\pi^2 \vep)^2} \,
\int dS \, \Bigl \{ - \half (1-\vep) V'''_{ijk} \,
g_{ijk\ell}^{\vphantom D} \, Q'_\ell + ( 1 - \half \vep - {\ts {1\over 6}}
\pi^2 \vep ) V'''_{ik\ell} \, V'''_{jk\ell} \, Q''_{ij} \cr
& ~~~~~~~~~~~~~~~~~~~~~~~~~- {\ts {1\over 6}} ( 1 - {\ts {1\over 12}} \vep -
\thir \pi^2 \vep ) V'''_{ijk} \, V'''_{ijk} \, K \Bigl \} ~. \cr}
\eqno (3.12) $$
For the remaining one particle reducible amplitudes we may use (B.27,28,31)
to give
$$ \eqalign {
\Gamma_c + \Gamma_d \sim {\mu^{-\vep}\over (16\pi^2 \vep)^2} \,
\int dS \, \Bigl \{& {\ts {1\over 8}} (1+\vep) V'''_{iik} \,
g_{jjk\ell}^{\vphantom D} \,
Q'_\ell + {\ts {1\over 8}}  V'''_{iik} \, V'''_{jj\ell} \, Q''_{k\ell} \cr
& - \half V'''_{ii\ell} \, V'''_{jk\ell} \, Q''_{jk}
+ {\ts {1\over 12}} (1+ {\ts {1\over 6}} \vep) V'''_{iik} \, V'''_{jjk} \, K
\Bigl \} ~. \cr}
\eqno (3.13) $$

Combining these we therefore obtain the somewhat lengthy result
$$ \eqalign {
Q^{(2)}_{\rm{c.t.}} = {1 \over (16\pi^2 \vep)^2} \,
\Bigl \{& g_{ijkk} \bigl ( 2 (Q''^3)_{ij} - (1-\vep) (Q''V'')_{ij} \cr
& ~~~ -{\ts{1\over 8}} (2+\vep) V'''_{ij\ell} Q'_\ell
- {\ts{2\over3}} (1-\vep) (Q''^2)_{ij} \, K -
\half (1 + {\ts {2\over 9}} \vep ) V''_{ij} \, K \bigl ) \cr
& + g_{ijk\ell} \bigl ( (3-\vep) Q''_{ij} V''_{k\ell} - 4 Q''_{ij}
(Q''^2)_{k\ell} + {\ts {4\over 3}} ( 2 - {\ts {1\over 3}} \vep) Q''_{ij}
Q''_{k\ell} \, K \bigl ) \cr
& + {\ts {1\over 8}} (1+\vep) V'''_{iik} \, g_{jjk\ell}^{\vphantom D} \,
Q'_\ell + {\ts {1\over 8}}  V'''_{iik} \, V'''_{jj\ell} \, Q''_{k\ell}
- \half V'''_{ii\ell} \, V'''_{jk\ell} \, Q''_{jk} \cr
& + {\ts {1\over 12}} (1+ {\ts {1\over 6}} \vep) V'''_{iik} \, V'''_{jjk} \,
K - \half (1-\vep) V'''_{ijk} \, g_{ijk\ell}^{\vphantom D} \, Q'_\ell \cr
& + ( 1 - \half \vep - {\ts {1\over 6}}
\pi^2 \vep ) V'''_{ik\ell} \, V'''_{jk\ell} \, Q''_{ij}
- {\ts {1\over 6}} ( 1 - {\ts {1\over 12}} \vep -
\thir \pi^2 \vep ) V'''_{ijk} \, V'''_{ijk} \, K \Bigl \} ~. \cr}
\eqno (3.14) $$

To obtain the renormalisation group equations we define, as in (2.24),
$$ \eqalign {
S_0 (\phi) = {}& S(\phi_0, g_0) + {\hat S} ( \phi_0, {\hat g}_0 ) +
{\tilde \C} (\lambda , {\hat \lambda}) ~, \cr
Q_0 (\phi_0) = {}& \mu^{-\vep} \bigl ( Q(\phi) +
Q_{\rm {c.t.}} (\phi) \bigl )  ~ , \cr}
\eqno (3.15) $$
for ${\tilde \C} (\lambda , {\hat \lambda})$ formed from the counterterms
$C(\lambda)$ and ${\hat C}({\hat \lambda})$ in this case. Since
$Q_{\rm {c.t.}}$ contains terms involving $K$, which is in general $\bhx$
dependent, it is necessary to regard the couplings ${\hat g}$ in $Q$ as also
$\bhx$ dependent to ensure multiplicative renormalisability or that $Q_0$
is obtained from $Q$ by ${\hat g}\to {\hat g}_0$.
In this case instead of (2.26) we may then define the $\beta$ functions by
$$ \eqalign {
\mu {d\over d\mu} S_0 = {}& \Bigl ( \D - \int_\M \! dv \, ({\hat \gamma}
\phi)_i
{\delta \over \delta \phi_i} \Bigl ) S_0 =
- {\tilde \C} (\beta^\lambda, \beta^{\hat \lambda} ) ~, \cr
& \D = \mu {\pr \over \pr \mu} + {\hat \beta}^V \! \cdot {\pr \over \pr V} +
{\hat \beta}^Q \! \cdot {\pr \over \pr Q} ~. \cr}
\eqno (3.16) $$
In particular ${\hat \beta}^Q$ is determined by
$$
\Bigl ( {\hat \beta}^V \! \cdot {\pr \over \pr V} +
{\hat \beta}^Q \! \cdot {\pr \over \pr Q} - ({\hat \gamma} \phi)_i
{\pr \over \pr \phi_i} - \vep \Bigl ) Q_0 =0 ~,~~
{\hat \beta}^Q (\phi) = \vep \bigl ( Q(\phi) - \half
Q'{}_{\! i} (\phi) \phi_i \bigl ) {}  + \beta^Q (\phi) ~.
(3.17) $$
Although $Q_0 , \, \beta^Q$ are both given as linear in $K$ it is possible
to generate $K^2$ terms in (3.17), with the assumed structure of counterterms
even if $Q'''=0$, but these may be neglected since they are not calculated
here. To two loops the $\beta$ function becomes
$$ \eqalignno {
\beta^Q - Q'_i (\gamma \phi)_i^{\vphantom D}= {}& {1\over 16 \pi^2} \,
\Bigl \{  - \half V'''_{iij} Q'_j + 2 (V'' Q'')_{ii}^{\vphantom D} -
{\ts {4\over 3}} (Q''^3)_{ii}^{\vphantom D} + {\ts {1\over 3}}\bigl (
4 (Q''^2)_{ii}^{\vphantom D} - V''_{ii} \bigl )  K \Bigl \}
\cr
& + {1 \over (16\pi^2)^2} \,
\Bigl \{ g_{ijkk}^{\vphantom D} \bigl ( 2 (Q''V'')_{ij}^{\vphantom D}
-{\ts{1\over 4}} V'''_{ij\ell} Q'_\ell \bigl ) {} -
2 g_{ijk\ell}^{\vphantom D} Q''_{ij} V''_{k\ell} \cr
& ~~~~~ + {\ts {1\over 4}} V'''_{iik} \, g_{jjk\ell}^{\vphantom D} \, Q'_\ell
+ V'''_{ijk} \, g_{ijk\ell}^{\vphantom D} \, Q'_\ell
- ( 1 + {\ts {1\over 3}} \pi^2) V'''_{ik\ell} \, V'''_{jk\ell} \, Q''_{ij} \cr
& ~~~~~ + {\ts {2\over 9}}g_{ijkk}^{\vphantom D} \bigl (
6 (Q''^2)_{ij}^{\vphantom D} - V''_{ij} \bigl )  K
- {\ts {8\over 9}} g_{ijk\ell}^{\vphantom D} Q''_{ij} Q''_{k\ell} \, K \cr
& ~~~~~ + {\ts {1\over 36}} \bigl ( V'''_{iik} \, V'''_{jjk}
+  ( 1 + 4 \pi^2 ) V'''_{ijk} \, V'''_{ijk} \bigl )  K
\Bigl \} ~. & (3.18) \cr}
$$
Instead of (2.27) the renormalisation group equation is now
$$ \Bigl ( \D + \int_\M \! dv \, ({\hat \gamma} J)_i {\delta \over \delta J_i}
\Bigl ) W (J) =  {\tilde \C} (\beta^\lambda , \beta^{\hat \lambda}) ~,
\eqno (3.19) $$
with $\D$ as in (3.16).

For the $O(n)$ symmetric case considered at the end of the last section, where
there is just a single coupling $g$, we may write
$$ \eqalign {
Q(\phi) = {}& c \, \half \phi^2 - {\hat h}_i \phi_i ~, ~~~~
{\beta}^Q (\phi) = \beta_c \half \phi^2 - {\hat \eta} \, {\hat h}_i
\phi_i  + {\hat \chi} ~, \cr
{\beta}_c  = {}&  \eta_c c  + {\hat \rho} K ~, ~~~
{\hat \chi} = {\hat \si}_m c m^2 + {\hat \si}_c  c^3
+ ( {\hat \rho}_m  m^2  + {\hat \rho}_c c^2 ) K
+ {\hat \omega}\, \pr_n m^2 ~, \cr}
\eqno (3.20) $$
where the term proportional to $\pr_n m^2$ is present if $m^2$ is dependent on
$x$ and is discussed further in the next section. The renormalisation group
equation (3.19), setting the curvature on $\M$ to zero, becomes
$$ \eqalignno {
\biggl (  \mu {\pr \over \pr \mu}& + {\hat \beta} {\pr \over \pr g} +
\int_\M \! dv \Bigl ( \gamma_m m^2 {\delta \over \delta m^2} +
{\hat \gamma} J_i {\delta \over \delta J_i} \Bigl ){}  +
\int_{\pr \M} \! \! dS \Bigl ( \beta_c {\delta \over \delta c}
+ (\half \vep + {\hat \eta}) {\hat h}_i {\delta \over \delta {\hat h}_i }
\Bigl ) \biggl ) W (J) \cr
& =  \mu^{-\vep}\int_\M \! dv \, \half p \, m^4 + \mu^{-\vep}\int_{\pr \M}
\! \! dS \, \bigl ( {\hat \chi} + C (\beta^{\hat \lambda}) \bigl ) ~,
& (3.21) \cr}
$$
where ${\hat \beta}^V (\phi)
= {1\over 24} {\hat \beta} \phi^4 + \half \gamma_m m^2 \phi^2 + \half p \,
m^4$.

The results of our calculations then imply to two loop order
$$ \eqalign {
{\eta}_c = {}& {\ts {1\over 3}} (n+2) u + {\ts {1\over 18}}
(1-4\pi^2) (n+2) u^2 ~, ~~
{\hat \eta} = - {\ts {1\over 6}} (n+2) u + {\ts {13\over 36}}
(n+2) u^2 ~, \cr
{{\hat \rho}\over {\eta}_c} = {}& -{{1\over 3}}\Bigl( 1 + {\ts {1\over 6}}
n u \Bigl ) ~, ~~~~~~
{\hat \rho}_m = -{1\over 16\pi^2}\, {n\over 3} \Bigl (1 + {\ts {2\over 9}}
(n+2) u \Bigl ) ~, \cr
{\hat \rho}_c = {}& {1\over 16\pi^2}\,
{4n\over 3} \Bigl (1 + {\ts {1\over 9}} (n+2) u \Bigl ) ~, ~~
{\hat \si}_m = {1\over 16\pi^2}\, 2n  ~, ~~
{\hat \si}_c = -{1\over 16\pi^2}\, {4n\over 3} ~, \cr}
\eqno (3.22) $$
where $\eta_c$ and ${\hat \eta}$ are in accord with the results of Diehl and
Dietrich [5] with $\eta_c \to \eta_c , \, {\hat \eta} = \half \eta_1 +
\gamma$. In a general regularisation scheme
$c$ is linearly divergent but as usual such contributions are absent with
dimensional regularisation.
\vfill\eject

\vskip 5pt
\leftline{\bf 4 Conformal Invariance and Consistency Relations}

The essential new results of the previous two sections has been to
calculate, to two loops, those counterterms necessary for a
renormalisable quantum scalar field theory which depend  on the
extrinsic curvature $K$ of the boundary. In this section we show how
consistency conditions resulting from considering arbitrary  local
Weyl rescalings of the metric on $\M$ and hence also of the induced metric
on $\pr \M$ may determine the $K$ dependent
counterterms without undertaking any calculations beyond those for a
flat spatial background and with a non curved boundary. The method is
an extension of that used to derive consistency relations which
determine the dependence of counterterms on the Riemann curvature for
general renormalisable field theories on manifolds without a boundary [18].

Under an infinitesimal Weyl rescaling the essential results here are
$$
\de_\si g^{\mu \nu} = 2 \si g^{\mu \nu} , \;\; \de_\si R = 2 \si R + 2
(d-1) \nab^2 \si\, , \;\; \de_\si n^\mu = \si n^\mu \, ,
\;\; \de_\si K = \si K + (d-1) \pr_n \si\, .
\eqno(4.1)
$$
The consistency conditions are derived by considering relations
involving the finite local scalar composite operators constructed from
the basic fields. To define these the essential couplings $g_I$,
$\hg_I$ parameterising the theory are extended to arbitrary local
functions $g_I(x)$, as well as $\hg_I(\bhx)$, or effectively
$V\to V(\phi,x)$, $Q\to Q(\phi,\bhx)$, so that local operators may be
obtained by functional differentiation. The quantum field theory
remains finite, in renormalised perturbation theory, so long as
appropriate additional counterterms depending on derivatives of the
couplings are introduced. In particular on $\pr \M$ there are possible
counterterms depending on $\pr_n g_I$.

Here we consider initially the basic scalar quantum field theory with
classical action (1.18) with Neumann boundary conditions so that there
is also a surface term as in (3.1). If $V(\phi) = \sum_I g_I v_I(\phi)$ and
$Q(\phi) = \sum_I \hg_I q_I(\phi)$, for some basis of quartic polynomials
$v_I(\phi) $ and quadratic polynomials $q_I(\phi)$,
then we define $\Del g_I$ in terms of $\Del
V(\phi) = 4 V(\phi) - V' \! {}_i(\phi) \phi_i$ and $\Del \hg_I$ by $\Del
Q(\phi)= 3 Q(\phi) - Q'\! {}_i(\phi) \phi_i$ so that if $\Del g_I = \Del_I
g_I$ and $\Del {\hat g}_I = {\hat \Del}_I {\hat g}_I$ then
$\Del_I$, $\hDel_I$ are the dimensions of the associated couplings.
Defining
$$ \Del_\si^W =  \int_\M \! dv \, \si \Bigl ( 2g^{\mu \nu} {\de \over
\de g^{\mu \nu}} + \Del g_I {\de \over
\de g_I} \Bigl ) ~,~~ {\hat \Del}_\si^W = \int_{\pr \M} \! \! dS \, \si
\, \Del {\hat g}_I {\de \over \de {\hat g}_I} ~, ~~
\Del_\si^\phi =  \int_\M \! dv \, \si \, \phi_i {\de \over \de \phi_i} ~,
\eqno (4.2) $$
then for $d=4$ by simple scale
invariance it is easy to see from (1.18) and (3.1) that
$$
\bigl ( \Del_\si^W + {\hat \Del}_\si^W + \Del_\si^\phi \bigl ) ( S + {\hat S})
= - \half \int_\M \! dv \, \phi^2 \, \nab^2 \si
- \half \int_{\pr \M} \! \! dS \, \phi^2 \, \pr_n \si ~.
\eqno(4.3)
$$
The $\nab^2 \si , \, \pr_n \si$ terms appearing on the r.h.s. of (4.3)
may be cancelled by introducing appropriate contributions involving $R\phi^2$
in $S$ and $K\phi^2$ in $\hat S$ which ensures a classically Weyl invariant
theory under local rescalings of the metric as in (4.1) if there are no
couplings with dimension, $\Del g_I = \Del {\hat g}_I = 0$. Later we show
how this may be extended to the quantum theory at a critical point ${\hat
\beta} (g) =0$.

In the full quantum field theory, for general $d$, where $S+ \hS \to S_0$
the corresponding equations representing local changes in scale
involve the $\beta$ functions of the various couplings and the anomalous
dimension of $\phi$. Defining in addition to (4.2)
$$
\Del_\si^\beta =  \int_\M \! dv \, \si \, {\hat \beta}^g_I {\de \over
\de g_I} \, ,~~
{\hat \Del}_\si^\beta = \int_{\pr \M} \! \! dS \, \si \,
{\hat \beta}^{\hat g}_I {\de \over \de {\hat g}_I} \, , ~~
\Del_\si^{{\hat \gamma}\phi} = \int_\M \! dv \, \si \,
({\hat \gamma}\phi)_i {\de \over \de \phi_i} \, .
\eqno(4.4)
$$
then the local renormalisation equation may be expanded in the form
$$ \eqalignno {
\bigl ( \Del_\si^W + {\hat \Del}_\si^W + \Del_\si^\phi +
\Del_\si^{{\hat \gamma}\phi} - & \Del_\si^\beta
- {\hat \Del}_\si^\beta \bigl )  S_0  = \mu^{-\vep} \! \int_\M \! dv \, \si \,
C(\beta^\lambda) + \mu^{-\vep}\!  \int_{\pr \M} \! \! dS \, \si \,
{\hat C}(\beta^{\hat \lambda}) \cr
& - \mu^{-\vep} \! \int_\M \! dv \, \bigl (\pr_\mu \si \Z^\mu +
\nab^2 \si \, \T \bigl ) {}
- \mu^{-\vep} \! \int_{\pr \M} \! \! dS \, \pr_n \si \, \Y ~, & (4.5) \cr}
$$
where $\Z^\mu,\, \T$ and $\Y$ may depend locally on $\phi$ but not its
derivatives. Since $g_I$ is required to be $x$ dependent for a consistent
treatment it is necessary to include additional terms in
$\beta^V(\phi)=\sum_I \beta_I^g v_I(\phi)$, and also in $C(\beta^\lambda)$,
depending on $\pr_\mu g_I$ which have been discussed in detail in ref. [18].
To absorb all counterterms by a redefinition of couplings
in this case it is also necessary to introduce an
external gauge field coupled to $\phi$ but this is omitted here since it is
unnecessary for obtaining the essential results of our discussion.
When $\si$ is constant then (4.3) is equivalent to combining the usual
renormalisation group equation (3.16) describing variations in $\mu$
together with a simple scaling relation. Since $S_0$ defines a
finite theory for arbitrary $g_{\mu \nu} (x)$ and
$g_I (x), \, {\hat g}_I(\bhx)$ the l.h.s. of (4.3) represents a
linear combination of finite local operators and hence the new quantities
$\Z^\mu$, $\T$ and on the boundary $\Y$ appearing on the r.h.s. of (4.3) must
therefore also be expressible in terms of finite local operators. In the first
paper of [18]
$\Z^\mu$ and $\T$ have been calculated to two loops and consistency
relations following from the requirement of finiteness have been analysed.

Here we consider just the surface contributions depending on $\pr_n
\si$. To calculate $\Y$ in (4.5) it is convenient to decompose the
counterterms for $Q$ in the form
$$
Q_{\rm {c.t.}} (\phi) = q(\phi) +  \rho(\phi) K + w_I (\phi) \pr_n g_I ~,
\eqno(4.6)
$$
where the last term is new in this case when the couplings are allowed
to be $x$ dependent. It is then easy to see from the defining
equations (4.3) and (4.1) that in general
$$ \Y (\phi)= u(\phi) - (3- \vep) \rho (\phi) - w_I (\phi)
(\Delta g_I - {\hat \beta}^g_I ) ~,
\eqno(4.7)
$$
if $u(\phi)$ is the surface contribution resulting
when the volume integrals are integrated by parts to achieve the
desired form (4.5).

The results for $\rho$ to two loops are contained in (3.7) and (3.14).
For $\omega_I(\phi)$ then a straightforward extension of the
calculations in section $3$ gives
$$ \eqalignno{
w_I^{(1)}\pr_n g_I = {}& - {1\over 16\pi^2 \vep} \, \half \pr'\! {}_n
V''_{ii} ~, & (4.8a) \cr
w_I^{(2)}\pr_n g_I = {}& {1\over (16\pi^2 \vep)^2} \Bigl \{ - \quar
(1+\half \vep) g_{ijkk} \pr'\! {}_n V''_{ij} + \eight (1+\vep) V'''_{iik}
\pr'\! {}_n V'''_{jjk} \cr
& ~~~ - \half (1-\vep) V'''_{ijk} \pr'\! {}_n V'''_{ijk}
- {\ts {3\over 4}} \pr_n g_{ijkk} \, V''_{ij} +
(1+\vep) \pr_n g_{ijkk} (Q''^2)_{ij} \Bigl \} \, , & (4.8b) \cr}
$$
where $\pr'\! {}_n = n^\mu \pr'\! {}_\mu$ with $\pr'\! {}_\mu$ denoting
the derivative at constant $\phi$. To obtain $u(\phi)$
then taking into account (4.3) and the
derivative terms which appear at two loops in the volume integral as
in (3.12) to this order gives
$$ u(\phi) = \half (1-\half \vep) \phi^2 - {1\over (16\pi^2)^2 \vep} \,
{\ts {1\over 24}}(1+\half\vep)\, V'''_{ijk} (\phi)  V'''_{ijk} (\phi)
+ \dots ~. \eqno(4.9)
$$

As already remarked the consistency conditions stem from the fact that
each term in (4.5) must separately be finite as operators, or on insertion in
the regularised functional integral defined by $S_0$. Since arbitrary finite
local composite operators corresponding to $h(\phi)= h_I q_I(\phi)$ on
$\pr \M$, for $h(\phi)$ at most quadratic in $\phi$, may be defined by
$$
h_I {\de \over \de {\hat g}_I(\bhx)} \, S_0 ~,
\eqno (4.10) $$
then we may write
$$
\mu^{-\epsilon} \Y (\phi) = h_I {\de \over \de {\hat g}_I} \, S_0 ~,
\eqno(4.11)
$$
for suitable $h(\phi)$, non singular as $\vep\to 0$, in this case.
As a consequence of minimal subtraction (4.11) all terms involving poles in
$\vep$ in $\Y(\phi)$ are determined in terms of $h(\phi)$ and the $Q$
dependent counterterms in $S_0$. $h(\phi)$ also is determined by the $\O(1)$
and $\O(\vep)$ terms in (4.11). It is easy to read off from (4.7) and (4.9)
the lowest order contributions
$$ h(\phi) = \half (1-\half \vep) \phi^2 + {1\over 16\pi} \,
{\ts {2\over 3}}\, \bigl (V''_{ii} (\phi) -
Q''_{ij}(\phi)Q''_{ij}(\phi) \bigl ) {} + \dots ~,
\eqno(4.12) $$
where the one loop contribution arises from $\vep
\rho^{(1)}(\phi)$ in (4.7) and also from the $-\quar\vep\phi^2$ term in
$h(\phi)$ in conjunction with $q^{(1)}$. (4.11) then provides the crucial
consistency relation which in effect determines
$\rho$ in (4.6) in terms of lower order results in the loop expansion.
We have verified that this confirms the one and two loop results of
section 3. As an example from (3.7)
$$
3\rho^{(1)} = {1\over 16\pi^2\vep} \bigl( 4(Q''^2)_{ii} - V''_{ii} \bigl )
{}~,~~~
q^{(1)} = {1\over 16\pi^2\vep} \bigl( 2(V''Q'')_{ii} - \half V''_{iij} Q'_j
- {\ts {4\over 3}} (Q''^3)_{ii} \bigl ) ~,
$$
and from (4.8a)
$$
w_I^{(1)} \Del g_I = - {1\over 16\pi^2\vep} \bigl( V''_{ii} - \half
V'''_{iij} \phi_j \bigl ) ~,
$$
so that it is easy to verify (4.11) at the lowest non trivial order since
$$
3\rho^{(1)} + w_I^{(1)} \Del g_I = - h^0_I {\pr \over \pr {\hat g}_I} q^{(1)}
{}~, \eqno (4.13) $$
for $h^0 (\phi) = \half \phi^2$. The same procedure has also been used to
check $\rho^{(2)}$ using $h$ in (4.12) although many more terms contribute.

{}From (4.5) we may derive a local renormalisation group equation for the
functional $W$. Restricting to the $O(n)$ symmetric model considered at the
end of sections 2 and 3 and also assuming the coupling $g=g_*$ is at the
critical point where ${\hat \beta} (g_*)=0$ then defining
$$ \eqalign {
\Del_\si = {}& \int_\M \! dv \, \si \, \Bigl ( 2g^{\mu \nu} {\de \over
\de g^{\mu \nu}}
+ \bigl( 2-\gamma_{m*}\bigl ) m^2 {\de \over \de m^2}
+ \bigl(3-\vep- {\hat \gamma}_{*}\bigl ) J_i {\de \over \de J_i} \Bigl ) \cr
& + \int_{\pr \M} \! \! dS \, \si \, \Bigl (
\bigl( 1-\eta_{c*}\bigl ) c {\de \over \de c}
+ \bigl(2- \half \vep- {\hat \eta}_{*}\bigl ) {\hat h}_i {\de \over \de {\hat
h}_i} \Bigl ) ~, \cr
\Del'\! {}_\si = {}& \int_\M \! dv \, \bigl ( \nab^2 \si + \half \si \, R'
\gamma_{m*} \bigl ) \tau {\de \over \de m^2}
+ \int_{\pr \M} \! \! dS \, \bigl ( \pr_n \si + \si \, K'
\eta_{c*} \bigl ) {\hat \theta} {\de \over \de c} ~, \cr}
\eqno (4.14) $$
where
$$
R' = {1\over d-1} \, R ~, ~~~~~ K' = {1\over d-1} \, K ~,
$$
the local renormalisation group equation has the general form
$$
\bigl ( \Del_\si + \Del' \! {}_\si \bigl ) W(m^2,c,J,{\hat h}) =
X ( m^2 - \half \tau R', c - {\hat \theta} K' ) ~,
\eqno (4.15) $$
with $X$ a volume or surface integral of local scalar functions of $m^2$ and
$c$ proportional to $\si$ or its derivatives of dimension 4 or 3 respectively.
The arguments of $X$ are written in the form shown in (4.15), without any loss
of generality, for later convenience. From (4.5) and (4.10) the coefficients
${\hat \theta}, \, \iota_c , \, \iota_m$ are determined by writing
$h(\phi)$ as
$$ h(\phi) = \half {\hat \theta}\, \phi^2 + \iota_c c^2 + \iota_m m^2 ~,
\eqno (4.16) $$
with the coupling $g$ evaluated at the critical point. Using (3.20,21) with
$g=g_*$ and the consistency relations (4.7), together with previous
results in the case of theories without boundary [18], gives
$$ \eqalign { \! \! \! \! \! \!
X(m^2,c) = {}& - \mu^{-\vep} \int_\M \! dv \, \Bigl \{ \si \bigl ( \half p
m^4 -\beta_a F - \beta_b G \bigl ) {} + \bigl ( \nab^2 \si - \half \si R'
(\gamma_{m*} -\vep ) \bigl ) km^2 \Bigl \} \cr
& - \mu^{-\vep} \int_{\pr \M} \! \! dS \, \Bigl \{ \si \bigl (
{\hat \si}_c c^3 + {\hat \si}_m cm^2 + {\hat \rho}'\! {}_m K' m^2  +
{\hat \omega} \, \pr_n m^2 \bigl ) \cr
& ~~~~~~~~~~~~~~~~~~~  -\pr_n \si \, \iota_m m^2 - \bigl ( \pr_n \si - \si K'
(2\eta_{c*} - \vep ) \bigl ) \iota_c c^2 + \dots \Bigl \} ~, \cr}
\eqno (4.17) $$
with
$$
{\hat \rho}'\! {}_m = (d-1) {\hat \rho}_m + {\hat \theta}\, {\hat \si}_m =
(\gamma_{m*} - \vep) \iota_m - (2-\gamma_{m*}) {\hat \omega} ~.
\eqno (4.18) $$
The neglected terms in the surface integral in (4.17) are linear in $c$ and
involve invariants constructed from $K_{ij}$ and the Riemann tensor on $\M$
for which the corresponding counterterms are not calculated here.

The restricted form of the functional operator $\Del'\! {}_\si$ in (4.14) and
also $X$ in (4.17) is a consequence of the consistency relations such as
given by (4.7) and (4.11) at the critical point. Alternatively the
expressions given by (4.14) and (4.17) may be regarded as determined  by the
integrability condition $[\Del_\si + \Del'\! {}_\si , \, \Del_{\si'} +
\Del'\! {}_{\si'} ] W =0$, without any commitment to a specific regularisation
scheme [18]. Such conditions also require the absence of a $R^2$ contribution
in addition to the $F,G$ curvature terms in (4.14). From (4.12) and (4.16)
we may find to one loop
$$
{\hat \theta} = 1 - \half \vep + {\ts {2\over 9}} (n+2) u_* ~,~~~
\iota_c = - {2\over 3} \, {n\over 16\pi^2} ~,~~~
\iota_m = {2\over 3} \, {n\over 16\pi^2} ~,
\eqno (4.19) $$
while from (4.8a,b) to two loops
$$
{\hat \omega} = - {1\over 2} \, {n\over 16\pi^2} \bigl ( 1 + {\ts {1\over 6}}
(n+2) u_* \bigl ) ~.
\eqno (4.20) $$
In terms of the definition of $\beta_c$ in (3.20) then (4.14) requires that
at the critical point $(d-1){\hat \rho} = - \eta_c {\hat \theta}$ which along
with (4.18) may be verified with the results (3.22) and (4.19,20) using
$ \gamma_{m*} = {1\over 3} (n+2)u_* + {\rm O} ( u_*^2)$. Even at the critical
point $W$ is arbitrary up to the addition of local functionals of $m^2$ and
$c$. For
$$ \de W = \mu^{-\vep} \int_{\pr \M} \! \! dS \, \bigl ( r K' m^2 + s \, \pr_n
m^2 \bigl ) {} - \mu^{-\vep} \int_{\M} \! dv \, t \, R' m^2 ~,
$$
then
$$
\de \iota_m = r + (2-\gamma_{m*}) s ~,~~~ \de {\hat \omega} = (\gamma_{m*} -
\vep ) s ~, ~~~ \de k = 2t ~,
$$
so that in the $\vep $ expansion it is possible to set $\iota_m$ and $k$ to
zero but not ${\hat \omega}$.

Given the form of the renormalisation group equation (4.14,15) at the
critical point it is straightforward to see that it may be rewritten as
$$
\Del_\si W(m^2 + \half \tau R',c+{\hat \theta} K',J,{\hat h}) =
X ( m^2 , c ) ~.
\eqno (4.21) $$
Manifestly $W(\half \tau R' , {\hat \theta}K', 0,0)$ is invariant under local
Weyl transformations, apart from the contributions involving $X$ on the
r.h.s. of (4.21). This is a consequence of the result that, at least for
simple field theories, invariance with respect to constant scale
transformations leads to symmetry under the full conformal group [19]. Using
previous results [18] in the $\vep$ expansion $\tau = 1- \half \vep +
{1\over 54} (n+2) u_*^3 + {\rm O}(\vep^4)$ so there is a modification of the
usual coefficient of the $\phi^2 R$ term in a free scalar theory for conformal
invariance in the interacting case at three loops.

{}From (4.21) it is straightforward to determine the behaviour of the composite
operator $\phi^2$ in the neighbourhood of the boundary at the critical point.
Defining
$$
\mu^{-\vep} \langle \phi^2 (x) \rangle = - {\de \over \de m^2(x)}
W(m^2+ \half \tau R' , {\hat \theta}K', 0,0) \Bigl |_{m^2=0} ~,
\eqno (4.22) $$
then from (4.21) we may obtain, with the standard coordinates $x=(\bx,y)$,
$$ \eqalign {
\Bigl ( \int_\M \! dv \, \si \, 2g^{\mu \nu} {\de \over \de
g^{\mu \nu}}& - x_m \, \si(\bx,y) \Bigl ) \langle \phi^2 (\bx,y) \rangle \cr
= {}& - {\hat \omega} \, \si^0(\bx) \bigl ( \de'(y) - (x_m -1) K'(\bx) \,
\de (y) \bigl ) \cr
& - \iota_m  \bigl ( \pr_n \si (\bx) - (x_m -2) \si^0(\bx) K'(\bx)
\bigl ) \de (y) ~, \cr}
\eqno (4.23) $$
where $x_m = d -2 + \gamma_{m*}$ and $\si^0(\bx) = \si(\bx,0),\, \pr_n \si
(\bx) = \pr _y \si (\bx,y)|_{y=0}$. Using that under a Weyl rescaling as in
(4.1)
$$
\de_\si y = - \si^0(\bx) \, y - \half \pr_n \si(\bx) \, y^2 - {\ts {1\over
6}} \nab_n^2 \si (\bx) \, y^3 + \dots
\eqno (4.24) $$
then we may solve (4.23) for the leading singular behaviour as $y\to 0$,
$$ \eqalign {
\langle \phi^2 (\bx,y) \rangle \sim {}& \bigl ( y^{-x_m} + \half x_m K'(\bx)
\, y^{-x_m +1} \bigl ) C \cr
& - {{\hat \omega} \over 2-x_m} \, \de '(y) - \Bigl ( \iota_m + {1-x_m \over
2-x_m} \, {\hat \omega} \Bigl )K'(\bx) \de (y) ~. \cr}
\eqno (4.25) $$
For $ y>0$ this is in agreement with a general formula of Cardy [9],
apart from a sign. The $\de'(y), \, \de(y)$ terms reflect the regularisation
of the singularity for $x_m \to 2$, when it is necessary that $C\to - {\hat
\omega}$.

A similar discussion to the above case can also be essentially carried out in
the Dirichlet case, corresponding to the calculations of section 2, with
similar consequences but it is not presented because of some technical
problems.

\vfill\eject

\vskip 5pt

\leftline{\bf 5 Schr\"odinger Equation}

The results of section 2 verify the feasibility of defining in perturbation
theory a finite wave functional ${\hat \Psi} (\hphi)$ for renormalisable
scalar quantum field theories on a manifold $\M$ where the quantum fields
$\phi$ are
specified by an arbitrary smooth field $\hphi$ on the boundary $\pr \M$.
For such wave functionals there is a corresponding Schr\"odinger equation
which describes the variation of ${\hat \Psi}$ under smooth local deformations
of $\pr \M$. A discussion related to that presented here was also given by
Symanzik [10] although only constant global shifts of the boundary were
considered. The relevant Hamiltonian is a functional differential operator
requiring careful regularisation as is discussed subsequently.

The wave functional $\hat \Psi$ is defined formally by
$$ {\hat \Psi} ( \hphi) = \int_{\phi |_{\pr \M} = \hphi} \! \! \! \! \! \!
\! \! \! \! \! \! \! \! d[\phi]\, \, e^{ -{1\over \hbar}S_0(\phi)} ~,
\eqno(5.1)
$$
where we have introduced Planck's constant $ \hbar$ in order to set up a
semiclassical expansion in powers of $\hbar$. In general the boundary of
$\M$ is supposed to be determined by $x^\mu (\bhx)$ for $\bhx$ coordinates
on $\pr \M$ and we then consider variations along the normal to $\pr \M$ by
taking
$$ \de x^\mu (\bhx) = - n^\mu (\bhx) \, \de t (\bhx) ~.
\eqno (5.2) $$
The implications of such variations on the Green functions $G_\Delta $
defined for operators of the form (1.2) with Dirichlet boundary conditions on
$\pr \M$ and other quantities, such as the extrinsic curvature $K_{ij}$,
whose definition also
depends on the specification of $\pr \M$ is considered in appendix C. The
functional Schr\"odinger equation for $\hat \Psi$ has the general form
$$
\Bigl ( \hbar {\de \over \de t (\bhx)} + \H_0 (\bhx) \Bigl ) {\hat \Psi}
(\hphi) =0 ~, ~~~ {\de \over \de t (\bhx)} \equiv - n^\mu (\bhx)
{\de \over \de x^\mu (\bhx)} ~,
\eqno (5.3) $$
where $\H_0$ is the corresponding Hamiltonian operator. At lowest order in the
loop expansion, as shown subsequently, it is sufficient to let
$$
\H_0 \to \H = - \half \mu^\vep \hbar^2 \, {\de^2 \over \de \hphi^2} +
\mu^{- \vep} \V (\hphi) ~,
\eqno (5.4) $$
where $\V$ is a potential depending locally on $\hphi$ and its derivatives in
$\pr \M$. Although $\hat \Psi$
is finite for arbitrary $\hphi$ and smooth boundary surfaces
$\pr \M$ it is necessary to take account also of additional divergences
arising from the second order functional derivative in $\H$, in general
$ \de^2 /\de \hphi (\bhx) \de \hphi (\bhx') $ is singular as $\bhx' \to
\bhx$. In our calculations using dimensional regularisation is also sufficient
to ensure a well defined Hamiltonian operator $\H_0$ may be found, satisfying
(5.3), which differs from $\H$ in (5.4) by local counterterms which are just
poles in $\vep$. These additional counterterms have essentially
a similar form to the terms already present in $\H$. As shown later the
detailed
structure of $\H_0$ is strongly constrained by the renormalisation group
equations for $\hat \Psi$ expressing independence of the arbitrary scale $\mu$.

The semiclassical expansion to the wave functional ${\hat \Psi}(\hphi)$ is
derived by a loop expansion of the functional integral (5.1) where the field
$\phi$ is also expanded about a background classical solution $\vphi_\c$,
with boundary conditions $\vphi_\c|_{\pr \M} = \hphi$.
If the classical action $S_\c (\hphi) = S(\vphi_\c,g)$, since the boundary
term ${\hat S}$ in (2.1) vanishes for $\vphi_\c$,
then to one loop order the semiclassical approximation to $\hat \Psi$ is,
from (2.10) with ${\hat \Psi}(\hphi) = \exp \Gamma (\hphi;\vphi_\c)$, given by
$$
{\hat \Psi}_{\rm {s.c.}}(\hphi) = \bigl ( \det \Delta_{\rm {reg}}
\bigl )^{- \hh} \, e^{- {1\over \hbar} \mu^{-\vep} S_\c (\hphi) } ~,
\eqno (5.5) $$
where $\vphi_\c$ is the classical solution, with boundary value $\hphi$, and
$\det \Delta_{\rm {reg}}$ is the dimensional regularised functional determinant
for Dirichlet boundary conditions defined as in (C.18). In order to discuss
the Schr\"odinger equation (5.3) it is necessary to determine the effect on
the classical solution $\vphi_\c$ of independent variations in the boundary
value $\hphi$ and also of deformations of $\pr \M$ as in (5.2). In both cases
the variation $\de \vphi_\c$ satisfies $\Delta \de \vphi_\c =0$. For changes
in $\hphi$ then the boundary condition is simply $\de \vphi_\c |_{\pr \M} =
\de \hphi$ and hence
$$ {\de \vphi_{\c i} (x)\over \de \hphi_j (\bhx) } = G_{ij} (x,x')
{\overleftarrow \pr} {}' {}_{\! \! n} \bigl |_{x' = x(\bhx)} ~,
\eqno (5.6) $$
with $G_{ij}(x,x')$ the Dirichlet Green function for $\Delta$. For changes in
the boundary $\pr \M$, as represented by (5.2), then we require
$$
(\vphi_\c + \de \vphi_\c ) \bigl |_{x = x(\bhx) + \de x(\bhx)} =
\hphi (\bhx) ~~\Rightarrow ~~\de \vphi_\c (x)\bigl |_{x=x(\bhx)} =
\de t (\bhx) \, \pr_n \vphi_\c (\bhx)  ~,
\eqno (5.7) $$
and thus
$$ {\de \vphi_{\c i} (x)\over \de t (\bhx) } = G_{ij} (x,x')
{\overleftarrow \pr} {}' {}_{\! \! n}\bigl |_{x' = x(\bhx)}
\pr_n \vphi_{\c j} (\bhx) ~.
\eqno (5.8) $$
{}From (5.7) and (5.8) it is easy to see that
$$
\Bigl ( {\de \over \de t} - \pr_n \vphi_{\c i} {\de \over \de \hphi_i}
\Bigl ) \vphi_\c (x) = 0 ~,
\eqno (5.9)$$
which is important in later calculations.

The verification that the semiclassical form for the wave function in (5.5)
is an approximate solution of the Schr\"odinger equation, which becomes exact
for quadratic Hamiltonians, is well known in conventional quantum mechanics
[20].
Here we extend the analysis to the next leading, or two loop, approximation
which is straightforward but also pay special attention to the divergences
which arise in a field theory context. To start it is necessary to consider
how the classical action $S_\c$ depends on $\hphi$ and also its dependence
on changes in $\pr \M$ as given by (5.2). By integration by parts and
using the classical equation for $\vphi_\c$, derived by requiring $S_\c$
to be stationary at $\phi=\hphi$,
$$
S_\c (\hphi) = \int_\M \! dv \, \bigl ( V(\vphi_\c) - \half V'{}_{\! i}
(\vphi_\c) \vphi_{\c i} \bigl ){} - \half \int_{\pr \M} \! \! dS \, \hphi_i
\pr_n \vphi_{\c i} ~.
$$
With this result then under arbitrary variations in $\hphi$ and also
$\vphi_\c$, subject to $\Delta \de \vphi_\c = 0$, we find
$$
\de S_\c (\hphi) = - \half \int_{\pr \M} \! \! dS \, ( \de \vphi_{\c i} +
\de \hphi_i ) \pr_n \vphi_{\c i} ~.
$$
Hence, taking $\de \vphi_\c = \de \hphi$,
$$
{\de \over \de \hphi_i (\bhx) } S_{\c} (\hphi) = - \pr_n \vphi_{\c i}
(\bhx) ~,
\eqno (5.10) $$
while, using (5.7) and also (C.8a,b) with (C.5) and the equation of motion for
$\vphi_\c$ on the boundary, we derive a local version of the Hamilton-Jacobi
equation
$$
{\de \over \de t (\bhx) } S_{\c} (\hphi) = - \half \pr_n \vphi_{\c i}
(\bhx) \pr_n \vphi_{\c i} (\bhx) + V(\hphi(\bhx)) + \half {\hat \gamma}
^{pq} \pr_p \hphi_i (\bhx) \pr_q \hphi_i (\bhx) ~.
\eqno (5.11) $$

The essential results (5.10,11) ensure that at lowest order in (5.3) it is
sufficient to take for $\V$ in (5.4)
$$
\V (\hphi) =  V(\hphi) +
\half {\hat \gamma}^{pq} \pr_p \hphi_i \pr_q \hphi_i ~,
\eqno (5.12) $$
since then
$$
\Bigl ( \hbar {\de \over \de t} + \H \Bigl )  e^{- {1\over \hbar}
\mu^{-\vep} S_\c (\hphi) } = \half \hbar \, {\de^2 \over \de \hphi^2}
S_\c (\hphi)\, e^{- {1\over \hbar} \mu^{-\vep} S_\c (\hphi)} ~,~~
{\de^2 \over \de \hphi(\bhx)^2} S_\c (\hphi) = -\K_{ii} (\bhx,\bhx) ~,
\eqno (5.13) $$
from (5.6) and (5.10) where $\K$ is defined as in (C.22),
$$
\K_{ij} (\bhx,\bhx') = \pr_n G_{ij} (x,x') {\overleftarrow \pr}{}'{}_{\! \! n}
\bigl |_{x=x(\bhx),x'=x(\bhx')} ~.
\eqno (5.14) $$

{}From (5.9) and (C.21) we may find
$$
\Bigl ( {\de \over \de t} - \pr_n \vphi_{\c i} {\de \over \de \hphi_i}
\Bigl )\, {\rm ln} \det \Delta_{\rm {reg}} = {\de^2 \over \de \hphi^2} S_\c
(\hphi) + {2 \mu^{-\vep} \over 16\pi^2\vep} \, \tr ({\hat \B}_4 ) ~,
\eqno (5.15) $$
where the pole term in $\vep$ is necessary for the subtraction of divergences
arising from $\de^2/\de \hphi^2$ in this case. Explicitly from (C.17),
considering only terms which depend on $V$ and are at most linear in $K$,
and using the equation of motion on $\pr \M$
$\nab_n {}^{\! 2} \vphi_\c - K \pr_n \vphi_\c + \hnab^2 \hphi = V'(\hphi)$,
$$ \eqalign {
{\hat \B}_{4ij} = {}& \half V''_{ik} V''_{kj} - \half \nab_n {}^{\! 2}
V''_{ij} - {\ts {1\over 6}} \hnab^2 V''_{ij} + {\ts {5\over 6}} K \pr_n
V''_{ij} +\dots ~, \cr
\tr ( {\hat \B}_{4}) = {}& \half V''_{ij}(\hphi) V''_{ij} (\hphi) -
\half V'''_{iij}  (\hphi) V'_j (\hphi) - \half g_{ijkk} \pr_n \vphi_{\c i}
\pr_n \vphi_{\c j} \cr
& - {\ts {1\over 6}} g_{ijkk} {\hat \gamma}^{pq} \pr_p \hphi_i \pr_q \hphi_i
+ {\ts {1\over 3}} V'''_{iij}(\hphi) \hnab^2 \hphi_j +
{\ts {1\over 3}} K V'''_{iij}(\hphi) \pr_n \vphi_{\c j} ~, \cr}
\eqno (5.16) $$
which is a local function of $\hphi$ and $\pr_n \vphi_\c$ on $\pr \M$.
In consequence from (5.13) and (5.15)
$$ \eqalign {
\Bigl ( \hbar {\de \over \de t} + \H \Bigl ) & {\hat \Psi}_{\rm {s.c.}}(\hphi)
= - {\hbar \mu^{-\vep} \over 16\pi^2\vep} \, \tr ({\hat \B}_4 ) \,
{\hat \Psi}_{\rm {s.c.}}(\hphi) + \hbar^2 \X_1 \, {\hat \Psi}_{\rm {s.c.}}
(\hphi) ~, \cr
& \X_1 = {\ts {1\over 4}} \mu^\vep \Bigl ( {\de^2 \over \de \hphi^2} \,
{\rm ln} \det \Delta_{\rm {reg}} - \half \,
{\de \over \de \hphi_i} \, {\rm ln} \det \Delta_{\rm {reg}} \,
{\de \over \de \hphi_i } \, {\rm ln} \det \Delta_{\rm {reg}} \Bigl ) ~. \cr}
\eqno (5.17) $$
To first order in $\hbar$ $\X_1$ can be neglected and so
${\hat \Psi}_{\rm {s.c.}}$ satisfies the Schr\"odinger equation (5.3) in the
leading semiclassical approximation if
$$ \eqalign { \! \! \! \!
\H_0^{(1)} = {}& {\hbar \over 16\pi^2 \vep} \Bigl (
- \half \mu^\vep \hbar^2 \, g_{ijkk} {\de^2 \over \de \hphi_i \de \hphi_j} +
{\ts {1\over 3}}\hbar K \, V'''_{iij}(\hphi) {\de \over \de \hphi_j} \Bigl ){}
+ \mu^{- \vep} \V_0^{(1)} (\hphi) ~, \cr
\! \! \! \! \V_0^{(1)} (\hphi) = {}& {\hbar \over 16\pi^2\vep}\bigl ( \half
V''_{ij}(\hphi) V''_{ij} (\hphi) - \half V'''_{iij}  (\hphi) V'_j (\hphi)
- \half g_{ijkk} {\hat \gamma}^{pq} \pr_p \hphi_i \pr_q \hphi_i
+ {\ts {1\over 3}} \hnab^2 V'''_{ii}(\hphi) \bigl ) ~, \cr}
\eqno (5.18) $$
since, with these expressions, it is easy to see that
$$ \eqalign {
\H_0^{(1)} {\hat \Psi}_{\rm {s.c.}}(\hphi)
= {}& {\hbar \mu^{-\vep} \over 16\pi^2\vep} \, \tr ({\hat \B}_4 ) \,
{\hat \Psi}_{\rm {s.c.}}(\hphi) + \bigl ( \hbar^2 (\X_2 + \X_3) +
{\rm O} (\hbar^3) \bigl ) {\hat \Psi}_{\rm {s.c.}}(\hphi) ~, \cr
\X_2 = {}& \half \, {1 \over 16\pi^2 \vep} \bigl (
g_{ijkk} \pr_n \vphi_{\c i} -
{\ts {1\over 3}} K \, V'''_{iij}(\hphi) \bigl ) {\de \over \de \hphi_j} \,
{\rm ln} \det \Delta_{\rm {reg}} ~, \cr
\X_3 = {}& \half \, {1 \over 16\pi^2 \vep} \, g_{ijkk}
{\de^2 \over \de \hphi_i \de \hphi_j} S_\c (\hphi) ~,~~~
{\de^2 \over \de \hphi_i (\bhx) \de \hphi_j (\bhx)} S_\c (\hphi)
= - \K_{ij} (\bhx,\bhx) ~. \cr}
\eqno (5.19) $$

At the next order in the semiclassical expansion of the wave functional
we may write
$$
{\hat \Psi}(\hphi) = {\hat \Psi}_{\rm {s.c.}}(\hphi) \,
e^{\hbar \Gamma_{\rm {reg}}^{(2)} + {\rm {O}}(\hbar^2)} ~, ~~~~~
\Gamma_{\rm {reg}}^{(2)} = \Gamma^{(2)} - S_0^{(2)} ~,
\eqno (5.20) $$
where $\Gamma^{(2)} = \Gamma_a + \Gamma_b + \Gamma_c + \Gamma_d$ as given in
(2.13), (2.15), (2.17) and (2.18) while $S_0^{(2)}(\vphi_\c)$ is the overall
local two loop counterterm which subtracts the divergent poles in $\vep$.
The calculations of section 2 determine the specific boundary terms.
{}From (2.7,9) and using the equations of motion for $\vphi_\c$ this
may be expressed in the convenient form
$$
S_0^{(2)}(\vphi_\c)  =  \int_{\M} \! dv \, V_0^{(2)} (\mu^{-\hh \vep} \vphi_\c
)
+ \int_{\pr \M} \! \! dS\, \bigl ( - \mu^{-\hh \vep}\hphi^{(2)}_{0i}(\hphi)
\pr_n \vphi_{\c i} + \mu^{- \vep}\rho^{(2)}(\hphi) K \bigl )
{} + \dots ~,
\eqno(5.21) $$
where $V_0 (\phi_0) = \mu^{-\vep} ( V(\phi) + V_{\rm {c.t.}} (\phi) )$
and $\hphi_0^{(2)}$ is given in terms of $Z^{(2)}$ and
$\hphi_{\rm {c.t.}}^{(2)}$ by (2.25).
In order to evaluate $\X_1 , \, \X_2$ in (5.16,18) we need to obtain an
analogous formula to (C.19) expressing the dependence of
$\det \Delta_{\rm {reg}}$ on $\hphi$. Using (5.6) and also the explicit form
of the counterterms given by (1.17) in this case
$$ \eqalign {
{\de \over \de \hphi_i (\bhx) } \, {\rm ln} \det \Delta_{\rm {reg}}
= {}& \int dv \, {\bar G}_{mn}| (x)\, V'''_{mnj} (\vphi_\c) \, G_{ji} (x,\hx)
{\overleftarrow \pr} {}_{\! \! n} \bigl |_{\hx = x(\bhx)} \cr
 + {\mu^{-\vep}\over 16\pi^2\vep}& \Bigl \{ \int dS' \, V'''_{k'k'j'}
(\hphi') \K_{j'i}(\bhx', \bhx) + g_{kkij}\pr_n \vphi_{\c j}(\bhx)
- {\ts {2\over 3}}K(\bhx) V'''_{jji} (\hphi) \Bigl \} ~, \cr}
\eqno (5.22) $$
where ${\bar G}_{mn}|$ is defined by (2.14) and $\K$ is as in (5.14).
Furthermore since $\de G = - G \de \Delta G$
$$
{\de \over \de \hphi_i (\bhx)} G_{j_1 j_2} (x_1,x_2) = - \int dv \,
G_{j_1 j} (x_1,x) V'''_{jk\ell}(\vphi_\c) G_{kj_2} (x,x_2) \, G_{\ell i}
(x,\hx)
{\overleftarrow \pr} {}_{\! \! n} \bigl |_{\hx = x(\bhx)} ~,
$$
we may also obtain in turn from (5.22) (as usual with dimensional
regularisation we set $\de^{d-1} (0) = 0$ so that the $K$ term in (5.22) does
not contribute to the second functional derivative here)
$$ \eqalignno {
{\de^2 \over \de \hphi (\bhx)^2 }& \, {\rm ln} \det \Delta_{\rm {reg}}\cr
= {}& \int dv \, {\bar G}_{mn}| (x)\, g_{mnij} \, G_{jk} (x,\hx)
{\overleftarrow \pr} {}_{\! \! n} \pr_n G_{ki} (\hx,x)
\bigl |_{\hx = x(\bhx)} \cr
- &\int dv dv' \, {\bar G}_{mn}| (x)\, V'''_{mnj} (\vphi_\c) G_{jj'} (x,x')
V'''_{j'k'\ell'} (\vphi'{}_{\!\! \c}) G_{k'i} (x',\hx)
{\overleftarrow \pr} {}_{\! \! n} \pr_n G_{i\ell'} (\hx,x')
\bigl |_{\hx = x(\bhx)} \cr
- &\int dv dv' \, \pr_n G_{ij} (\hx,x) V'''_{jmn} (\vphi_\c)\cr
& ~~~~~~~~~~~ \times G_{mm'}(x,x') G_{nn'}(x,x')
V'''_{m'n'j'} (\vphi'{}_{\!\!\c})
G_{j'i} (x',\hx) {\overleftarrow \pr} {}_{\! \! n}\bigl |_{\hx = x(\bhx)} \cr
+ &{}  {2\mu^{-\vep}\over 16\pi^2 \vep} \int dv \, \pr_n G_{ij} (\hx,x)
V'''_{jmn} (\vphi_\c) V'''_{mnk} (\vphi_\c) G_{ki} (x,\hx)
{\overleftarrow \pr} {}_{\! \! n} \bigl |_{\hx = x(\bhx)} \cr
- &{}  {\mu^{-\vep}\over 16\pi^2 \vep} \int dS'dv \, V''_{k'k'j'}(\hphi')
\pr' {}_{\! \! n} G_{j'k} (x',x) V'''_{k\ell m} (\vphi_\c)
G_{mi} (x,\hx) {\overleftarrow \pr} {}_{\! \! n} \pr_n G_{i\ell} (\hx,x)
\bigl |_{\hx = x(\bhx),x'=x(\bhx')} \cr
+ &{}  {2\mu^{-\vep}\over 16\pi^2 \vep} \, g_{kkij} \K_{ji} (\bhx,\bhx)
{}~. & (5.23) \cr}
$$

In order to verify that (5.20) is a valid approximation to two loop order
it is sufficient to show
$$
\Bigl ( {\de \over \de t} - \pr_n \vphi_{\c i} {\de \over \de \hphi_i}
\Bigl )\, \Gamma^{(2)}_{\rm {reg}} = -\X_1 - \X_2 - \X_{\rm {loc}} ~,
\eqno (5.24) $$
since this entails
$$
\Bigl ( \hbar {\de \over \de t} + \H + \H_0^{(1)} \Bigl ) {\hat \Psi}(\hphi)
= - \bigl ( \hbar^2 \X_{\rm {loc}} + {\rm O}(\hbar^3) \bigl ) {\hat \Psi}
(\hphi) ~.
\eqno (5.25) $$
After using the equation of motion for $\vphi_\c$ to eliminate
$\nab_n {}^{\! 2} \vphi_\c$ we assume that
$\X_{\rm {loc}}(\hphi,\pr_n \vphi_\c)$ is local function,
which for the renormalisable theory considered here is of dimension 4.
This may then be cancelled by an appropriate choice for
$$
\H_0^{(2)} = \hbar^2 \X_{\rm {loc}}(\hphi,{\hat p}) ~,~~~~
{\hat p} = \mu^\vep\hbar{\de \over \de \hphi} ~,
\eqno (5.26) $$
by virtue of (5.10). We therefore endeavour to demonstrate (5.24) by
considering each contribution to $\Gamma^{(2)}$ in turn, with $\vphi
\to \vphi_\c$, and using (5.9) with (C.8a,b), (C.11).
The surface terms arising according to (C.8a)
from the differentiation of the volume
integrals expressing $\Gamma_a,\Gamma_b,\Gamma_c$ vanish
since $G$ satisfies Dirichlet boundary conditions. Beginning with the
expression for $\Gamma_a$ in (2.13) it is simple to obtain
$$ \eqalign {
\Bigl ( {\de \over \de t (\bhx)} &{} - \pr_n \vphi_{\c i}(\bhx)
{\de \over \de \hphi_i (\bhx)} \Bigl ) \Gamma_a \cr
& = - {\mu^\vep \over 4} \, \int dv \, g_{ijk\ell} \, {\bar G}_{ij}
| (x) \, G_{km} (x,\hx) {\overleftarrow \pr} {}_{\! \! n}
\pr_n G_{m\ell} (\hx,x) \bigl |_{\hx = x(\bhx)} ~. \cr}
\eqno (5.27) $$
For $\Gamma_b$ given by (2.15) the corresponding result is
$$ \eqalign {
\Bigl ( {\de \over \de t (\bhx)} &{} - \pr_n \vphi_{\c i}(\bhx)
{\de \over \de \hphi_i (\bhx)} \Bigl ) \Gamma_b \cr
= {}& {\mu^\vep\over 4} \, \int \! \int dv dv' \, V'''_{ijk}(\vphi_\c)
G_{ii'}(x,x') G_{jj'}(x,x') \cr
& ~~~~~~~~~~~~~~~~~~ \times G_{k\ell} (x,\hx) {\overleftarrow \pr}
{}_{\! \! n} \pr_n G_{\ell k'} (\hx,x')  V'''_{i'j'k'} (\vphi'{}_{\!\!\c})
\bigl |_{\hx = x(\bhx)} \cr
& - {1\over 2}\, {1\over 16\pi^2 \vep} \int dv \,
V'''_{ijk}(\vphi_\c)  V'''_{ij\ell}(\vphi_\c)
G_{km} (x,\hx) {\overleftarrow \pr} {}_{\! \! n}
\pr_n G_{m\ell} (\hx,x) \bigl |_{\hx = x(\bhx)} ~.\cr}
\eqno (5.28) $$
For the one particle reducible amplitude given by (2.17) we may obtain in this
case
$$ \eqalignno {
\Bigl ( {\de \over \de t (\bhx)} &{} - \pr_n \vphi_{\c i}(\bhx)
{\de \over \de \hphi_i (\bhx)} \Bigl ) \Gamma_c  \cr
={} &{\mu^\vep \over 8} \, \int \! \int dvdv'\, {\bar G}_{jk}| (x)
V'''_{ijk}(\vphi_\c) \, G_{i\ell} (x,\hx) {\overleftarrow \pr}
{}_{\! \! n} \pr_n G_{\ell i'} (\hx,x') V'''_{i'j'k'}(\vphi'{}_{\!\!\c})
{\bar G}_{j'k'}| (x') \bigl |_{\hx = x(\bhx)} \cr
& + {\mu^\vep \over 4} \, \int \! \int dvdv'\, {\bar G}_{jk}| (x)
V'''_{ijk}(\vphi_\c) \, G_{ii'} (x,x') \cr
&~~~~~~~~~~~~~~~~~~~~~~~~~~~ \times
V'''_{i'j'k'}(\vphi'{}_{\!\! \c}) G_{j'\ell} (x',\hx) {\overleftarrow \pr}
{}_{\! \! n} \pr_n G_{\ell k'} (\hx,x') \bigl |_{\hx = x(\bhx)} ~.& (5.29) \cr}
$$
For the remaining term in (2.18), using now (C.23) and (C.24) as well as
(C.11) and also ${\bar G}_{k\ell}| =
2\mu^{-\vep} {\tilde V}_{k\ell}(\hphi)/(16\pi^2\vep)$ on $\pr\M$.
$$ \eqalignno {
\Bigl ( {\de \over \de t (\bhx)} &{} - \pr_n \vphi_{\c i}(\bhx)
{\de \over \de \hphi_i (\bhx)} \Bigl ) \Gamma_d  \cr
={} & {1\over 4}\, {1\over 16\pi^2 \vep} \biggl \{ - g_{ijkk}
\pr_n \vphi_{\c j}(\bhx) \int dv' \, \pr_n G_{ii'}(x,x')\bigl |_{x = x(\bhx)}
\,
V'''_{i'j'k'}(\vphi'{}_{\! \! \c}) {\bar G}_{j'k'}|(x') \cr
& + \int dS'' dv' \, V'''_{k\ell\ell}(\hphi'') \K_{ki} (\bhx'', \bhx)
\pr_n G_{ii'}(\hx,x')\bigl |_{\hx = x(\bhx)}
V'''_{i'j'k'}(\vphi'{}_{\! \! \c}) {\bar G}_{j'k'}|(x') \cr
& + \int dS'' dv' \, V'''_{j\ell\ell}(\hphi'')
\pr_n G_{ji'}(\hx'',x') V'''_{i'j'k'}(\vphi'{}_{\! \! \c}) \cr
& ~~~~~~~~~~~~~~~~~~~~~~ \times
G_{j'i} (x',\hx) {\overleftarrow \pr} {}_{\! \! n}
\pr_n G_{ik'} (\hx,x') \bigl |_{\hx = x(\bhx),\hx'' = x(\bhx'')} \biggl \} \cr
& + {1\over 8}\, {\mu^{-\vep}\over(16\pi^2 \vep)^2} \biggl \{
4V'''_{ijj}(\hphi) V'''_{ik\ell}(\hphi) {\tilde V}_{k\ell}(\hphi) - 2 g_{ijkk}
\pr_n \vphi_{\c j}(\bhx) \int dS' \, \K_{ii'} (\bhx,\bhx') V'''_{i'j'j'}
(\hphi') \cr
& ~~~~~~~~~~~~~~~~~~ + \int dS'dS'' \, V'''_{ijj} (\hphi'') \K_{ik}
(\bhx'',\bhx) \K_{ki'} (\bhx,\bhx') V'''_{i'j'j'} (\hphi') \cr
& ~~~~~~~~~~~~~~~~~~ - {\hat \gamma}^{pq} \pr_p V'''_{ijj}(\hphi) \pr_q
V'''_{ikk} (\hphi) - V'''_{ikk}(\hphi) V''_{ij} (\hphi) V'''_{j\ell\ell}
(\hphi) \biggl \} ~. & (5.30) \cr}
$$

Combining the results for the individual contributions we therefore find
$$
\Bigl ( {\de \over \de t} - \pr_n \vphi_{\c i} {\de \over \de \hphi_i}
\Bigl )\, \Gamma^{(2)} = -\X_1 - \X_2 -\X_3 - \X_{\rm {loc},1} ~,
\eqno (5.31) $$
where, neglecting $K^2$ terms,
$$ \eqalign {
\X_{\rm {loc},1} =
{\mu^{-\vep}\over(16\pi^2 \vep)^2} \Bigl \{& -{\ts {3\over 8}} g_{ik\ell\ell}
g_{jkmm} \, \pr_n \vphi_{\c i} \pr_n \vphi_{\c j}  + {\ts {1\over 3}}
g_{ijkk} V'''_{j\ell \ell} (\hphi) \, K \pr_n \vphi_{\c i} \cr
& - \half V'''_{ijj}(\hphi) V'''_{ik\ell}(\hphi) {\tilde V}_{k\ell}(\hphi)\cr
& + {\ts {1\over 8}} g_{ikkm} g_{j\ell\ell m}\, {\hat \gamma}^{pq} \pr_p
\hphi_i
\pr_q \hphi_j + {\ts {1\over 8}} V'''_{ikk}(\hphi) V''_{ij} (\hphi)
V'''_{j\ell\ell} (\hphi) \Bigl \} ~.
\cr}
\eqno (5.32) $$
This result is the critical part of verifying
(5.24) but in addition we need to consider the contribution arising from the
overall two loop counterterms as in (5.21)
$$ \eqalign {
\Bigl ( {\de \over \de t} - \pr_n & \vphi_{\c i}
{\de \over \de \hphi_i} \Bigl ) S_0^{(2)}(\vphi_\c) = \X_{\rm {loc},2} \cr
= {}& \mu^{-\hh \vep}\hphi^{(2)\prime}_{0i,j} (\hphi) \bigl ( \pr_n
\vphi_{\c i}
\pr_n \vphi_{\c j} + {\hat \gamma}^{pq} \pr_p \hphi_i \pr_q \hphi_j \bigl )
{} + V_0^{(2)} (\mu^{-\hh \vep} \hphi ) \cr
& + \mu^{-\hh \vep} V'_i (\hphi) \hphi^{(2)}_{0i}
- \mu^{- \vep} \bigl ( \rho^{(2)\prime}_i (\hphi) \pr_n \vphi_{\c i} K
+ \hnab^2  \rho^{(2)} (\hphi) \bigl ){} + \dots ~. \cr}
\eqno(5.33) $$
Hence (5.31) and (5.33) together ensure the validity of the crucial equations
(5.24) or (5.25) with $\X_{\rm {loc}} = \X_{\rm {loc},1} + \X_{\rm {loc},2}$
which may then be used to determine $\H_0^{(2)}$, as described in (5.26),
so that ${\hat \Psi} (\hphi)$ satisfies (5.3) at two loop order.

Together with (5.17) the result of the calculations here are compatible
with the general form
$$ \eqalign {
\H_0 = {}& - \half \hbar^2 \, {\de^2 \over \de \hphi_0^2} -
\hbar K \, \rho'_{0i}(\hphi_0) {\de \over \de \hphi_{0i}} + \V_0(\hphi_0) ~,
\cr
& \V_0(\hphi_0) = V_0(\hphi_0) + \half {\hat \gamma}^{pq} \pr_p \hphi_{0i}
\pr_q \hphi_{0i} - \hnab^2  \rho_0(\hphi_0) +\dots ~,\cr}
\eqno (5.34) $$
for $ \rho_0(\hphi_0) = \mu^{-\vep} \rho(\hphi)$ and where the neglected terms
represent contributions to $\V_0$ involving scalars formed from the metric
of dimension 2 or 4. The essentially complete expression for $\H_0$ to all
orders, in terms of bare quantities,
shown in (5.34) is a direct result of the requirement of consistency of the
Schr\"odinger equation (5.3) with the renormalisation group equation (2.27)
for the wave functional $\hat \Psi$ which is here written as
$$
\D {\hat \Psi}(\hphi) = {\ts {1\over \hbar}} \bigl ( \C (\beta^\rho) +
{\tilde \C} (\beta^\lambda,\beta^{\hat \lambda})\bigl ) {\hat \Psi}(\hphi) ~.
\eqno (5.35) $$
Applying the renormalisation group operator $\D$, as given by (2.26), to
(5.3) then since from (2.29a,b) $\D \hphi_0 =0 , ~ \D \rho_0 = - \mu^{-\vep}
\beta^\rho$ we may use
$$
\bigl [ \D , \, {\de \over \de t} \bigl ] {} = 0 ~,~~~~
\bigl [ \D , \, \H_0 \bigl ]{}= \hbar \mu^{-\vep} K {\pr \over \pr \hphi_{0i}}
\beta^\rho (\hphi) \, {\de \over \de \hphi_{0i}} + \mu^{-\vep} \hnab^2
\beta^\rho (\hphi) + \dots ~,
\eqno (5.36) $$
and also, with $\C$ defined by (2.24),
$$ \eqalign {
\bigl [ \hbar {\de \over \de t}, \, {\ts {1\over \hbar}} \C (\beta^\rho)\bigl ]
{} = {}& - \mu^{-\vep} \hnab^2 \beta^\rho (\hphi) + \dots ~, \cr
\bigl [ \H_0 , \, {\ts {1\over \hbar}} \C (\beta^\rho)\bigl ]
{} = {}& - \hbar \mu^{-\vep} K {\pr \over \pr \hphi_{0i}}
\beta^\rho (\hphi) \, {\de \over \de \hphi_{0i}} + \dots ~, \cr}
\eqno (5.37) $$
to show how the form of $\H_0$ is dictated by imposing (5.35). The simple
structure of $\H_0$ is at least in part a consequence of the restrictions on
counterterms obtained by using dimensional regularisation with minimal
subtraction.
\vfill\eject

\vskip 5pt
\leftline{\bf 6 Conclusion}

In this paper we have been primarily concerned with calculational techniques
for quantum field theories with a boundary in the simplest case of a scalar
field theory. The necessary labour is significantly increased beyond that for
field theories without boundary. Even if the manifold on which the quantum
field theory is defined is flat, as is usual for applications in statistical
physics, the machinery associated with covariant treatments on curved space
are a necessary ingredient in our method of calculating divergences for
general curved boundaries. In spite of this complexity there
are nevertheless potential formal
developments associated with the functional Schr\"odinger equation discussed
in the previous section. This may offer a non perturbative approach to
discussing quantum field theories, such as in variational methods applied to
the
effective potential, but it is clearly important to ensure that the
functional Hamiltonian operator is well understood in a perturbative context
first. A further area of possible relevance of these calculations is to
critical phenomena in real systems with boundary, as was the motivation for
the work of Diehl and Dietrich [5] whose results for plane boundaries are
reproduced here. Even for simple systems there is a rich variety of phase
transitions associated with boundaries [4] whose universal properties may be
calculated using renormalisable quantum field theories. For the $O(n)$
symmetric model discussed in sections 2 and 3 there is a fixed point for
$g=g_*, \, m^2 =0$ and on the boundary in the Dirichlet case $\hphi =0$
while in the Neumann case it is necessary to require $\beta_c =0$ or
$c={\rm O} (K)$ (for a plane boundary $c=0$). More generally if there is
a cubic coupling $g_{ijk}$ in $Q(\phi)$ in the treatment of section 3 then
from (3.18) to one loop
$$ {\hat \beta}^{\hat g}_{ijk} = -\half \vep {\hat g}_{ijk}^{\vphantom T} +
{1\over 16\pi^2} \bigl ( - 8 {\hat g}_{i\ell m}^{\vphantom T}
{\hat g}_{jmn}^{\vphantom T} {\hat g}_{kn\ell}^{\vphantom T} +
6 g_{\ell m(ij}^{\vphantom T} {\hat g}_{k)m\ell}^{\vphantom T} -
{\ts {3\over 2}} g_{\ell \ell m(i}^{\vphantom T} {\hat g}_{jk)m}^{\vphantom T}
\bigl ) ~,
\eqno (6.1) $$
so that there is a possible fixed point for ${\hat g}_*
= {\rm O}(\vep^{1\over 2})$,
as discussed by Diehl and Ciach [22] for a single component field.

In connection with the discussion in section 5 it is interesting to note that
recently L\"uscher {\it et al} [23] have considered the Schr\"odinger wave
functional for gauge theories with a view to applying finite size techniques
to lattice gauge calculations. Although in other respects more complicated,
due to requirements of gauge fixing and introducing ghost fields, the gauge
theory case is simpler to the extent that there are no gauge invariant
operators with dimensions $<4$ which can occur as boundary counterterms.
Lattice calculations of wave functionals are of course well defined non
perturbatively but perturbative results, such as obtained here, provide
useful constraints on the approach to the continuum limit.
\vskip 15pt

One of us (DMM) would like to thank the Royal Commisssion for the Exhibition
of 1851 for an overseas scholarship.

\vfill\eject

\leftline{\bf Appendix A}

The  short distance behaviour of the Green function $G_\Del$ near
the boundary may be found from an asymptotic expansion of the heat kernel
$\G_\Del$ since
$$G_\Del(x,x') = \int^\infty_0 \! d\tau\, \G_\Del(x,x';\tau)  ~,
\eqno(A.1) $$
where $\G_\Delta$ satisfies
$$ \Bigl ( {\pr \over \pr \tau} + \Del_x \Bigl ) \G_\Del(x,x';\tau) =0 ~,~~~~~
\G_\Del(x,x';0) = \delta^d (x,x') ~,
\eqno (A.2) $$
with boundary conditions corresponding to (1.3)
$$
\bigl ( 1 - \P(\bhx) \bigl ) \G_\Del (x,x';\tau) \bigl |_{x=x(\bhx)} = 0~,~~~
\bigl(\P(\bhx) n^\mu(\bhx) D_\mu + \psi(\bhx) \bigl)
\G_\Del (x,x';\tau) \bigl |_{x=x(\bhx)} = 0 ~.
\eqno(A.3) $$

The extension of the DeWitt ansatz for $\G_\Del$ takes the form [13]
$$ \G_\Del (x,x';\tau) \simeq {1\over (4\pi \tau)^{d\over 2}}
\Bigl ( e^{- {\si(x,x')/ 2\tau}} \Omega (x,x';\tau )
+ e^{- {\bsi(x,x')/ 2\tau}} {\bar \Omega} (x,x';\tau )\Bigl ) ~,
\eqno(A.4)
$$
where $\si(x,x')$ is the geodetic  interval based on the geodesic path from
$x'$ to $x$ while $\bsi(x,x')$ is the corresponding biscalar resulting
from the geodesic which undergoes reflection on the boundary.
The conventional DeWitt asymptotic expansion [14] is obtained by writing
$\Omega (x,x';\tau) \simeq \Del^\hh (x,x') \sum_{n=0} a_n(x,x')\tau^n$,
with $\Del^\hh (x,x')$, $a_n(x,x')$ regular for $x \approx x'$ and
$\Del^\hh (x,x) = a_0 (x,x) = 1$. This expansion determines the singular
behaviour of $G_\Del (x,x')$ for $x'\to x$, as in (1.5,6). The presence of the
additional term ${\bar \Omega}$ allows the boundary conditions
(A.3) to be satisfied by the full expression in (A.4), while maintaining the
usual form for $\Omega$. An appropriate expansion of ${\bar \Omega}$ then
enables the singular terms of $G_\Del$ near the boundary to be obtained.

To achieve this in the neighbourhood of the boundary, with coordinates so
that the metric takes the form (1.1), both $\si$  and $\bsi$ are expanded in
powers of $\epsilon$, according to (1.9), where to order $\epsilon^3$ from I
$$
\eqalignno {
\si ={}& \half(y-y')^2 + \half \hga_{ij} \hsi^i \hsi^j
- (y+y')\half K_{ij}\hsi^i \hsi^j ~, & (A.5a) \cr
\bsi = {}& \half(y+y')^2 + \half \hga_{ij} \hsi^i \hsi^j
- {y^2 + y'^2 \over y+y'} \half K_{ij} \hsi^i \hsi^j ~. & (A.5b) \cr}
$$
Using these results ${\bar \Omega}$ may be obtained as a formal power series
in $\epsilon$ with ${\bar \Omega}_n = \O (\epsilon^n)$,
taking also $\tau = \O(\epsilon^2)$, by solving at each order
sets of coupled differential equations in $y$ and $\tau$ with appropriate
boundary conditions determined by $\Omega_n$ in the analogous expansion of
$\Omega$. The solutions of these equations may be expressed in terms of the
functions
$$ \eqalign {
f_n (u,\tau) ={}& {1\over n!} \, {1\over (2\tau)^n} \int_0^\infty dz \, z^n \,
e^{- {z^2\over 4\tau} -  {zu \over 2\tau}} ~ ,\cr
& 2\tau (n+1) f_{n+1}(u,\tau) + uf_n(u,\tau) = f_{n-1}(u,\tau) +
2\tau \delta_{n0} ~. \cr}
$$
Using (A.5a,b) the results are more conveniently expressed in the form
$$ \G_\Del (x,x';\tau) \simeq {1\over (4\pi \tau)^{d\over 2}}
e^{- \hsi(\bhx,\bhx')/ 2\tau}\Bigl ( e^{- v^2/4\tau} \Omega' (x,x';\tau )
+ e^{- u^2/4\tau} {\bar \Omega}' (x,x';\tau )\Bigl ) ~,
\eqno(A.6)
$$
where as before $v=y-y', \, u=y+y'$.

The first two orders in the expansion are straightforward to obtain in terms
of the extrinsic curvature of the boundary $K_{ij}(\bhx)$ and also $\psi
(\bhx)$ and the projection operator $\P (\bhx)$ appearing in the boundary
condition (A.3)
$$ \eqalign {
\Omega'_0 = {}& {\hat I} ~,~~~~\Omega'_1 = u \, \quar
K_{ij} \hsi^i \hsi^j \, {\hat I} ~ , ~~~~
{\bar {\Omega}}'_0 = \P_- {\hat I} ~,~~~\P_- = 2\P -1 ~, \cr
{\bar \Omega}'_1 = {}& \Bigl ( K - {1\over \tau} \, \half K_{ij} \hsi^i
\hsi^j \Bigl ) \Bigl ( {yy' \over 2\tau }\, f_0 (u,\tau)\P_-
- 2\tau \, f_2 (u,\tau)\P \Bigl ) {\hat I} \cr
& + {1\over \tau} \, u \, \half K_{ij} \hsi^i \hsi^j \,
\P_- {\hat I} + f_0 (u,\tau)( K \P + 2\psi ) {\hat I}  \cr
& - 2\P \hsi^i \hD_i \P \, {\hat I} +
{1\over \tau} \, f_0 (u,\tau)\bigl ( - y (1-\P) \hD_i \P + y'
\P \hD_i \P \bigl ) \hsi^i\,  {\hat I} ~. \cr}
\eqno (A.7)
$$
Using (A.1) and
$$ \int_0^\infty \! d\tau \, {1\over (4\pi\tau)^{d\over 2}}\, \tau^{r+s-1} \,
e^{-\hsi/2\tau - u^2/4\tau}
f_s (u,\tau) = {1\over s!} \, 2^{2-2r-s} \, G_{r,s} (u) ~, $$
from the definition in (1.12), these results are sufficient to obtain (1.13).

At order $\epsilon^2$ the number of terms proliferate and the calculation
becomes tedious. Here we consider only those terms which depend on $X$, which
represents a mass$^2$ term in the operator $\Delta$, and also $\psi$, which
has dimensions of mass, appearing in the
boundary condition. For the conventional DeWitt expansion the relevant result
is very simple
$$
\Omega'_2 = - \tau X^0 \, {\hat I} ~,
\eqno (A.8) $$
where $X^0 = X|_{y=0}$. From the results of calculations carried out
elsewhere we obtain
$$ \eqalign {
{\bar {\Omega}}'_2 ={}&
\tau \bigl (X^0 - 2 \P X^0 \P \bigl){\hat I} -
\bigl ( y (1-\P ) X^0 \P + y' \P X^0 (1-\P ) \bigl ) f_0(u,\tau) \, {\hat I}
\cr
& + \bigl ((yy' + 2\tau ) f_1(u,\tau) -10\tau^2 f_3(u,\tau) \bigl) \Bigl (
2K - {1\over\tau} K_{ij}\hsi^i \hsi^j \Bigl )  \psi \, {\hat I} \cr
& + 2\tau f_1 (u,\tau) (K\psi + 2\psi^2) \, {\hat I} +
K_{ij}\hsi^i \hsi^j \psi \, {\hat I} ~. \cr}
\eqno (A.9) $$
The corresponding contributions to the Green function in the expansion
(1.11) are
$$ \eqalign {
G_2^B = {}& - \quar G_2^R (v) \, X^0 \, {\hat I} ~, \cr
{\bar G}_2^B ={}& \quar G_2^R (u) \bigl (X^0 - 2 \P X^0 \P \bigl){\hat I} -
G_{1,0}(u)\bigl ( y (1-\P ) X^0 \P + y' \P X^0 (1-\P ) \bigl ) \, {\hat I} \cr
& + G_{1,1}^R (u) ({\ts {4\over 3}}K\psi + 2\psi^2) \, {\hat I} +
\bigl ( 4yy'G_{0,1} (u) + {\ts {5\over 3}} u G_{0,2}(u) \bigl ) K\psi \,
{\hat I} \cr
& - \bigl ( 8yy'G_{-1,1} (u) + {\ts {2\over 3}} G_{0,1}(u) +
{\ts {10\over 3}} u G_{-1,2}(u) - G_1(u) \bigl )
K_{ij}\hsi^i \hsi^j \psi \, {\hat I} ~. \cr}
\eqno (A.10) $$
In (A.10)
$$ G_2^R (u) = G_2^{\vphantom R} (u) + {\mu^{-\vep}\over 2\pi^2 \vep} ~,~~~~
G_{1,1}^R (u) = G_{1,1}^{\vphantom R} (u) + {\mu^{-\vep}\over 4\pi^2 \vep} ~,
$$
which ensures that $G_2^R , \, G_{1,1}^R $ have a smooth limit as $\vep\to 0$.
\vfill\eject

\leftline{\bf Appendix B}

In this appendix we describe the essential calculational details
associated with determining the singular terms arising from the
presence of a boundary, as $\vep \to 0$, of the two loop amplitudes
considered in sections 2 and 3. For the graph in fig.1a we need to
consider the singular behaviour of the product of two coincident Green
functions, i.e. $G | (x) \times G |(x)$, where $G |(x) = G (x,x)$.
{}From the results of appendix A the regular part of the coincident limit
$\bG| (x)$, as given by (1.5), has the general form near the boundary of the
region on which $G$ is defined
$$\eqalign {
\bG |(x) = {}& G | (x) + {\mu^{-\vep}\over 8\pi^2 \vep} \bigl ( X(x) - \six
R(x) \bigl ) \cr
= {}& y_+^{2-d} e_d (\bx) + y_+^{3-d} f_d (\bx) + {2\over \vep}\bigl
(y_+^{4-d}g_d (\bx) - \mu^{-\vep} \theta(y) g_4 (\bx) \bigl )\cr
& + {2\over \vep} \bigl ( y_+^{5-d} h_d (\bx)- \mu^{-\vep} y_+^{\vphantom d}
h_4 (\bx) \bigl ) {} + H (x) ~ , \cr }
\eqno(B.1)
$$
with $H, \pr_y H$ regular at $y=0$ for $d \approx 4$. $y_+^{\lambda
-1}$ is a distribution or generalised function with support on $y\ge0$ and
which has a simple pole at $\lambda =-p,\, p=0,1,2, \dots$ [21],
$$
y_+^{\lambda-1} \sim {1\over \lambda  + p }{(-1)^p \over p!} \de^{(p)} (y)~ .
\eqno(B.2)
$$
Furthermore it is important to note that
$$ {1\over \vep} \bigl ( y_+^{\alpha \vep - p -1} - y_+^{\beta \vep -
p -1} \bigl ) {} \sim {1 \over \vep^2} \Bigl ( {1\over \alpha } -{1 \over
\beta } \Bigl ) {(-1)^p \over p!}\de^{(p)} (y) ~ ,
\eqno(B.3)
$$
with no simple pole in $\vep$,
as may be verified  by considering Fourier transforms.

Writing for convenience
$$ e_d = {\Ga (\half d -1 )\over (4\pi)^{\hh d}} (\te_0 + \te_1 \vep
+ \dots ) ~,
$$
as an expansion in $\vep$, and similarly for $f_d , \, g_d , \, h_d$,
then from (B.2) and (B.3) the singular contributions are, suppressing indices
and assuming on the r.h.s. the products are symmetrised so that for instance
$\te_0 \times \tf_0 = \tf_0 \times \te_0$,
$$ \eqalignno {
\bG | \times \bG | \sim {\mu^{-2 \vep } \over (16 \pi^2 \vep)^2} \Bigl \{& -
\vep \, \te_0 \times \te_0 \tw \de''' (y) + \vep \, \te_0 \times \tf_0
\half \de'' (y) - \vep \, \tf_0 \times \tf_0  \half \de'(y) & \cr
& + 2 \bigl (  \te_0 \times \tg_0  + \vep (
\te_1 \times \tg_0 - \te_0 \times \tg_1 ) \bigl ) \de' (y) & \cr
&- 2 \bigl ( \te_0 \times \th_0 + \tf_0 \times \tg_0 & \cr
&~ + \vep ( \te_1 \times \th_0 - \te_0 \times \th_1 + \tf_1 \times \tg_0
- \tf_0 \times \tg_1 ) \bigl ) \de (y) \Bigl \} & \cr
& \hskip-25pt + {2\mu^ {-\vep} \over 16 \pi^2 \vep } \,
\bigl ( -\te_0 \de'(y) + \tf_0 \de(y)\bigl ){} \times H ~ . & (B.4) \cr }
$$
If we introduce the notation $G|^0 = G|_{y\to 0}$ then from (1.6,7) and (B.1)
we may write by analytic continuation in $d$ from $d<2$
$$ \eqalign {
G |^0 ={}& - {2\mu^{-\vep} \over 16\pi^2 \vep}\,
\bigl ( X^0 - \six R^0 + \tg_0 \bigl ){} + H^0 ~, \cr
\pr_n (G |)^0 ={}& - {2\mu^{-\vep} \over 16\pi^2 \vep}\,
\bigl (\pr_n X^0 - \six \pr_n R^0 + \th_0 \bigl ){} + \pr_n H^0 ~. \cr}
\eqno (B.5) $$
In calculations of the divergent parts of amplitudes where $G| \times G|$
appear the non local singular terms present in (B.4) are cancelled by
additional counterterms which involve contributions of the form
$$ \eqalign {
- & {2\mu^{-\vep} \over 16\pi^2 \vep} \,\bigl ( -\te_0 \de'(y) +
\tf_0 \de(y)\bigl ){} \times G | = - {2\mu^{-\vep} \over 16\pi^2 \vep}\,
\bigl ( -\te_0 \de'(y) + \tf_0 \de(y)\bigl ) {}\times H  \cr
& - {4\mu^{-2\vep} \over (16\pi^2 \vep)^2}\, \Bigl \{ \te_0
\times (\tg_0 + X^0) \, \delta'(y) - \bigl ( \te_0 \times (\th_0 + \pr_n X^0)
+ \tf_0 \times (\tg_0 + X^0) \bigl ) \delta (y)\Bigl \} ~. \cr}
\eqno (B.6) $$

Using the results from calculations of the asymptotic expansion of the heat
kernel we obtain explicitly
$$\eqalign {
\te_0 = {}& \P_-  ~ , \hskip40pt \te_1 = 0 ~, \cr
\tf_0 = {}& 4 \psi + {\ts {2\over 3}} K + K \P_-,
\hskip25pt \tf_1 = 4\psi +{\ts {5 \over 9}} K + \half K \P_- ~, \cr
\tg_0 = {}& - X^0 + 2 \P X^0 \P - 4 \psi^2 - {\ts {8 \over 3}} \psi K ~,
\cr
\tg_1 = {}&-(1-\P) X^0 \P - \P X^0 (1-\P) - 4 \psi^2 - {\ts {11 \over
9}} \psi K ~, \cr
\th_0 ={}& - \pr_n X^0 + (1-\P) \pr_n X^0 \P + \P \pr_n X^0 (1- \P) -2
\P_- X^0 \psi - 2\psi X^0 \P_- \cr
& +8 \psi^3 + {\ts {16 \over 3 }}\psi^2 K -2 \P X^0 (1-\P) K -2(1-\P)
X^0 \P K ~, \cr
\th_1 = {}& -\half ( \pr_n X^0 -2 \P \pr_n X^0 \P ) + \half (1-\P)
\pr_n X^0 \P + \half \P \pr_n X^0 (1-\P) \cr
& +2 X^0 \psi + 2\psi X^0 - {\ts {38\over 9}} \psi^2 K\cr
& +\half X^0 K + \P X^0 \P K - {\ts {13 \over 3 }} \bigl (
\P X^0 (1-\P) + \P X^0 (1-\P)\bigl ) K ~, \cr}
\eqno (B.7)
$$
where in $\tg_{0,1} , \, \th_{0,1}$ we have given
only those terms depending on $X, \, \psi$. For Dirichlet boundary conditions,
$ \P, \psi = 0$ and from (B.5)
$$ G|^0 = H^0 = 0 ~, ~~~~~ \pr_n (G|)^0 = \pr_n H^0 = 0 ~,
\eqno (B.8) $$
so that the last line of (B.4) vanishes in this case. For Neumann
boundary conditions $\P=1$ and
$$ \pr_n (G|)^0 = -\psi G|^0 - G|^0 \psi ~,~~~~
\pr_n H^0 = -\psi H^0 - H^0 \psi ~ .
\eqno (B.9) $$
When integrating by parts in integrals over $dv$ it is useful to note that
$$ dv = dS\, dy \bigl ( 1-yK + \O (y^2) \bigl ) ~,~~~~
\de'(y) = \nab_\mu \bigl( n^\mu \de(y)\bigl ) {} + K \de (y) ~,
\eqno (B.10) $$
with a corresponding formula for $\de''(y)$.

For the amplitude corresponding to the graph in fig.1b it is
necessary to determine the singular behaviour of
$$ \eqalign {
\int \! \! &\int \! dy\, dy' \, f(y) f(y') \, G_\Delta (x,x')^3 \cr
&- {2\mu^{-\vep}\over 16\pi^2 \vep} \, \int \! dy \, {\sqrt {\hat \gamma}\over
\sqrt g} \, f(y)^2 \bigl (1^2 G_\Delta (x,x) + 1 G_\Delta (x,x) 1 +
G_\Delta (x,x) 1^2 \bigl ) \delta_{\pr \M} (\bx, \bx') ~, \cr}
\eqno (B.11) $$
where $\sqrt g = \sqrt {\hat \gamma} (1-yK + \dots )$ and the second term
subtracts the subdivergence as usual. $f(y)$ is an arbitrary smooth test
function vanishing rapidly as $y\to \infty$ and $\de_{\pr \M}(\bx,\bx') =
\de^{d-1} (\bx-\bx') / \sqrt {{\hat\gamma} (\bx)}$ denotes the surface $\de$
function on $\pr \M$. The singular terms as $\vep \to 0$
arise for $\bx \approx \bx'$ and it is convenient to use directly the
representation of $G_\Delta$ provided by the heat kernel, as in (A.1), where
writing
$$ {1\over (4\pi \tau)^{\hh(d-1)}} \, e^{- \hsi/2\tau}
\buildrel {\tau \to 0} \over \sim \de_{\pr \M}
+ \tau (\hnab^2 - {\ts{1\over 3}}{\hat R}) \de_{\pr \M} + \dots ~,
\eqno (B.12) $$
is equivalent to a Taylor expansion of the Fourier transform. For the leading
term in the expansion (1.11) it is necessary in (B.11) to take $G_\Delta \to
G_1 (v) + \P_- G_1 (u)$ with $G_1 (v)| =0$. Using (B.12) gives
$$
G_1(v)^3 \sim {\Gamma (\half d -1 )^3 \over 64\pi^d } \, {\pi^{-\hh} \over
\Gamma ( {\ts {3\over 2}}d - 3) } \Bigl \{ \Gamma ( d - {\ts {5\over 2}} )
|v|^{-2d + 5} \delta_{\pr \M} + \quar \Gamma ( d - {\ts {7\over 2}} )
|v|^{-2d + 7} \hnab^2 \delta_{\pr \M} + \dots \Bigl \} ~,
\eqno (B.13) $$
dropping the ${\hat R}$ term. This is then easily shown, by considering
specific test functions such as $f(y)= e^{-\rho y}$, to lead to a simple pole
in $\vep$,
$$ \eqalign {
\int & \! \! \int \! dy\, dy' \, f(y) f(y') \, G_1 (v)^3  \cr
& \sim {\mu^{-2\vep}\over 2\vep} \, {1\over (16\pi^2)^2 } \Bigl \{
\int_0^\infty \! \! dy \, f(y)^2 \hnab^2 - \int_0^\infty \! \! dy \, f'(y)^2
- f(0) f'(0) \Bigl \} \delta_{\pr \M} ~, \cr}
\eqno (B.14) $$
where the singular contributions arise both from the usual short distance
limit $y\to y'$, as for manifolds without boundary, and also $y,y' \to 0$.
Similarly for $G_1 (u)$, although only the latter contribution arises
in this case,
$$
\int \! \! \int \! dy\, dy' \, f(y) f(y') \, G_1 (u)^3
\sim {\mu^{-2\vep}\over 2\vep} \, {1\over (16\pi^2)^2 } \,
f(0) f'(0) \delta_{\pr \M} ~.
\eqno (B.15) $$
In order to discuss the $G_1(v)^2 G_1(u) $ contribution we write, using
(B.12) again,
$$ \eqalign {
\Bigl ( & G_1(v)^2 - {\mu^{-\vep} \over 8\pi^2 \vep} \, \delta (v) \,
\delta_{\pr \M} \Bigl ) G_1 (u) \cr
& \sim {\Gamma (\half d -1 )^2 \over 64\pi^d } \, \lim_{\ep \to 0}
\Bigl \{ {\pi^{-\hh} \over \Gamma ( d - 2) }
\int_\ep^\infty \! dt \, \int_0^\infty \! ds \, {s^{d-3} t^{\hh d -2} \over
(s+t)^{\hh(d-1)} } \, e^{-sv^2 -tu^2} \cr
& ~~~~~~~~~~~~~~~~~~~~~~~~~~~~~~ - {\mu^{-\vep} \over \vep} \, \delta (v) \,2
\pi^{-\hh \vep} \!
\int_\ep^\infty \! dt \, t^{\hh d -2} \, e^{-4ty^2} \Bigl \} ~, \cr}
\eqno (B.16) $$
where $\ep$ plays the role of an infra red regulator so that for $\ep > 0$
the r.h.s. of (B.16)
has a rapid fall off for $y,y' \to \infty$ and we may therefore take $ f(y)
\to a + by$ which allows all integrals to be explicitly evaluated. When
$\ep\to 0$ the $\ep$ dependence cancels in the terms containing poles in
$\vep$, since they arise only from short distance effects, and we find
$$
\int \! \! \int \! dy\, dy' \, f(y) f(y') \,
\Bigl ( G_1(v)^2 - {\mu^{-\vep} \over 8\pi^2 \vep} \, \delta (v) \,
\delta_{\pr \M} \Bigl ) G_1 (u)
\sim - {\mu^{-2\vep}\over (16\pi^2\vep)^2 }(2-\vep) \,f(0) f'(0)
\delta_{\pr \M} ~.
\eqno (B.17) $$
For the final leading order contribution we obtain
$$
\int \! \! \int \! dy\, dy' \, f(y) f(y') \,
G_1(u)^2 \, G_1 (v)
\sim {\mu^{-2\vep}\over \vep} \, {1\over (16\pi^2)^2 } \, f(0) f'(0)
\delta_{\pr \M} ~.
\eqno (B.18) $$
For pure Neumann, Dirichlet boundary conditions the surface contributions
combine to give
$$ \eqalign {
\int \! \! \int \! dy\, dy' \, f(y)& f(y') \,
\Bigl ( \bigl (G_1(v) \pm G_1(u) \bigl )^3 {} \mp {3\mu^{-\vep} \over 8\pi^2
\vep} \, \delta (v) \, \delta_{\pr \M} G_1 (u)\Bigl ) \cr
& \sim - {6\mu^{-2\vep}\over (16\pi^2\vep)^2 }\bigl ( \pm 1 -
{\ts {5\over 12}} \vep \mp {\ts {7\over 12}}
\vep \bigl ) f(0) f'(0) \delta_{\pr \M} ~. \cr}
\eqno (B.19) $$

For the next order in the expansion in (1.11) we need to calculate similar
results involving $G^B_1$ and ${\bar G}^B_1$ as given in (1.13). Although the
labour involved increases significantly similar methods may be
applied, in particular when there is a subdivergence introducing the
regulator $\ep$ as in (B.16) so that we can take now $f(y) \to 1$ and evaluate
all integrals before letting $\ep \to 0$.
$$ \eqalignno {
\int \! \! \int & \! dy\, dy' \, f(y) f(y') \,
G_1(u)^2 \, {\bar G}^B_1 \cr
& \sim {\mu^{-2\vep}\over \vep} \, {1\over (16\pi^2)^2 } \, f(0)^2
\bigl ( {\ts {1\over 9}} K\P_- + {\ts {2\over 3}} K\P + 2\psi \bigl)
\delta_{\pr \M} ~, & (B.20a) \cr
\int \! \! \int & \! dy\, dy' \, f(y) f(y') \,
G_1(u)G_1(v) \, {\bar G}^B_1 \cr
& \sim {\mu^{-2\vep}\over \vep} \, {1\over (16\pi^2)^2 } \, f(0)^2
{\ts {1\over 3}}\bigl ( {\ts {5\over 8}} K\P_- + {\ts {1\over 3}}\pi^2
K\P + \pi^2 \psi \bigl) \delta_{\pr \M} ~, & (B.20b) \cr
\int \! \! \int & \! dy\, dy' \, f(y) f(y') \,
\Bigl ( G_1(v)^2 - {\mu^{-\vep} \over 8\pi^2 \vep} \, \delta (v) \,
\delta_{\pr \M} \Bigl ) {\bar G}^B_1 \cr
& \sim - {\mu^{-2\vep}\over (16\pi^2\vep)^2 } \, f(0)^2
\bigl ( {\ts {1\over 3}}(1-{\ts {11\over 12}}\vep) K\P_- + {\ts {4\over 3}}
(1 - {\ts {1\over 6}}\vep) K\P + 4 \psi \bigl)
\delta_{\pr \M} ~, & (B.20c) \cr
\int \! \! \int & \! dy\, dy' \, f(y) f(y') \, G_1(u)^2 \, G^B_1 \cr
& \sim {\mu^{-2\vep}\over \vep} \, {1\over (16\pi^2)^2 } \, f(0)^2 \,
\quar K \, \delta_{\pr \M} ~, & (B.20d) \cr
\int \! \! \int & \! dy\, dy' \, f(y) f(y') \, G_1(v)^2 \, G^B_1 \cr
& \sim {\mu^{-2\vep}\over 6\ep} \, {1\over (16\pi^2)^2 } \Bigl \{
\int_0^\infty \! \! dy \, yf(y)^2 \bigl ( K\hnab^2 + 2 K^{ij} \hnab_i
\hnab_j \bigl ){} - \int_0^\infty \! \! dy \, yf'(y)^2 K \Bigl \}
\delta_{\pr \M} \, , \cr
& & (B.20e) \cr
\int \! \! \int & \! dy\, dy' \, f(y) f(y') \,
\Bigl ( G_1(v) G_1^B - {\mu^{-\vep} \over 16\pi^2 \vep} \, yK\, \delta (v) \,
\delta_{\pr \M} \Bigl ) G_1 (u) \cr
& \sim - {\mu^{-2\vep}\over (16\pi^2\vep)^2 } \, f(0)^2 \,
\half ( 1 - \quar \vep ) K \, \delta_{\pr \M} ~. & (B.20f) \cr}
$$
The terms on the r.h.s of (B.20e) correspond to the expansion of $\nab^2
\delta^d$ around $y,y'=0$ to this order in accord with (1.8), $\de^d (x,x')
= (1+yK + \dots ) \de (y-y') \de_{\pr \M} (\bx,\bx')$.

For the remaining one particle reducible graph in fig.1c it is
necessary because of the form of the counterterms to consider the Neumann and
Dirichlet cases separately. The basic Green function may be written as
$$
G_\Del (x,x') = (1- \P ) G^D (y,y') (1- \P) +
\P G^N (y,y') \P  + \dots ~,
\eqno (B.21) $$
where other contributions are irrelevant in the following discussion and
the explicit dependence on $\bx,\bx'$ is suppressed. In the Neumann case
it is sufficient to investigate integrals over $y,y'$ of the
essential form, assuming $\rho_d(y) = f(y) (e_d + y f_d ) $ where $e_d (\bx) ,
\, f_d (\bx)$ are given by (B.7) for $\P\to 1$ to the first order in an
expansion in
$\vep$ and $f(y)$ is a suitable test function as considered previously,
$$ \eqalign {
I^N  = {}& \int \! \! \int \! dy\, dy' \, y_+^{2-d} \, y^{\prime 2-d}_+
\, \rho_d(y)  \, G^N (y,y') \,\rho_d (y') \cr
& - {\mu^{-\vep}\over \vep}\biggl \{ \int \! dy \, y_+^{2-d} \rho_d (y) \,
G^N (y,0) \bigl ( \rho_4 \! {}'(0) - \rho_4 (0) \psi\bigl ) \cr
& ~~~~~~~~~~~~ + \int \! dy' \, y^{\prime 2-d}_+ \,
\bigl ( \rho_4 \! {}'(0) - \rho_4 (0) \psi\bigl )
G^N (0,y') \, \rho_d (y') \biggl \} \cr
& + {\mu^{-2\vep}\over \vep^2}\,
\bigl ( \rho_4 \! {}'(0) - \rho_4 (0) \psi\bigl ) G^N (0,0)
\bigl ( \rho_4 \! {}'(0) - \rho_4 (0) \psi\bigl ) ~. \cr}
\eqno (B.22) $$
As remarked in section 3 for the regular part of $G^N$, using the
appropriate boundary conditions, the counterterms in (B.22) may be integrated
by
parts to give
$$
I^{N,{\rm {reg}}} = \int \! \! \int \! dy\, dy' \, D(y)
\, G^{N,{\rm {reg}}} (y,y')\,  D(y') \, ,~~ D(y) =  y_+^{2-d}\rho_d (y) +
{\mu^{-\vep} \over \vep} \de'(y) \rho_4(y)  ~,
\eqno (B.23) $$
where by virtue of (B.2) $D(y)$ is regular, regarded as a distribution
on smooth test functions, as $\vep \to 0$. In order to calculate the
poles in $\vep$ arising from this one particle reducible amplitude then
it is necessary to only consider the leading singular contributions to $G^N$
when $y,y'\to0$ and also $\bx'\to \bx$, as exhibited in (1.13) or (1.15).

The most singular term in (1.13) corresponds to taking
$G^N(y,y') \to G_1(v) + G_1(u)$ and also letting $\psi \to 0$ in (B.22).
Analysing the integrals over $y,y'$ in this case gives an expansion in
powers of $\hsi$ of the form
$$ \eqalign {
I^N_a = {}& \alpha \, \rho_d(0)^2 (2\hsi)^{4 - {3\over 2}d}\cr
& - {1\over \vep} \, {1\over S_{d-1}} \Bigl ( \gamma
\bigl ( \rho_d(0) \rho_d \! {}' (0) + \rho_d \! {}' (0) \rho_d(0) \bigl )
(2\hsi)^{{9\over 2} - {3\over 2}d} \cr
&~~~~~~~~~~~~~~~~~~ - \delta  \bigl ( \rho_d(0) \rho_4 \! {}' (0) +
\rho_4 \! {}' (0) \rho_d(0) \bigl) (2\hsi)^{{5\over 2} - d} \Bigl ) {} + \dots
{}~, \cr}
\eqno (B.24)
$$
for
$$ S_d = {2\pi^{\hh d}\over \Gamma ( \hh d)} ~.
$$
The remaining terms not shown in (B.24) are less singular as $\hsi\to 0$
and do not lead to poles in $\vep$. The numerical coefficients
$\alpha , \, \gamma $ and  $\delta$ may be easily
evaluated in terms of gamma functions for general $d$ and as $\vep \to 0$
$$
\alpha = {\rm O} (1) ~,~~~~\gamma \sim 1 - \vep ~,~~~~
\delta \sim \mu^{- \vep}  (1 + \O(\vep^2)) ~.
\eqno (B.25) $$
Hence the poles in $\vep$ appear to cancel as expected by virtue of the
counterterms present in (B.22). Nevertheless if
in the expression obtained in (B.24) $(2\hsi)^{- s}$ is  regarded
as a distribution on $\pr \M$ defined by analytic continuation in $s$ from
$s< \half (d-1)$ then
it is non singular when $s\approx 2$, as in the first term of (B.24),
but for $s \to \half (d-1)$ there is a pole given by
$$ \eqalign {
(2\hsi)^{-\hh (d-1) + \hh \lambda} \sim {}& {1\over \lambda} \, S_{d-1} \,
\delta_{\pr \M} ~, \cr
{1\over \vep} \bigl ( (2\hsi)^{-\hh (d-1) + \hh a \vep } -
(2\hsi)^{-\hh (d-1) + \hh b \vep } \bigl ) {} \sim {} & {1\over \vep^2} \,
\Bigl ( {1\over a} - {1\over b} \Bigl ) S_{d-1} \, \delta_{\pr \M} + \O (1) ~.
\cr }
\eqno (B.26) $$
This result is a direct extension of (B.2) and (B.3).
Hence although $I^N_a$ is a one particle reducible amplitude including
counterterms for all sub-divergences it
still results in a local divergence represented here by poles in
$\vep$ of the form
$$
I^N_a \sim {\mu^{-2\vep}\over (16\pi^2\vep)^2} \Bigl ( f(0) f'(0) \, (1+\vep)
\te_0 \te_0 + \half f(0)^2 \bigl ( (1+\vep) ( \te_0 \tf_0 + \tf_0 \te_0)
- \vep ( \te_0 \tf_1 + \tf_1 \te_0)\bigl ) \Bigl ) \de_{\pr \M} ~,
\eqno (B.27) $$
where we have inserted the explicit form of $\rho_d(y)$, with $f(y)$ an
arbitrary smooth test function, and also used $\te_1 =0$ from (B.7).
For the terms proportional to $\psi$ then in (B.22) from (1.13) we let
$G^N(y,y') \to 2\psi G_{1,0} (u)$ or in the $\psi$ dependent counterterms
$G^N(y,0) \to 2G_1(y), \, G^N(0,y') \to 2G_1(y')$.
In a similar fashion to the analysis leading to (B.24) and (B.27)
$$ \eqalignno {
I^N_b \sim {}&  {1\over \vep} \, \psi  \, {1\over S_{d-1}}
\Bigl ( 2 {\tilde {\gamma}} \, \rho_d(0)^2
(2\hsi)^{{9\over 2} - {3\over 2}d} - \delta \bigl ( \rho_d(0) \rho_4(0) +
\rho_4(0) \rho_d(0)\bigl ) (2\hsi)^{{5\over 2} - d} \Bigl ){} + \dots \cr
\sim {}& - {\mu^{-2\vep}\over (16\pi^2\vep)^2} \, \psi  \, f (0)^2 \,
\te_0 \te_0 \, \de_{\pr \M} ~,
& (B.28) \cr}
$$
where we have used (B.26) again with ${\tilde {\gamma}}\, , \,
\mu^{\vep}\delta = 1 + \O(\vep^2)$.
The analogous results for the $K$ dependent terms are obtained in (B.22) by
taking
$$ \eqalign {
G^N (y,y') \to &
{K\over d-1} \Bigl ( 2\hsi \bigl (G_0(v) + G_0(u)\bigl ) u  + 2 G_0(u) u
yy' + (d-2) G_{1,0}(u) \Bigl ) \cr
& +  X \Bigl ( K - (d-1) {K_{ij} \hsi^i \hsi^j \over 2\hsi} \Bigl )  ~. \cr}
\eqno (B.29) $$
The various additional terms represented by $X$, which may easily be found
from (1.15), may be discarded here as they do not result in any overall
divergence since on using
$$
(2\hsi)^{-\hh (d+1) + \hh \lambda} \hsi_i \hsi_j \sim {1\over \lambda} \,
{S_{d-1}\over d-1} \, {\hat \gamma}_{ij} \, \delta_{\pr \M} ~.
\eqno (B.30) $$
in conjunction with (B.26) ensures that the poles in $\vep$ cancel.
In this case the counterterms in (B.22) are also irrelevant in obtaining the
leading singular term as $\hsi \to 0$ and we obtain
$$
I^N_c = {K \over d-1} \, {1\over S_{d-1}} \,
\kappa \, \rho_d(0)^2\, (2\hsi)^{{9\over 2} - {3\over 2}d}
\sim {\mu^{-2\vep}\over (16\pi^2\vep)^2} \, \half \vep  K \, f(0)^2 \,
\te_0 \te_0 \, \de_{\pr \M} ~,
\eqno (B.31) $$
since $\kappa = 3 + {\rm O}(\vep)$ (individual contributions to $\kappa$
possess poles in $\vep$ but these cancel when combined together).
Combining this with (B.27) and (B.28) gives finally for the singular part of
the integral in (B.22) $I^N = I^N_a + I^N_b + I^N_c$.

In the corresponding Dirichlet case we are concerned with integrals
of the form
$$ \eqalign {
I^D  = {}& \int \! \! \int \! dy\, dy' \,
y_+^{2-d} y^{\prime 2-d}_+ \, \rho_d (y) \, G^D (y,y')\, \rho_d (y')  \cr
&  - {\mu^{-\vep}\over \vep} \biggl \{\int \! dy \, y_+^{2-d}\, \rho_d (y) \,
G^D(y,y') \overleftarrow{\pr} \! {}_{y'} \bigl |_{y'=0} \, \rho_4(0) \cr
& ~~~~~~~~ + \int \! dy' \, y^{\prime 2-d}_+ \, \rho_4 (0) \,
\pr_y G^D(y,y') \bigl |_{y=0} \, \rho_d (y') \biggl \}
\cr
& + {\mu^{-2\vep} \over \vep^2} \, \rho_4(0) \, \pr_y G^D (y,y')
\overleftarrow \pr \! {}_{y'} \bigl |_{y=y'=0}\, \rho_4(0) ~ , \cr}
\eqno (B.32) $$
where $\rho_d(y) = f(y) (e_d + y f_d ) $ as before but now $\P\to 0$.
The analysis of the singularities of $I^D$ follows  similarly to
the previous discussion for $I^N$. Again for the regular part of
$G^D$ it is possible to write (B.32) in the analogous form to (B.23),
assuming $G^{D,{\rm {reg}}} (0,y') = G^{D,{\rm {reg}}} (y,0) = 0$,
showing that poles in $\vep$ can only arise from the singular short
distance part of $G^D$ as can be obtained from (1.13) or (1.15).
In this case it is necessary to note that
$\pr_y G^D \overleftarrow{\pr}\! {}_{y'}|_{y=y'=0}$, which appears in the
$1/\vep^2$ counterterm in (B.32), involves terms proportional to
$(2\hsi)^{-\hh d}, \, (2\hsi)^{-\hh(d-1)}$ when $\bx'\to \bx$, as shown in
(C.25), which represent non integrable singularities on
integration over $\pr\M$. Furthermore the resulting divergences are not
regularised by analytic continuation in $d$. Elsewhere [17]
we have discussed how to deal with this problem, which also occurs in the
calculation of electrostatic energies on $\M$ when the potential is given on
$\pr \M$. It is sufficient to treat the singular short distance terms
appearing in $\pr_y G^D \overleftarrow{\pr}\! {}_{y'}|_{y=y'=0}$ as
distributions
but it is crucial if correct results are to be obtained to also include
an additional term involving $\de_{\pr \M}$ proportional to $K$ which is
shown in (C.25).

For the contributions arising for $G^D(y,y') \to G_1(v) - G_1(u)$ then instead
of (B.24)
$$ \eqalignno {
\! \! \! I^D_a = {1\over \vep^2} {2\over S_d} & \Bigl ( \alpha
\rho_d(0)^2  (2\hsi)^{4 - {3\over 2}d} \! - \beta
\bigl ( \rho_d(0) \rho_4(0)\! + \! \rho_d(0) \rho_4(0) \bigl ) (2\hsi)^{2 - d}
\! + \mu^{-2\vep} \! \rho_4(0)^2 (2\hsi)^{- {1\over 2}d} \Bigl ) \cr
+ {1\over \vep} {1\over S_{d-1}} & \Bigl ( \gamma \bigl ( \rho_d \! {}'(0)
\rho_d(0) \! + \! \rho_d(0) \rho_d \! {}'(0)\bigl )
(2\hsi)^{{9\over 2} - {3\over 2}d} \! - \delta \bigl ( \rho_d \! {}'(0)
\rho_4(0) \! + \! \rho_4(0)\rho_d \! {}'(0) \bigl )
(2\hsi)^{{5\over 2} - d} \Bigl ) \cr
+ \dots & ~, & (B.33) \cr }
$$
where the numerical coefficients here have the form
$$
\alpha , \, \gamma  \sim  1 - \vep  ~, ~~~~
\beta \sim \mu^{-\vep} (1 - \half \vep) ~,~~~~
\delta \sim \mu^{-\vep} (1 - \vep) ~.
\eqno (B.34) $$
In this case the counterterms contribute to both the leading singular parts
shown in (B.33). These are necessary to ensure that only poles in $\vep$ with
local residues may appear. Using equations (B.26) and (B.34), the result
becomes, with the explicit form of $\rho_d(y)$ again,
$$
I^D_a \sim - {\mu^{-2\vep}\over (16\pi^2\vep)^2} \Bigl ( f(0) f'(0) \,
(1-\vep) \te_0 \te_0 + \half f(0)^2 \bigl ( (1-\vep)
( \te_0 \tf_0 + \tf_0 \te_0) + \vep ( \te_0 \tf_1 + \tf_1 \te_0)\bigl )
\Bigl ) \de_{\pr \M} ~,
\eqno (B.35) $$
For the $K$ dependent terms then using instead of (B.29)
$$
G^D (y,y') \to
{K\over d-1} \Bigl ( 2\hsi \bigl (G_0(v) - G_0(u)\bigl ) u  - 2G_0(u) u
yy' \Bigl ) {}
+  X' \Bigl ( K - (d-1) {K_{ij} \hsi^i \hsi^j \over 2\hsi} \Bigl )  ~,
$$
yields, since no relevant contributions result from
$X'$, in the Dirichlet case
$$ \eqalign {
I^D_b = {}&  {1\over \vep} \,  {K\over d-1} \, {1\over S_{d-1}} \Bigl (
2 \kappa \, \rho_d(0)^2 (2\hsi)^{{9\over 2} - {3\over 2}d}
- \lambda \bigl ( \rho_d(0) \rho_4(0) +
\rho_4(0) \rho_d(0)\bigl ) (2\hsi)^{{5\over 2} - d} \Bigl ) \cr
& +  {\mu^{-2\vep}\over \vep^2} \, {d-2\over 2(d-1) }\,  K \, \rho_4(0)^2 \,
\de_{\pr \M}  \ , \cr }
\eqno(B.36)
$$
where the last term is is a consequence of the extra $\de_{\pr \M}$
contribution shown in (C.25). The coefficients are
$ \kappa, \mu^\vep \lambda  \sim  1 - 2 \vep$ so that (B.36) gives
$$ I^D_b \sim {\mu^{-2\vep}\over (16\pi^2\vep)^2}\,
\half\vep K \, f(0)^2 \, \te_0 \te_0 \, \de_{\pr \M} ~.
\eqno(B.37)
$$
The total singular contribution resulting from (B.32) is then $I^D = I^D_a +
I^D_b$ as given by (B.35) and (B.37).

\vfill\eject

\vskip 5pt

\leftline{\bf Appendix C}

For the derivation of the functional Schr\"odinger equation in section
5 we here summarise the necessary results for variations induced by a
deformation of the boundary surface $\pr \M$. For $\pr \M$
parameterised by coordinates $\hx^i$ and the embedding in terms of the
coordinates for $\M$ specified by $x^\mu(\bhx)$ we define
$$
 e^\mu {}_{\! i} (\hbx ) = {\pr x^\mu \over \pr \hx^i} ~ , ~~~~  n_\mu
e^\mu {}_{\! i} = 0 ~, ~~~~~~ n_\mu n^\mu = 1 ~, \eqno(C.1)
$$
with $n^\mu(\bhx)$ the unit inward normal on $\pr \M$. For a
metric $g_{\mu \nu}$ on $\M$ the induced metric on $\pr \M$ is
determined by
$$
 \hga_{ij} = g_{\mu \nu} e^\mu {}_{\! i} e^\nu {}_{\! j} ~, \eqno(C.2)
$$
and then
$$\eqalign {
\nab_i  e^\mu {}_{\! j}  = {}& \pr_i e^\mu {}_{\! j}  + \Gamma^\mu_{\si \rho}
e^\si {}_{\! i} e^\rho {}_{\! j} =  e^\mu {}_{\! k}
{\hat \Gamma}^k_{ij}+ n^\mu K_{ij} ~,\cr
\nab_i n^\mu = {}&- K_{ij} e^{\mu j} ~ , ~~~~~~~~~~ \pr_i = e^\mu {}_{\! i}
\pr_\mu \cr} ~, \eqno(C.3)
$$
defines the extrinsic curvature $K_{ij} = K_{ji}$ of $\pr \M$ for
$\Gamma^{\mu}_{\si \rho}$ the usual Christoffel connection formed from
$g_{\mu \nu}$. If $\hnab_i$ is defined with the connection
${\hat {\Gamma}}^k_{ij} = {\hat {\Gamma}}^k_{ji}$ acting on tensors on
$\pr \M$ then $\hnab_i \hga_{jk} = 0$ so that ${\hat {\Gamma}}^k_{ij}$
is the Christoffel connection formed from $\hga_{jk}$. From (C.3) it is
straightforward to derive the Gauss-Codazzi equations relating the
Riemann curvature tensor $R_{\mu \nu \si \rho}$ for $x\in \pr \M$, with
zero and one component along the normal $n^\mu$, to the intrinsic
Riemann curvature ${\hat R}_{ijk\ell}$ of $\pr \M$ associated with the
covariant derivative $\hnab_i$ and also the extrinsic curvature $K_{ij}$.

For a variation $\de x^\mu = - n^\mu \de t$, as in (5.1), we let
$$ \eqalign {
\de' e^\mu {}_{\! i} ={}& \pr_i \de x^\mu - \de t \, n^{\si} \Gamma^\mu_{\si
\nu}
e^\nu {}_{\! i} \cr
= {}& \de t \, K_{ij} e^{\mu j} - n^\mu \pr_i \de t ~, \cr }
\eqno (C.4) $$
from (C.3) and hence, regarding $n_\mu$ as defined by (C.1),
$$
\de' n_\mu = \de n_\mu - n_\nu\Gamma^\nu_{\si \mu}n^\si \de t =
e_\mu {}^{\! i} \pr_i \de t ~ . \eqno(C.5)
$$
With the induced metric given by (C.2) we have
$$
\de \hga_{ij} = 2 K_{ij} \, \de t ~, \eqno(C.6)
$$
and using $K_{ij} = n_\mu \nab_i e^\mu{}_{\! j}$ from (C.3)
$$\eqalign {
\de K_{ij} = {}& e_\nu {}^{\! k} \pr_k \de t \, \nab_i e^\nu {}_{\! j} - n_\mu
[\nab_n,\nab_i] e^\mu {}_{\! j} + n_\mu \nab_i \de' e^\mu {}_{\! j} \cr
= {}& -\hnab_i \pr_j \de t - \bigl ( R_{njni} - K_{jk} K^k {}_{\! i}\bigl )
\de t ~. \cr}
\eqno(C.7)
$$
For integrals over a local scalar function $f$ on $\M$ or, restricted
to the boundary, on $\pr \M$
$$\eqalignno {
 \de \int_\M \!  dv\, f ={}& \int_{\pr \M} \! \!  dS\, \de t \, f |_{\pr
\M} ~,  & (C.8a)\cr
\de \int_{\pr \M} \! \!  dS \, f |_{\pr \M} ={}& \int_{\pr \M}\! \!  dS \,
\de t \bigl ( -\pr_n f + K f \bigl ) \bigl |_{\pr \M} ~, & (C.8b) \cr }
$$
neglecting in (C.8b) any dependence of $f$ on $n_\mu, \hga_{ij}, K_{ij}
\dots$  where it would be necessary to use results such as (C.4,5,6,7) above.

The deformation of the boundary induces a variation in the Green
function $G_\Del$ and the heat kernel $\G_\Del$ since boundary
conditions are an essential part of their definitions in (1.4) and
(A.2).
In the Dirichlet case, with a choice of gauge so that $A_n =0$ as in (1.10),
$$
\bigl (\de \G_\Del (x,x';\tau ) - \de t(\hbx) \, \pr_n \G_\Del
(x,x';\tau) \bigl ) \bigl |_{x=x(\hbx)} = 0 ~, \eqno (C.9)
$$
and hence for arbitrary $x,x' \in \M$
$$
\de \G_\Del (x,x';\tau ) = \int_0^\tau \! d \tau' \,\int_{\pr \M} \! \!
dS \, \de t (\hbx) \, \G_\Del (x,\hx ; \tau -\tau'){\overleftarrow \pr}_{\! n}
\pr_n \G_\Del (\hx,x'; \tau') \bigl |_{\hx=x(\hbx)} ~. \eqno(C.10)
$$
Analogous results hold for  $G_\Del$ as a consequence of (A.1) and therefore
$$
{\de \over \de t (\hbx)} G_\Del (x,x') =
G_\Del (x,\hx) {\overleftarrow \pr}_{\! n}
\pr_n G_\Del (\hx,x') \bigl |_{\hx=x(\hbx)} ~.
\eqno(C.11) $$
Since by definition  $\G_\Del$ is the kernel for the operator $e^{-\tau
\Del}$ it follows from (C.10) that
$$
\de \Tr ( e^{- \tau \Del } ) = \tau \int_{\pr \M}\! \!  dS \, \de t(\hbx) \,
\tr \bigl ( \pr_n \G_\Del (x,x';\tau)  {\overleftarrow \pr}{}'{}_{\! \! n}
\bigl |_{x=x'=x(\hbx)} \bigl ) ~.
\eqno(C.12)
$$
In general the asymptotic form of the heat kernel is of the form
$$
(4 \pi \tau )^{\hh d} \Tr ( e^{- \tau \Del } ) \sim \int_\M  \! dv \,
\sum_{n=0} B_{2n} (x) \tau^n + \int_{\pr \M}\! \!  dS\, \sum_{n=1} {\hat
{B}}_n (\hbx)\tau^{\hh n} ~, \eqno(C.13)
$$
and we may also write
$$
\tau (4 \pi \tau )^{\hh d} \pr_n \G_\Del (x,x';\tau)
{\overleftarrow \pr} \!{}'{}_{\! \!  n}
\bigl |_{x=x'=x(\hbx)}{} \sim \sum_{n=0} \hB_n (\hbx) \tau^{\hh n} ~,
\eqno(C.14)
$$
defining $\hB_n(\bhx)$ on $\pr \M$. As a consequence of (C.12)
$$
B_{n} + {\de \over \de t} \int_{\pr \M}\! \!  dS\, {\hat {B}}_n = \tr
(\hB_n ) ~, \eqno(C.15)
$$
with $B_n =0 $ for $n$ odd. This result
may be verified using our previous results [13] for $n=0,1,2$,
$$
\hB_0 =1 ~,~~~~ \hB_1 = -\half \sqrt{\pi} K ~,~~~~ \hB_2 =
\six ( \hR + K^2 - K^{ij} K_{ij} ) - X |_{\pr \M} ~, \eqno(C.16)
$$
with ${\hat {R}}$ the scalar curvature on $\pr \M$. For the discussion
in section 5 we need the contribution to  $\hB_4$ given by
$$
\hB_4 = \bigl ( \half X^2 - \half \nab_n {}^{\! \! 2} X - \six \hD^2 X + {\ts
{5\over 6}} K \pr_n X \bigl ) \bigl |_{\pr \M} + \dots ~, \eqno (C.17)
$$
where on $\pr \M$ $\nab_n {}^{\! \! 2} X = n^\mu n^\nu \nab_\mu \nab_\nu X$.
This has been determined by direct calculation and also checked
by using (C.15).

For the functional determinant defined by (1.15) the finite regularised
form in four dimensions is given by
$$
{\rm ln} \det \Delta_{\rm reg} = {\rm ln} \det \Delta +
{2\mu^{-\vep} \over 16\pi^2\vep} \biggl \{ \int_\M  \! dv \,
B_{4} + \int_{\pr \M}\! \!  dS\, {\hat {B}}_4 \biggl \} ~,
\eqno(C.18) $$
where $B_4 , \, {\hat B}_4$ are given by (1.16). For general boundary
conditions, with $\Delta$ given by (1.2),
it is easy to see that, under smooth variations in $X$,
$$
\de \, {\rm ln} \det \Delta_{\rm reg} = \int_\M \! dv \, \tr ( D_\Del \de X) ~,
\eqno (C.19) $$
where $D_\Del(x)$, as given by (B.1), (B.7) and (B.12), has the form
$$\eqalign {
D_\Del(x) = G_\Del (x,x) & + {2\mu^{-\vep} \over 16\pi^2\vep} \bigl ( X(x) -
\six R (x) \bigl ) \cr
& + {\mu^{-\vep} \over 16\pi^2\vep} \Bigl ( \nab_\mu \bigl (
n^\mu (\bhx) \P_- (\bhx) \de_n (x) \bigl) - \bigl ( 4\psi(\bhx) +
{\ts {2\over 3}} K(\bhx) \bigl ) \de_n (x) \Bigl ) ~, \cr}
\eqno (C.20) $$
with $\de_n (x)$ the surface $\de$ function defined by $dv(x) \de_n(x)
= dS(\bhx)$, in the standard coordinates of (1.1) $\de_n(x) = \de (y)$,
$\nab_\mu \bigl (n^\mu (\bhx) \de_n (x) \bigl) = \de'(y) - K(\hbx) \de(y)$.
This represents the
regularised form of $G_\Delta (x,x)$ in dimensional regularisation with both
short distance and boundary divergences removed by subtraction of poles in
$\vep$. In the Dirichlet case, when $\P_- \to 1$, we may also find for
variations of the boundary, from (C.12) and (C.15),
$$
{\de \over \de t (\hbx)} {\rm ln} \det \Delta_{\rm reg} =
- \tr \bigl ( \K_\Del (\bhx, \bhx )  \bigl ) {} +
{2\mu^{-\vep} \over 16\pi^2\vep} \, \tr \bigl ( \hB_4 (\hbx) \bigl ) ~,
\eqno (C.21) $$
where formally
$$ \K_\Del (\bhx, \bhx' ) =  \pr_n G_\Del (x,x')
{\overleftarrow \pr}{}'{}_{\! \! n} \bigl |_{x=x(\hbx),x'=x(\hbx')} ~.
\eqno (C.22) $$

For applications in section 5 it is necessary to also consider variations of
the Green function when arguments lie on the boundary. By careful analysis,
assuming ${\hat f}(\bhx)$ and $f(x)$ are smooth test functions on $\pr \M$ and
$\M$,
$$ \eqalign {
{\de \over \de t (\bhx)} \int \! dS' dv'' & \, {\hat f} (\hbx')
\pr' {}_{\! \! n} G_\Del ( x',x'' ) \bigl |_{x'=x(\bhx')} f(x'') \cr
= {}&  \int \! dS' dv''  \, {\hat f} (\hbx') \K_\Del ( \bhx', \bhx) \,
\pr_n G_\Del ( x,x'' ) \bigl |_{x=x(\bhx)} f (x'')\cr
& ~~~~ + {\hat f} (\hbx) \, f(x(\hbx))  ~. \cr}
\eqno (C.23) $$
Also, in a similar fashion,
$$ \eqalign {
{\de \over \de t (\bhx)} \int \! dS'dS'' & \, {\hat f} (\hbx')
\K_\Del (\bhx',\bhx'') {\hat f} (\hbx'') \cr
= {}&  \int \! dS'dS'' \, {\hat f} (\hbx') \K_\Del ( \bhx', \bhx) \,
\K_\Del (\bhx, \bhx'') {\hat f} (\hbx'') \cr
& ~~~~ -{\hat \gamma}^{ij}(\hbx) \pr_i {\hat f} (\hbx) \pr_j {\hat f} (\hbx)
- {\hat f} (\hbx)  X(x(\hbx)) {\hat f} (\hbx) ~, \cr}
\eqno (C.24) $$
which is analogous to a result quoted by Symanzik\footnote{*}{See eq. (6.6c)
in ref. [10]}.

In general the coefficients $\hB_n$ are related to the short distance
expansion of $\K_\Del (\bhx, \bhx' )$ when $\bhx' \approx \bhx$. We obtain
$$\eqalign {
\K_\Del \sim {}&
{\Gamma(\hh d) \over \pi^{\hh d}} (2\hsi)^{-\hh d} {\hat {I}}
+ {d-2 \over 2(d-1)} \, K \, \de_{\pr \M} \cr
& - {\Gamma(\d2 -\half) \over 4 \pi^{\hh d -\hh}}
(2 \hsi)^{\hh - \hh d}\Bigl (
K -(d-1) {K_{ij}\hsi^i \hsi^j \over 2 \hsi} \Bigl ) {\hat {I}} ~. \cr }
\eqno(C.25) $$
The term proportional to $\de_{\pr \M}$ results from a careful treatment of
the singular behaviour of $\K$ at coincident points [17]. The result given is
compatible with treating $(2\hsi)^{-s}$ as a distribution [21] or in
integrating the kernel $\K(\bhx,\bhx')$ with a smooth function ${\hat
f}(\bhx')$
excluding a disc of radius $c$ centred on $\bhx$ and subtracting a term
$\propto  c^{-1}$ to ensure a finite result as $c\to 0$.

\vfill\eject

\leftline {\bf References}
\vskip 15pt
\item{[1]} H.B.G. Casimir, {\it Proc. Kon. Nederl. Akad. Wet.} {\bf B51}
(1948) 793;
\item{} H.B.G. Casimir and D. Polder, {\it Phys. Rev.} {\bf 73} (1948) 360.
\vskip 7pt
\item{[2]}G. Plumien, B. Muller and W. Greiner, {\it Phys. Reports} {\bf
134} (1986) 87.
\vskip 7pt
\item{[3]}S. Blau, M. Visser and A. Wipf,  {\it Nucl. Phys.} {\bf B310}
(1988) 163;
\item{} E. Elizalde and A. Romeo, {\it Int. J. Mod. Phys.} {\bf A5} (1990)
1653;  {\it J. Math. Phys.} {\bf 30} (1989) 1133, (E) {\bf 31} (1990) 31.
\vskip 7pt
\item{[4]}H.W. Diehl, `Phase Transitions and Critical Phenomena',
vol 10, p. 75, (C. Domb and J.L. Lebowitz eds.) Academic Press,
London (1986).
\vskip 7pt
\item{[5]}H.W. Diehl, and S. Dietrich, {\it Zeit. f. Phys. B} {\bf 42}
(1981) 65, (E) {\bf 43} (1981) 281; {\bf 50} (1983) 117.
\vskip 7pt
\item{[6]} A. Chodos, R.L. Jaffe, K. Johnson, C.B. Thorn and V.F.
Weisskopf, {\it Phys. Rev.} {\bf D9} (1974) 3471;
\item{} A. Chodos and  C.B. Thorn, {\it Phys. Lett.} {\bf 53B} (1974) 359;
\item{} T.H. Hansson and R.L. Jaffe, {\it Phys. Rev.} {\bf D28} (1983) 882;
\item{} S.N. Goldhaber, T.H. Hansson and R. L. Jaffe,  {\it Phys. Lett.}
{\bf 131B} (1983) 445;  {\it Nucl. Phys.} {\bf B277} (1986) 674;
\item{} J. Baacke and Y. Igarashi,  {\it Phys. Rev}. {\bf D27} (1983) 460.
\vskip 7pt
\item{[7]} R. Balian and C. Bloch, {\it Ann. Phys. (N.Y.)} {\bf 60}
(1970) 401; {\bf 64} (1972) 271; {\bf 69} (1974) 76; (E) (1974) {\bf 84} 554;
\item{} R. Balian and B. Duplantier,  {\it Ann. Phys.  (N.Y.)} {\bf 104}
(1977) 300; {\bf 112} (1978) 105;
\item{} T.H. Hansson and R.L. Jaffe, {\it Ann. Phys. (N.Y.)} {\bf 151}
(1983) 204;
\item{} R.L. Jaffe and L.C. Vintro,  {\it Ann. Phys. (N.Y.)} {\bf 162}
(1985) 212.
\vskip 7pt
\item {[8]} O. Alvarez,  {\it Nucl. Phys.} {\bf B216} (1983) 125;
\item{} H. Dorn and H-J. Otto, {\it Zeit. f. Phys.} {\bf C 32} (1986) 599;
\item{} A. Abouelsaood, C.G. Callan, C.R. Nappi and S.A. Yost, {\it
Nucl. Phys.} {\bf B280} [FS18] (1987) 599;
\item {} C.G. Callan, C. Lovelace, C.R. Nappi, and S.A. Yost, {\it Nucl.
Phys.} {\bf B288} (1987) 525;
\item{} A.A. Tseytlin, {\it Nucl. Phys.} {\bf B276} (1986) 391;
\item{} O.D. Andreev and A.A. Tseytlin, {\it Mod. Phys. Lett.} {\bf 3}
(1988) 1349;
\item{} C.G. Callan and L. Thorlacius, {\it Nucl. Phys.}
{\bf B319} (1989) 133; {\bf B329} (1990) 117;
\item{} H. Luckock, {\it Ann. Phys. (N.Y.)} {\bf 194} (1989) 113;
\item{} Z. Jask\'olski, {\it Comm. Math. Phys.} {\bf 128} (1990) 285.
\vskip 7pt
\item{[9]} J.L. Cardy, {\it Nucl. Phys.} {\bf B240} [FS12] (1984) 514;
\item{} J.L. Cardy, {\it Nucl. Phys.} {\bf B275} [FS17] (1986) 200;
\item{} J.L. Cardy, {\it Nucl. Phys.} {\bf B324} (1989) 581;
\item{} J.L. Cardy, {\it Phys. Rev. Lett.} {\bf 65} (1990) 1443;
\item{} J.L. Cardy and D.C. Lewellen, {\it Phys. Lett.} {\bf 259B} (1991)
274;
\item{} D.C. Lewellen, {\it Nucl. Phys.} {\bf B372} (1992) 654.
\vskip 7pt
\item{[10]} K. Symanzik, {\it Nucl. Phys.} {\bf B190} [FS3] (1981) 1;
\item{} M. L\"uscher, {\it Nucl. Phys.} {\bf B254} (1985) 52.
\vskip 7pt
\item{[11]} M. L\"uscher,  K. Symanzik and P. Weisz,  {\it Nucl. Phys.} {\bf
B173} (1980) 365.
\vskip 7pt
\item{[12]} J.B. Hartle and S.W. Hawking, {\it Phys. Rev.} {\bf D28} (1983)
2960;
\item{} J.L. Halliwell and J.B. Hartle, {\it Phys. Rev.} {\bf D43} (1991)
1170.
\vskip 7pt
\item{[13]} D.M. McAvity and H. Osborn, {\it Classical and Quantum Gravity},
{\bf 8} (1991) 603; 1445; (E) {\bf 9} (1991) 317;
\item{} D.M. McAvity, {\it Classical and Quantum Gravity}, in press,
Cambridge preprint, DAMTP 92-17.
\vskip 7pt
\item{[14]} B.S. DeWitt, `Dynamical Theory of Groups and Fields', Gordon and
Breach, London (1965).
\vskip 7pt
\item{[15]} D.I. Jack and H. Osborn, {\it Nucl. Phys.} {\bf B324} (1984) 331.
\vskip 7pt
\item{[16]} I.G. Moss and J.S. Dowker, {\it Phys. Lett.} {\bf 229B} (1989)
261;
\item{} J.S Dowker and J.P. Schofield, {\it J. Math. Phys.} {\bf 31} (1990)
808;
\item{} T.P. Branson and P.B. Gilkey, {\it Comm. Partial Diff. Eqs.} {\bf 15}
(1990) 245;
\item{} A. Dettki and A. Wipf, preprint ETH-TH/91-19.
\vskip 7pt
\item{[17]} D.M. McAvity and H. Osborn, {\it Journal of Physics A},
{\bf 25} (1992) 3287.
\vskip 7pt
\item{[18]} D.I. Jack and H. Osborn, {\it Nucl. Phys.} {\bf B343} (1990) 647.
\item{} H. Osborn, {\it Nucl. Phys.} {\bf B363} (1991) 486.
\vskip 7pt
\item{[19]} J. Polchinski, {\it Nucl. Phys.} {\bf B303} (1988) 226.
\vskip 7pt
\item{[20]} L.S. Schulman, `Techniques and Applications of Path Integrals',
John Wiley and Sons, New York (1981).
\vskip 7pt
\item{[21]} I.M. Gel'Fand and G.E. Shilov, `Generalised Functions', volume 1,
(Academic Press, 1964).
\vskip 7pt
\item{[22]} H.W. Diehl and A. Ciach, {\it Phys. Rev.} {\bf B44} (1991) 6642.
\vskip 7pt
\item{[23]} M. L\"uscher, R. Narayanan, P. Weisz and U. Wolff, Desy preprint,
DESY 92-025.
\bye